\documentclass{aa} 
\usepackage{graphicx}
\usepackage{txfonts}
\usepackage{subfigure}
\usepackage{natbib}
\usepackage{multirow}

\begin{document}
   \title{SiO emission from low- and high-velocity shocks in Cygnus-X massive dense clumps}

   \subtitle{}

\author{A. Duarte-Cabral\inst{1,2,3}
  \and S. Bontemps\inst{2,3}
  \and F. Motte\inst{4}
  \and A. Gusdorf\inst{5}
  \and T. Csengeri\inst{6}
  \and N. Schneider\inst{2,3}
  \and F. Louvet\inst{4}
  }

\offprints{Ana Duarte Cabral, \email{adc@astro.ex.ac.uk}}

 \institute{School of Physics and Astronomy, University of Exeter, Stocker Road, Exeter EX4 4QL, UK
 \and Univ. Bordeaux, LAB, UMR 5804, F-33270, Floirac, France
 \and CNRS, LAB, UMR 5804, F-33270, Floirac, France
 \and Laboratoire AIM, CEA/DSM-CNRS-Universit\'e Paris Diderot, IRFU/Service d'Astrophysique, C.E. Saclay, Orme de merisiers, 91191 Gif-sur-Yvette, France
 \and LERMA, UMR 8112 du CNRS, Observatoire de Paris, Ecole Normale Sup\'{e}rieure, 24 rue Lhomond, F-75005, Paris, France
 \and Max Planck Institute for Radioastronomy, Auf dem H\"{u}gel 69, 53121 Bonn, Germany}

\date{Received 19 February 2014; accepted 4 July 2014}

 
  \abstract
   {The SiO molecule is formed through interstellar shocks and is often used as a tracer of high-velocity shocks from protostellar outflows. However, recent studies have suggested that low-velocity shocks in the interstellar medium can be responsible for a significant amount of SiO emission observed in star-forming regions.}
   {We aim to investigate the existence of SiO emission that may or may not be associated with outflow shocks, within several massive dense clumps (MDCs), and explore how the SiO luminosities compare with the outflow properties estimated using CO.}
   {We used observations of SiO (2-1) from the Plateau de Bure Interferometre, towards a sample of six MDCs in Cygnus-X, and compared them to the CO high-velocity outflow emission, and to the velocity shears found in these regions.}
   {We find that most molecular outflows are detected in both SiO and CO, although there are some cases of CO outflows with no SiO counterpart. The data also shows narrow line SiO emission ($\sigma_{\varv} \lesssim 1.5$~km~s$^{-1}$) which, in some cases, appears to be unrelated to outflows. The kinematics of this narrow emission often differs from those found by other high-density tracers such as H$^{13}$CO$^{+}$, and its extent varies from rather compact ($\sim 0.03$\,pc) to widespread ($\sim 0.2$\,pc). We find that the least centrally concentrated clumps with the least massive protostellar cores have the most widespread narrow SiO emission. The fraction of the total SiO luminosity that is not associated with outflows is highly variable in the different MDCs (from 10\% to 90\%); this might be a problem when extrapolating outflow properties from SiO luminosities without resolving individual outflows.}
   {In line with previous {evidence of SiO emission associated with low-velocity shocks, we propose an evolutionary picture to explain the existence and distribution of narrow SiO line profiles. In this scenario, the least centrally condensed MDCs are at an early stage where the SiO emission traces shocks from the large-scale collapse of material onto the MDC. This could be the case of CygX-N40, a MDC with a low-mass protostar at its centre, a weak outflow, and where 90\% of the SiO emission is narrow and arises from the outskirts. As the MDC collapses, the SiO emission becomes more confined to the close surroundings of cores, tracing the post-shock material from the infalling MDC against the dense cores, such as in the small-scale converging flows of CygX-N3, N12, and N48. At later stages,} when single massive protostars are formed, as for instance in CygX-N53 and N63, the SiO luminosity is largely dominated by powerful outflows, and the weaker narrow component shows perhaps the last remnants of the initial collapse.} 

   \keywords{Stars: formation, massive, protostars; ISM: jets and outflows, kinematics and dynamics, molecules: SiO }

   \maketitle
%

\section{Introduction}
\label{intro}

The SiO molecule is the most abundantly observed silicon-bearing molecular species in the interstellar medium (ISM), and it can be produced by the photo-evaporation of icy grain mantles \citep[e.g.][]{1991ApJ...376..573T,1999A&A...342..542W,2001A&A...372..291S,2013A&A...550A...8G} or by the destruction of grain cores in shocks \citep[e.g.][]{1992A&A...254..315M,1997A&A...321..293S,2001A&A...372.1064L,2008A&A...482..809G,2008A&A...490..695G}. The high-velocity shocks from protostellar outflows can easily produce SiO, and therefore, observations of SiO emission are commonly used to trace such outflows \citep[e.g.][]{2002A&A...387..931B,2007ApJ...654..361Q,2011A&A...526L...2L}. Until recently, only fast molecular shocks were believed to be able to produce SiO by sputtering of the grain cores \citep[typically C-shocks with $\varv_{\rm s} > 25$~km~s$^{-1}$, but fast J-shocks can also form SiO, see e.g.][]{2009A&A...497..145G}. Nevertheless, we do observe significant SiO emission at low velocities and with a small velocity dispersion. For instance, single-dish SiO pointed observations towards the Cygnus-X massive dense clumps (MDCs) by \citet[][]{2007A&A...476.1243M} had shown evidence of two different gas components contributing to the SiO emission: a broad component tracing the shocked gas from protostellar outflows and a narrower one ($\sigma_{\varv} \lesssim 1.5$~km~s$^{-1}$, i.e. FWHM $\lesssim  3.5$~km~s$^{-1}$) interpreted at the time as possibly arising from the hot cores. More recently, some individual studies have detected narrow and wide-spread SiO emission in molecular clouds, even in regions where no feedback from the forming protostars has yet kicked in \citep[e.g.][Louvet et al. in prep.]{2010MNRAS.406..187J,2013ApJ...775...88N,2013ApJ...773..123S,2013ApJ...765L..35K}. The origin of SiO is in these cases enigmatic. 

\begin{table*}[!t]
\caption{Observing parameters, beam sizes, and rms.}
\label{obs_summary}
\renewcommand{\footnoterule}{}  
\begin{tabular}{c | c c c c c  | c c c}
\hline \hline
 Source          &    \multicolumn{2}{c}{Phase centre (J2000)}  & Synthesised beam & P.A. & r.m.s.$^{(a)}$   & \multicolumn{2}{c}{Number of fragments$^{(b)}$} & \multirow{2}{*}{Morphology}      \\
name     & R.A. ($^h$ $^m$ $^s$)     &  Dec. ($^\circ$ $'$ $''$)                   & [$'' \times ''$]      & [$^\circ$]    & [{\rm  mJy/beam}]  	& Total & \, \, High-mass &   \\
\hline
 CygX-N3    & 20 35 34.1   &  42 20 05.0      &  3.68$\,\times\,$3.02  &  63  &  12  &    4 	& \, \,  2 ($>$\,10\,$M_\odot$) 	& Elongated \\
 CygX-N12  & 20 36 57.4   &  42 11 27.5      &  4.09$\,\times\,$3.51  &  62  &  15  &    4	& \, \, 2 ($>$\,15\,$M_\odot$)   	& Elongated\\
 CygX-N40  & 20 38 59.8   &  42 23 42.0      &  3.64$\,\times\,$3.01  &  64  &  12  &    1 	& \, \, 0     					& Diffuse\\
 CygX-N48  & 20 39 01.5   &  42 22 04.0      &  4.25$\,\times\,$3.35  &  71  &  21  &    5	& \, \, 2 ($>$\,10\,$M_\odot$)	& Clustered \\
 CygX-N53  & 20 39 03.1   &  42 25 50.0      &  4.23$\,\times\,$3.33  &  70  &  21  &    7	& \, \,  2 ($>$\,20\,$M_\odot$)	& Centrally condensed\\
 CygX-N63  & 20 40 05.2   &  41 32 12.0      &  4.12$\,\times\,$3.69  &  46  &  15  &    3	& \, \,  1 ($>$\,40\,$M_\odot$) 	& Centrally condensed\\
\hline
\end{tabular}\\
$^{(a)}$ 1$\sigma$ r.m.s. estimated in 0.27 km\,s$^{-1}$ channels.\\
$^{(b)}$ From \citet[][]{2010A&A...524A..18B}.
\end{table*}

Some models suggest that the formation of SiO from low-velocity shocks could be linked to the presence of the element Si in the pre-shock medium.
If there is some Si already in the gas phase, shocks with velocities as low as $\sim$~5\,km\,s$^{-1}$ can produce SiO  \citep[e.g. ][]{2013ApJ...775...88N}. On the other hand, despite only considering silicates in grain cores, \citet[][]{1997A&A...322..296C} demonstrated that grain-grain collisions dominate over sputtering for evaporation of the ices on the grain mantles for low-velocity shocks ($10-15$\,km\,s$^{-1}$). In this context, if a fraction of Si is in the grain ice mantles, then shock velocities of $\sim$~10\,km\,s$^{-1}$ can desorb Si into the gas phase and effectively increase the abundances of SiO to match observations \citep[e.g.][]{2008A&A...490..695G,2008A&A...482..549J}. If the Si is solely in the grain cores, then shock velocities above 25\,km\,s$^{-1}$ are needed to form SiO \citep[e.g.][]{1997A&A...321..293S,2008A&A...482..809G}. However, none of these models include grain-grain interactions. To our knowledge, the most sophisticated account of these interactions has been given by the studies of \citet[][]{2007A&amp;A...476..263G,2009A&A...497..145G,2011A&A...527A.123G} and \citet[][]{2013A&A...556A..69A}. When included in shock models, these interactions (namely vaporisation, coagulation, and shattering) do significantly affect the structure of C-type shock waves propagating in high pre-shock density conditions ($n_{\rm H} > 10^5$\,cm$^{-3}$). Because they result in the production of numerous small grains, the modelled shock layers are both thinner and hotter. Ultimately, even if collisional vaporisation becomes more efficient than sputtering at destroying grains, especially at slightly lower shock velocities ($20-25$\,km\,s$^{-1}$), the overall result of these improved models is less bright and narrower SiO line profiles that do not match the observed intensities, unless a fraction of silicon-bearing material is placed in the grain mantles, similar to what \citet[][]{2008A&A...490..695G} had done.

It is vital to further investigate the origin of the measured SiO components, to be able to disentangle SiO outflows from other possible forms of shocks. However, most existing surveys observing SiO emission towards star-forming regions do not possess enough angular resolution or sensitivity to determine the exact origin of the SiO emission, and it becomes impossible to disentangle any emission that may not arise from outflows. On the one hand, ignoring other sources of shocks for the origin of SiO could introduce significant uncertainties on any outflow properties derived from SiO luminosities alone. On the other hand, pinpointing the exact origin of SiO emission can help to understand the physics and mechanisms for accretion at core-scales.

In this article, we present SiO $(2-1)$ interferometric observations towards a sample of IR-quiet MDCs in Cygnus-X, located at a distance of 1.4\,kpc \citep[][]{2012A&A...539A..79R}, which are known to have both broad and narrow SiO emission \citep[][]{2007A&A...476.1243M}, and for which we have detailed information about the protostellar content and the existing outflows \citep[][]{2013A&A...558A.125D}. We address the consequences of not resolving the SiO emission for inferring outflow properties, and we investigate the different possibilities for the origin of the observed SiO emission. Details on the source sample and observations are provided in Sect.~\ref{sample}. We describe our results in Sect.~\ref{sec:results}, which include a comparison of the spectral profiles of SiO with those obtained with single-dish observations, and a comparison of the SiO and CO outflow emission. We analyse the different trends observed with SiO and the emission morphologies in Sect.~\ref{sec:analyse}, and present our concluding remarks in Sect.~\ref{concl}.


\section{Source sample}
\label{sample}

\subsection{Cygnus-X massive dense clumps}
\label{sec:cygnusX}

We have studied the six IR-quiet MDCs in Cygnus-X, originally identified by \citet[][]{2007A&A...476.1243M} and more recently observed with the PdBI in the 1\,mm and 3\,mm continuum emission \citep[][]{2010A&A...524A..18B} and $^{12}$CO $(2-1)$ outflow emission \citep[][]{2013A&A...558A.125D}. Table~\ref{obs_summary} lists the name and properties of each MDC, including the number of fragments and morphology.

These MDCs are found in a range of different environments, and each field, with the exception of CygX-N40, is subfragmented into several cores with envelope masses larger than $\sim 10$ M$_{\odot}$  \citep[as estimated from the 1mm emission by][]{2010A&A...524A..18B,2013A&A...558A.125D}. The CygX-N3 and N12 MDCs are situated to the west of the DR21 ridge and they both show elongated morphologies, consistent with being shaped by the winds from the nearby OB clusters (to the N and NW). Both are fragmented into several compact cores, with two high-mass cores per field (see Table~\ref{obs_summary}). CygX-N40, N48, and N53 are located along the massive DR21 ridge, a highly dynamical star formation site \citep[][]{2010A&A...520A..49S,2012A&A...543L...3H}. The interferometric continuum data of CygX-N40 \citep[from][]{2010A&A...524A..18B} showed that this core is not centrally condensed, with its emission being extended and filtered out by the PdBI. \citet[][]{2010A&A...524A..18B} had detected a single low-mass fragment in this region, with $M_{\rm env} < 2 $\,M$_{\odot}$. CygX-N48 has a clustered star formation, not very centrally condensed, but with two massive compact sources, while CygX-N53 is centrally condensed and fragmented into two compact massive cores (with more than 20~M$_{\odot}$ each). Finally, CygX-N63 is a relatively isolated dense core to the south of DR21. It does not show signs of fragmentation and has nearly 50~M$_{\odot}$ within its inner $\sim$2500~AU, making it the most massive core of the sample.

In \citet[][]{2013A&A...558A.125D}, we confirmed the nature of these millimetre cores as high-mass Class 0-like protostars, based on their masses, luminosities, and outflow power/accretion rates. We identified the individual outflows powered by each massive protostar, and estimated their properties, concluding that the opening angles and morphologies are similar to those of low-mass objects, and they are 1-2 orders of magnitude more powerful, following a linear trend with the envelope mass.


\subsection{Observations}
\label{sec:obs}

\begin{figure*}[!t]
	\centering
	{\renewcommand{\baselinestretch}{1.1}
	\hspace{-0.7cm}
	\includegraphics[width=0.3\textwidth]{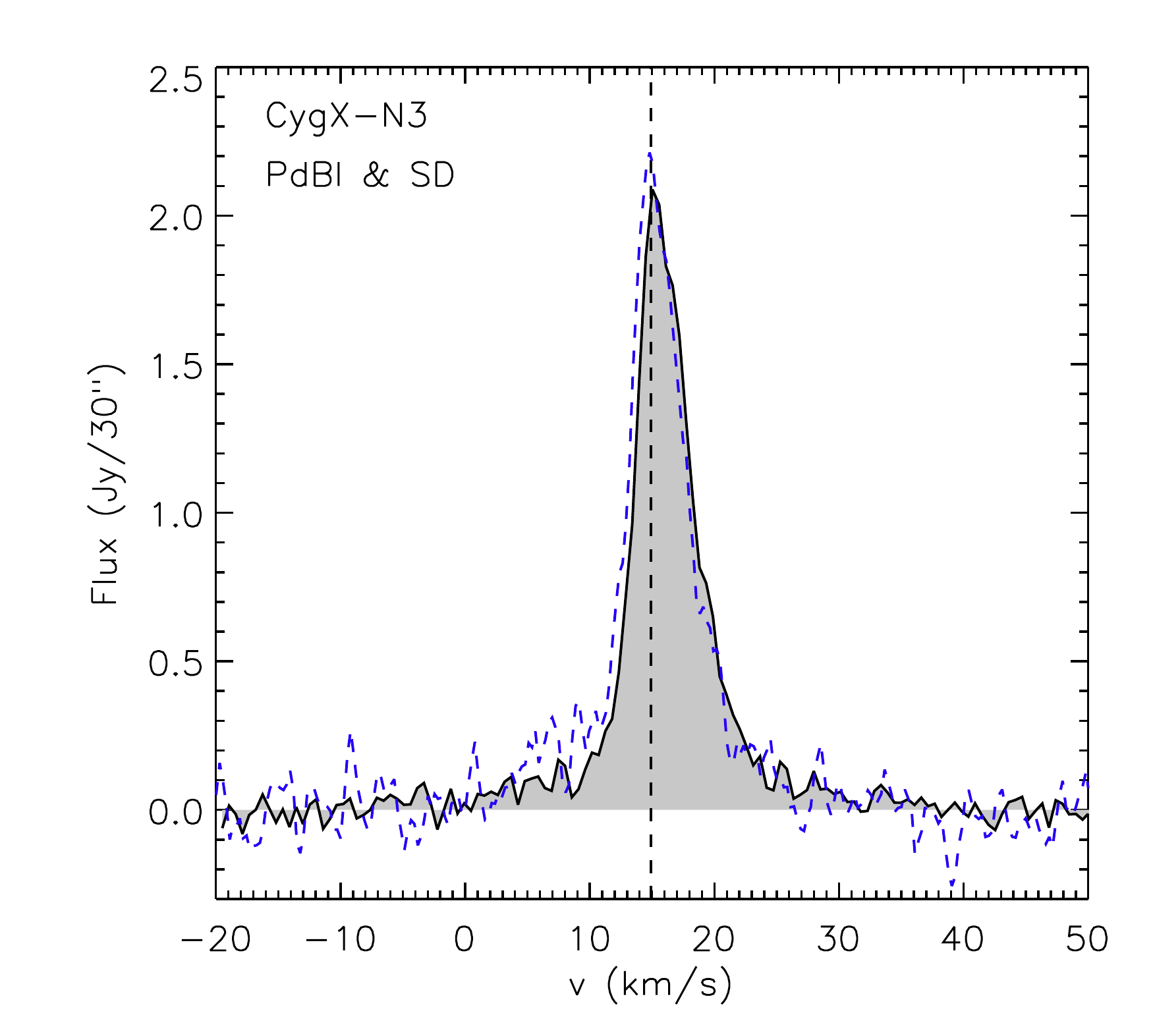}
	\hspace{-1.4cm}
	\includegraphics[width=0.3\textwidth]{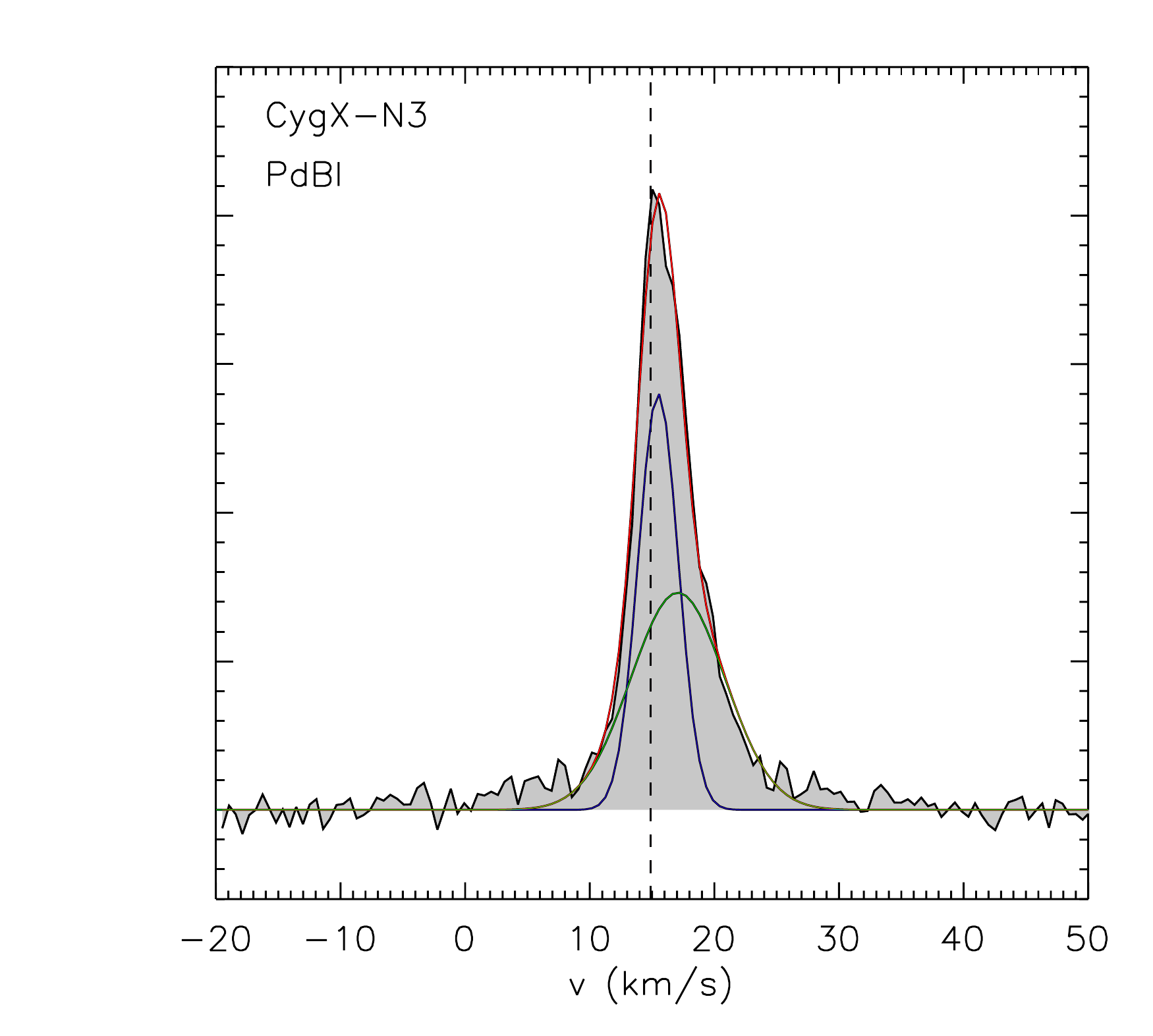}
	\hspace{-0.69cm}
	\includegraphics[width=0.3\textwidth]{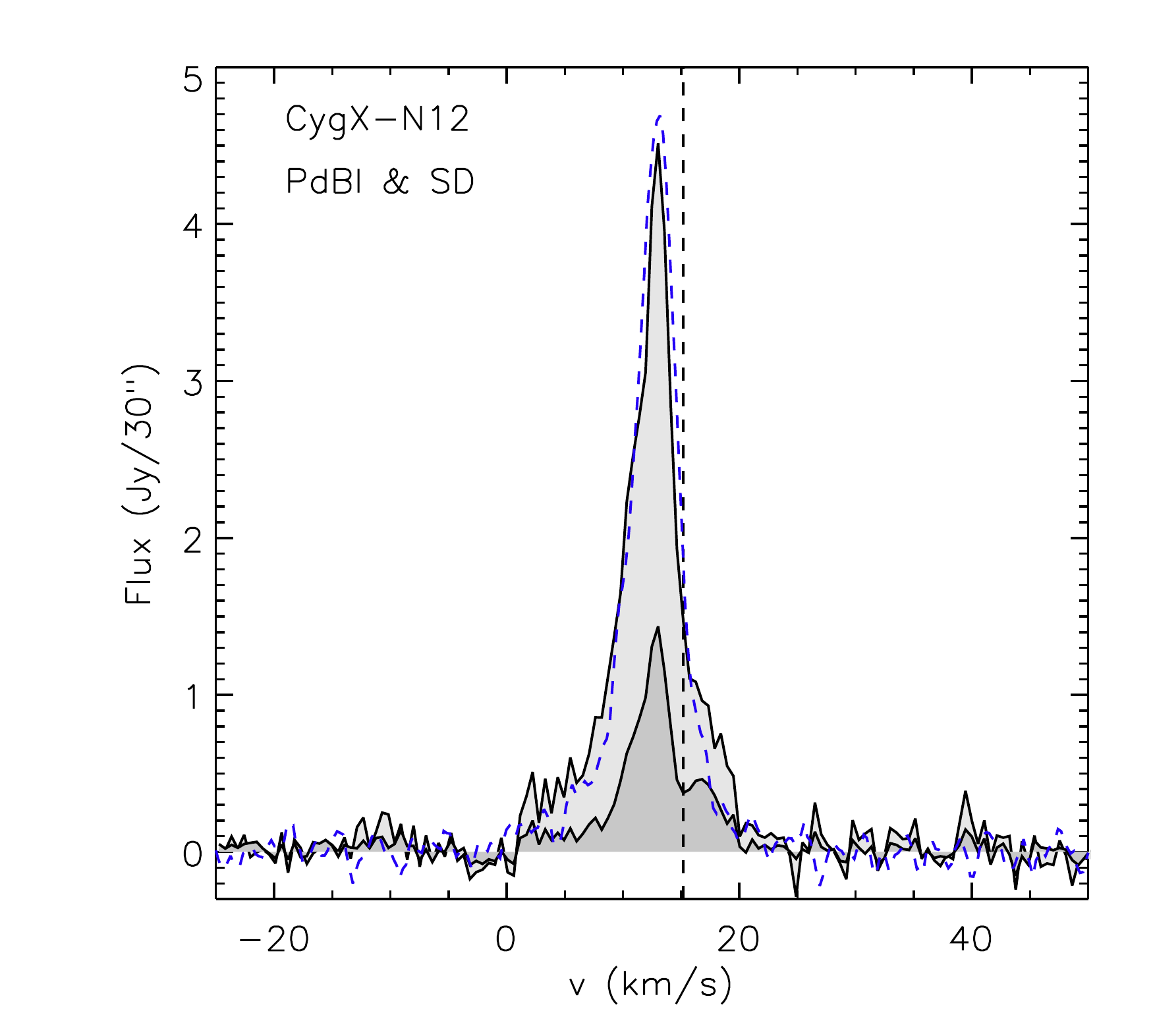}
	\hspace{-1.4cm}
	\includegraphics[width=0.3\textwidth]{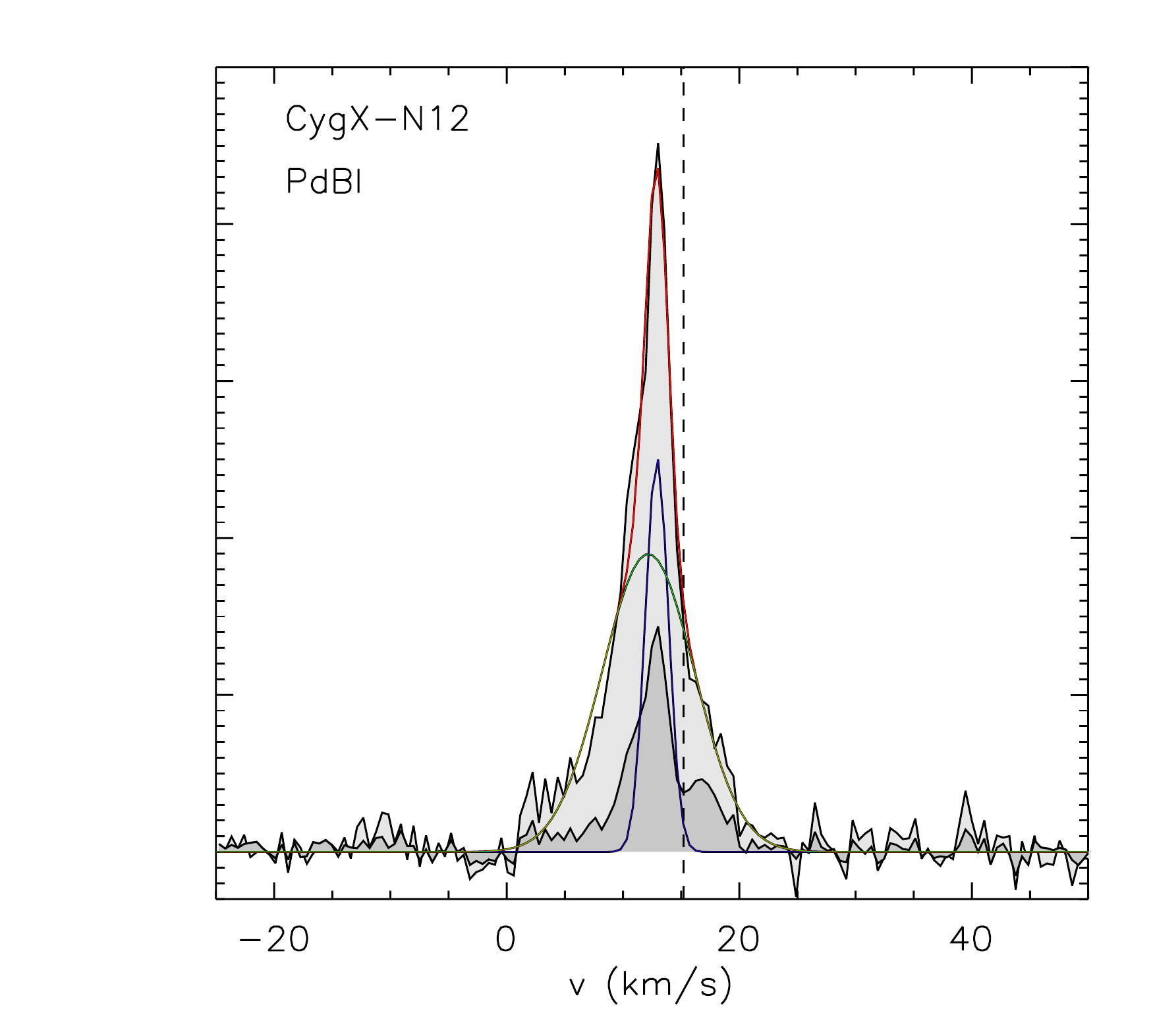}\\
	\hspace{-0.75cm}
	\includegraphics[width=0.3\textwidth]{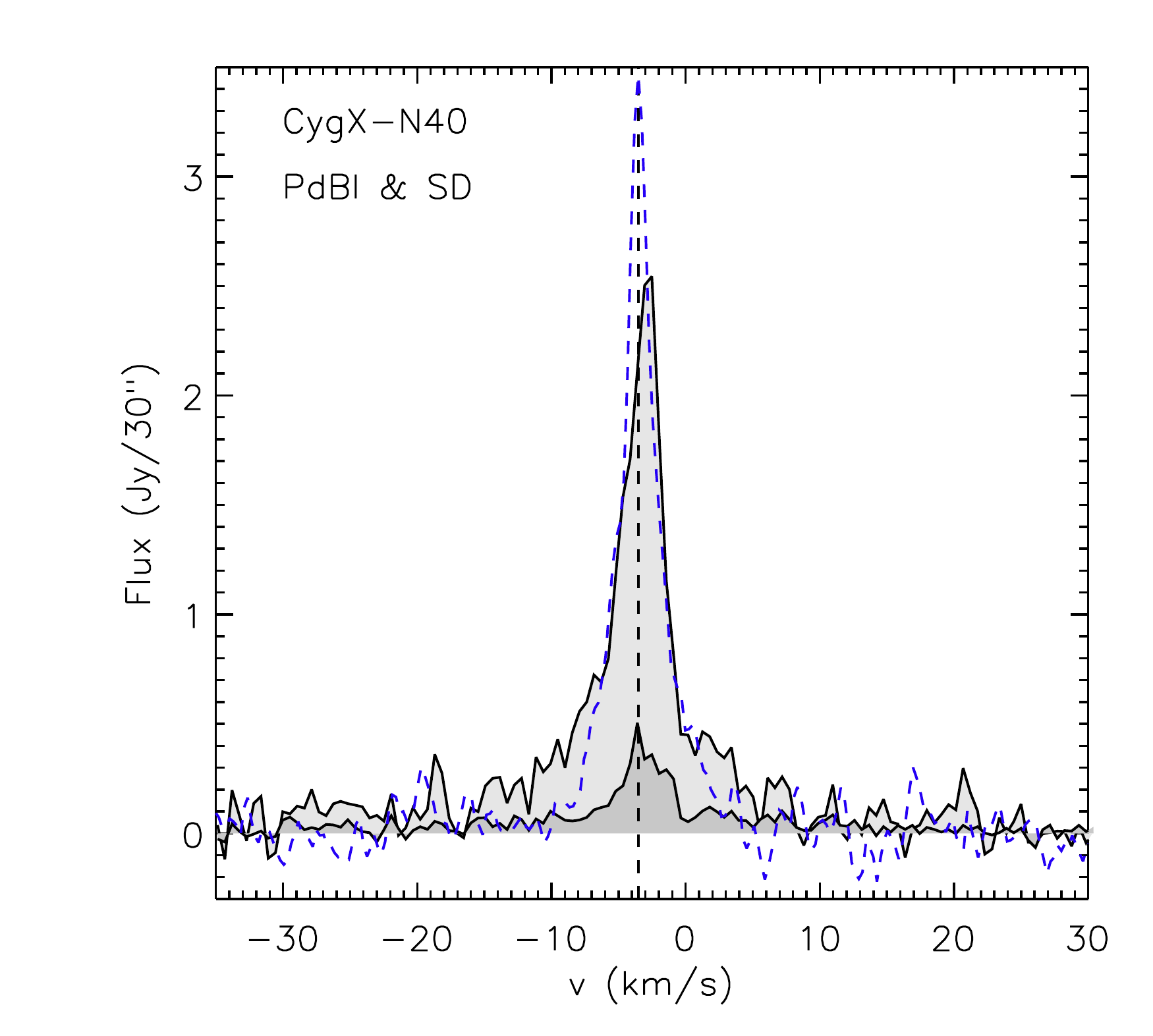}
	\hspace{-1.4cm}
	\includegraphics[width=0.3\textwidth]{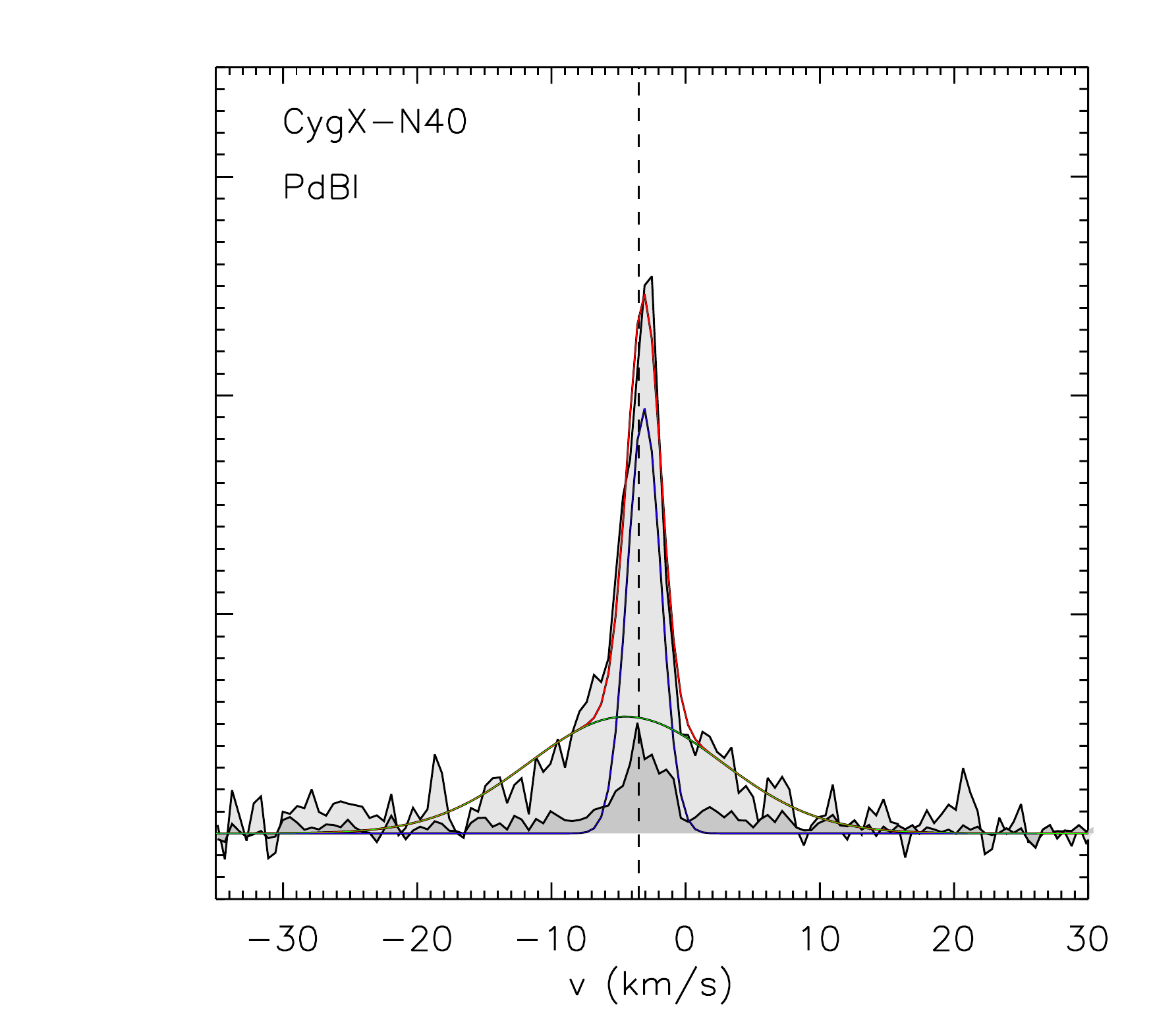}
	\hspace{-0.64cm}
	\includegraphics[width=0.3\textwidth]{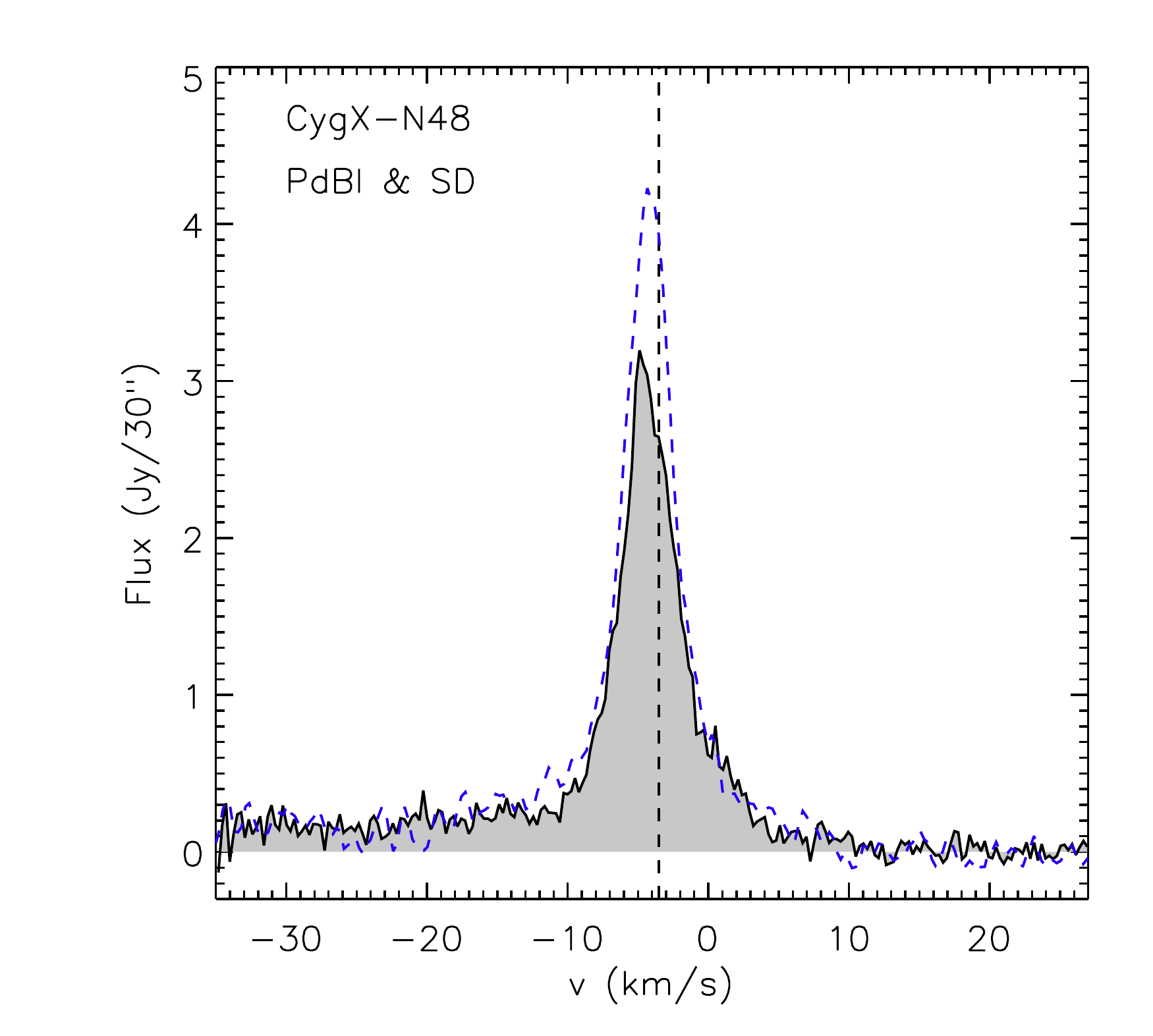}
	\hspace{-1.4cm}
	\includegraphics[width=0.3\textwidth]{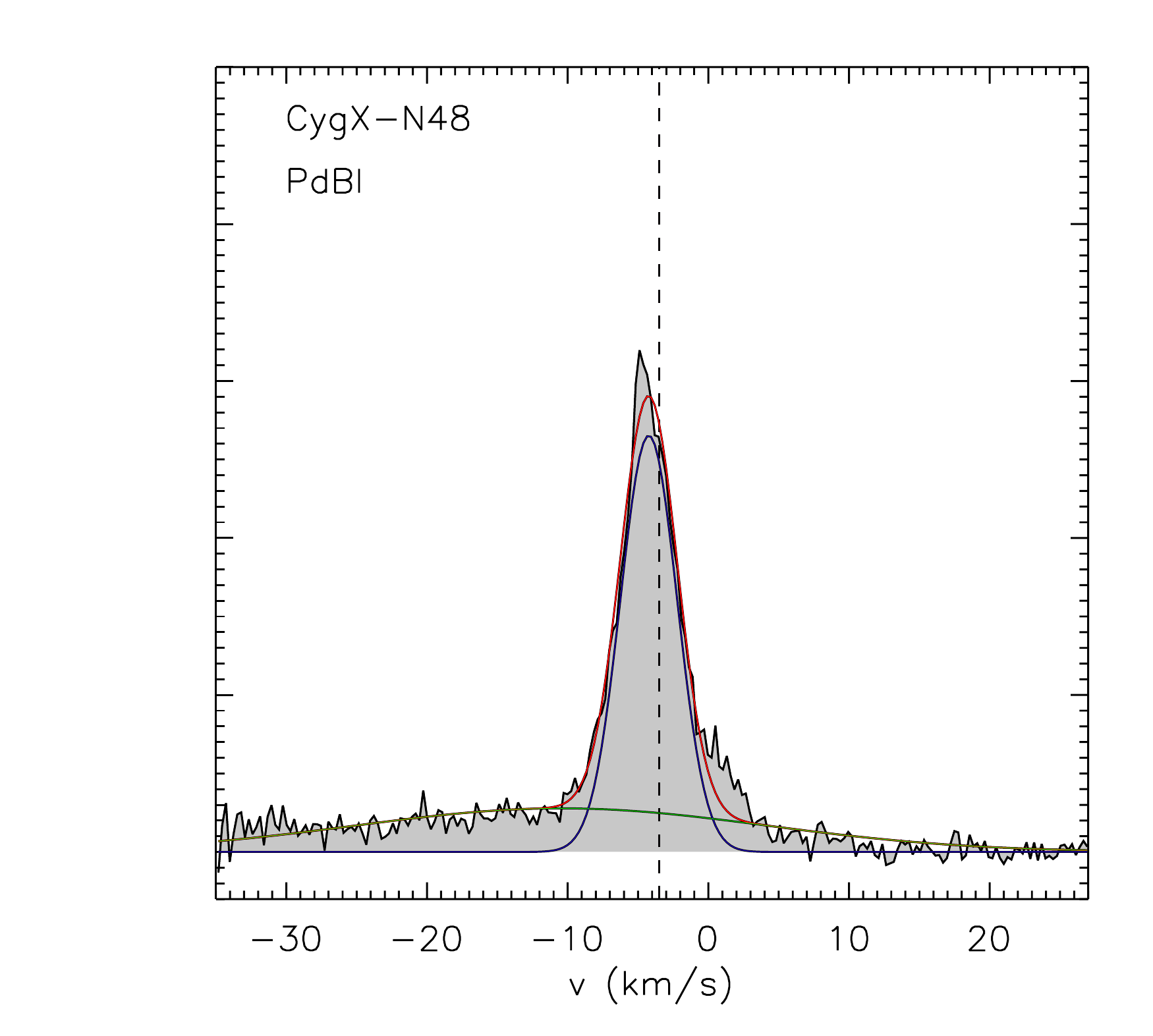}\\
	\hspace{-0.75cm}
	\includegraphics[width=0.3\textwidth]{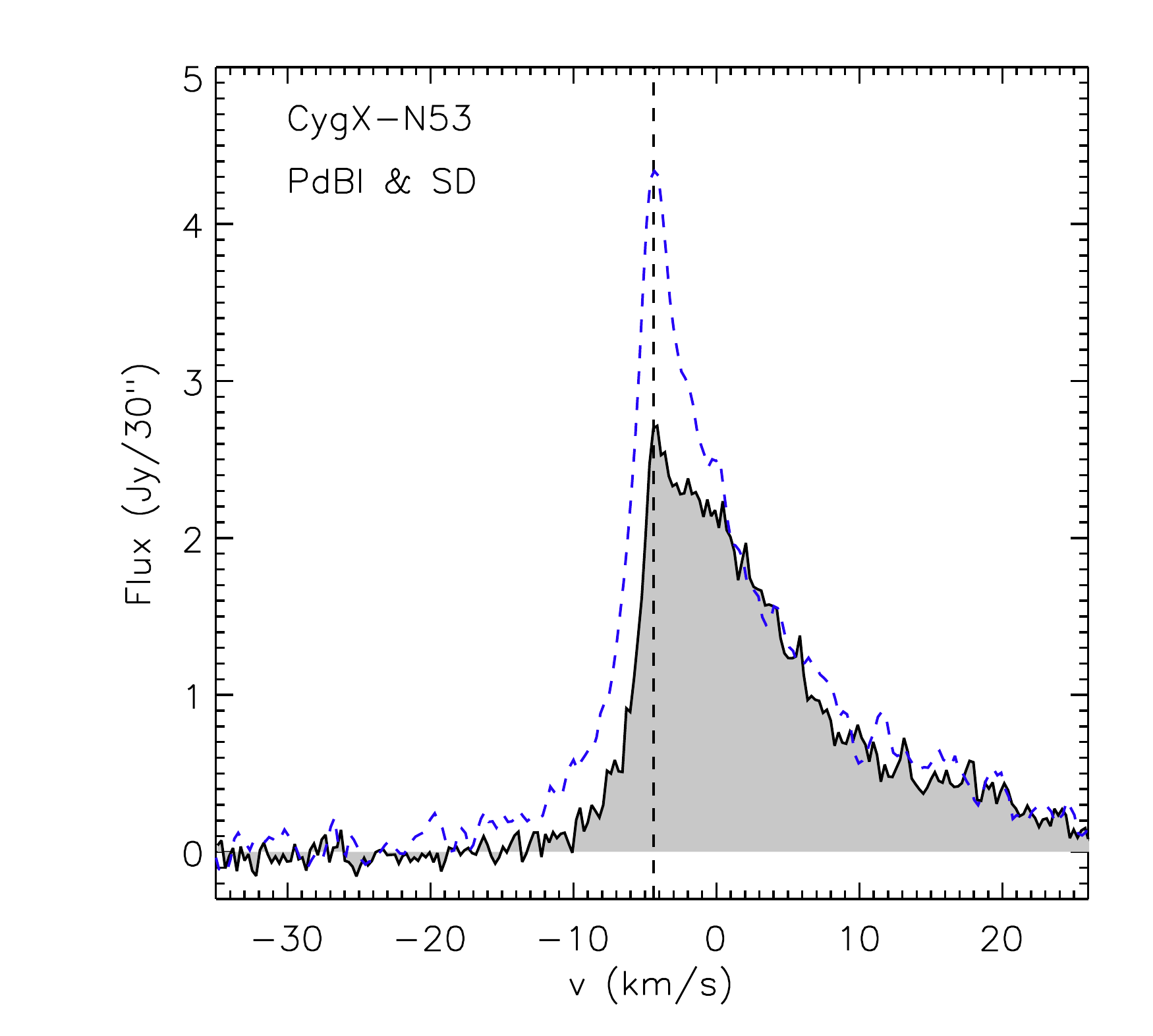}
	\hspace{-1.4cm}
	\includegraphics[width=0.3\textwidth]{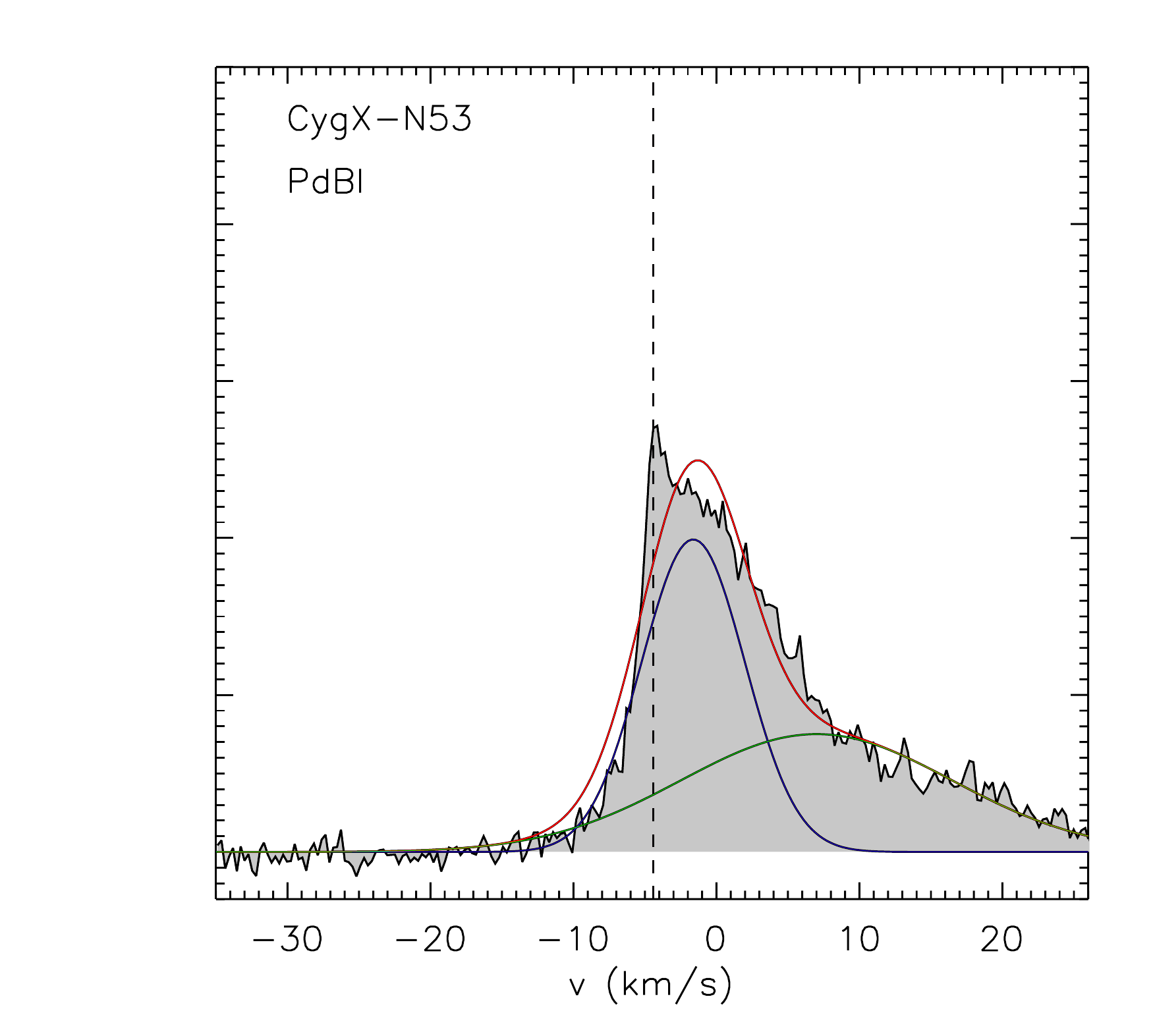}
	\hspace{-0.64cm}
	\includegraphics[width=0.3\textwidth]{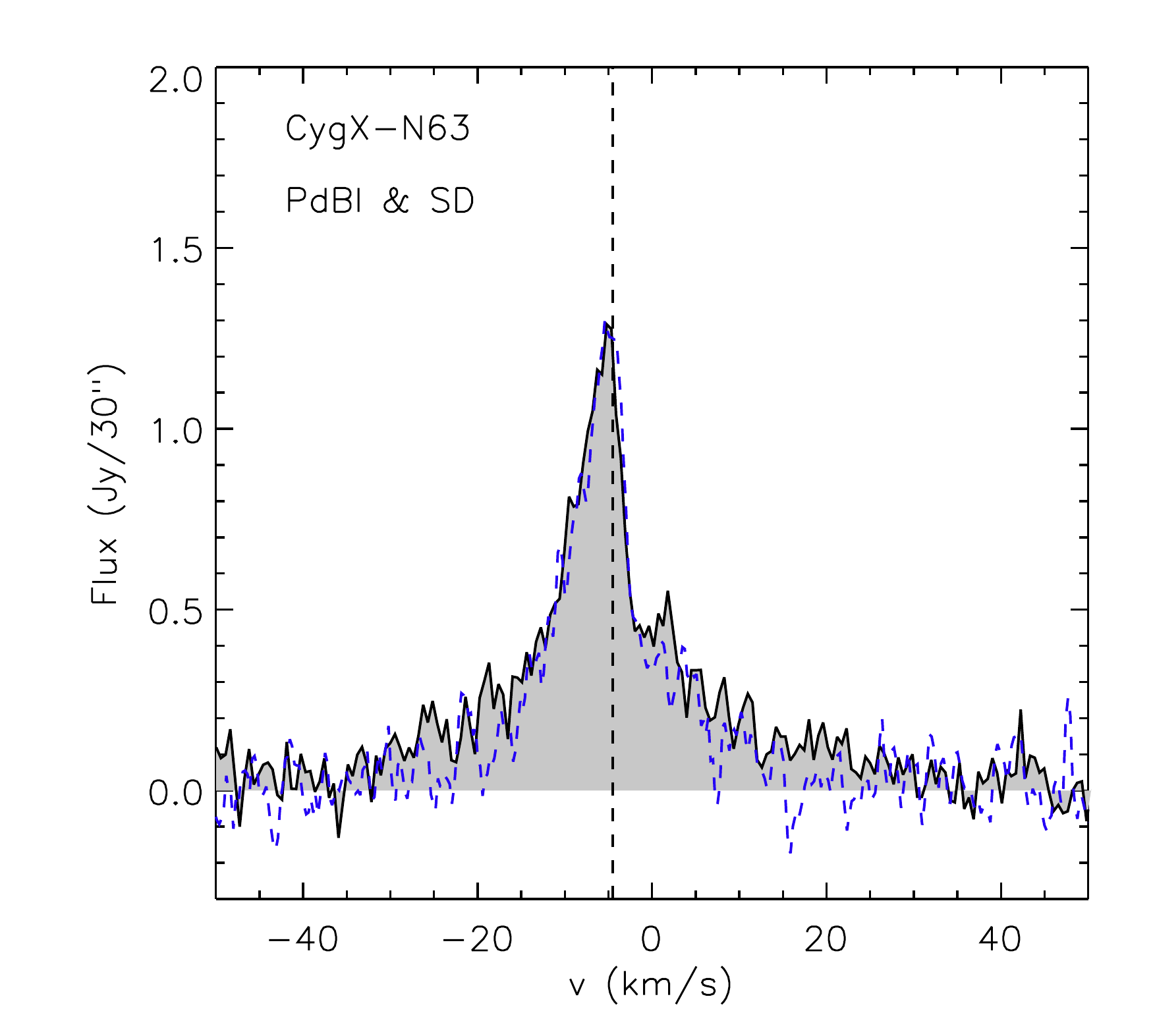}
	\hspace{-1.4cm}
	\includegraphics[width=0.3\textwidth]{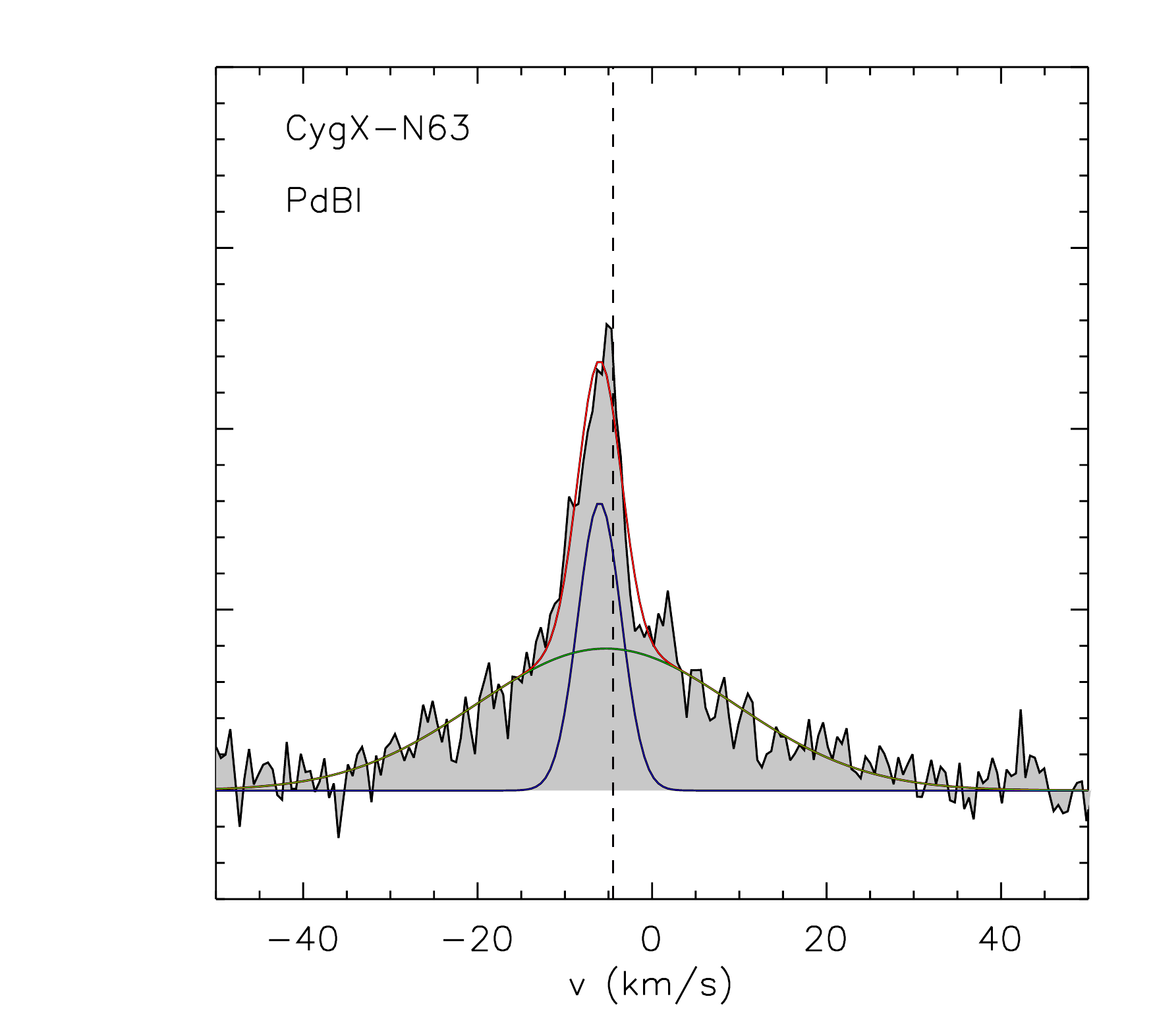}\\
	\caption[]{\small{Average spectra for the six MDCs of our sample (labelled on the top-left corners). The left side of each subfigure shows the comparison of the single-dish (SD) IRAM 30m pointed observations \citep[][]{2007A&A...476.1243M} in blue dashed lines, and the spectra of our PdBI SiO observations integrated over the central $30''\times30''$ (equivalent to the IRAM 30m beam) are shadowed in grey. The vertical black dashed lines show the systemic velocity of each core, as measured with N$_{2}$H$^{+}$ by \citet[][]{2010A&A...524A..18B}. The right side shows the same PdBI SiO emission as the left, overplotted with a two-component Gaussian fit to illustrate the existence of narrow (in blue) and broad (in green) components of emission (whose sum is shown in red). For CygX-N12 and N40, we show a shadowed spectrum (in light grey) which corresponds to the PdBI emission integrated over an area of $50''\times50''$ (to assess a possible contamination from strong SiO emission outside the SD primary beam).}}
	\label{fig:sio_average_spectra}}
\end{figure*} 

The six MDCs were observed in 2004 with the PdBI in the 1.3\,mm and 3.5\,mm continuum emission and in four spectral units covering the $^{12}$CO\,($2-1$), SiO\,($2-1$), H$^{13}$CO$^+$\,($1-0$), and H$^{13}$CN\,($1-0$) lines. The observations in the continuum and in the H$^{13}$CO$^+$\,($1-0$) and H$^{13}$CN\,($1-0$) lines are reported in \citet[][]{2010A&A...524A..18B} and \citet[][]{2011A&A...527A.135C}. The $^{12}$CO\,($2-1$) observations \citep[shown in][]{2013A&A...558A.125D} have been short-spaced using IRAM 30m data, and will be used here to aid the identification of individual outflows. This article presents the SiO\,($2-1$) observations, at 86.85 GHz,  with an angular resolution of $\sim 3''$ (i.e.  $\sim$0.02pc at 1.4\,kpc distance), tracing interstellar shocks at core-scales (typical protostellar outflows being $\sim$0.1-0.4\,pc in length). 

The observations were performed in track-sharing mode with two targets per track for the following pairs: CygX-N48/N53, CygX-N3/N40, and CygX-N12/N63. The D configuration track observations were performed between June and October 2004 (five antennas with baselines between 24\,m and 82\,m). The C configuration tracks were obtained in November and December 2004 (six antennas in 6Cp with baselines from 48\,m to 229\,m). The bright nearby quasar 2013+370 was used as a phase calibrator, and the evolved star MWC349 as a flux calibrator.

\begin{table*}[!t]
\caption{Systemic velocities and SiO line properties.}
\begin{tabular}{l | c | c c c | c c c | l l | c c}
\hline 
\hline 
	    & N$_{2}$H$^{+}$	 & \multicolumn{3}{c |}{Narrow component}  & \multicolumn{3}{c |}{Broad component}  &   \multicolumn{2}{c |}{Total flux (in $30''$)$^{*}$} & \multicolumn{2}{c}{High-$\varv$ range} \\
Source  & $\varv_{0}$ & $I_{\rm peak}^{n}$ & $\varv_{\rm{peak}}^{n}$ & $\sigma_{\varv}^{n}$ & $I_{\rm peak}^{b}$ & $\varv_{\rm{peak}}^{b}$ & $\sigma_{\varv}^{b}$ & PdBI & SD &  Blue	& Red \\
			& \scriptsize{(km\,s$^{-1}$)} & \scriptsize{(Jy)} 	& \scriptsize{(km\,s$^{-1}$)} & \scriptsize{(km\,s$^{-1}$)}	& \scriptsize{(Jy)}	& \scriptsize{(km\,s$^{-1}$)} & \scriptsize{(km\,s$^{-1}$)} & \scriptsize{(Jy)} & \scriptsize{(Jy)}	& \scriptsize{(km\,s$^{-1}$)} & \scriptsize{(km\,s$^{-1}$)} \\
\hline
CygX-N3   & 14.9 	& 1.4 & $15.5\pm0.2$ 	& $1.6\pm0.3$ 	& 0.7 & $17.0\pm0.8$ 	& $3.8\pm0.6$ 		& 13.6 			& 14.4  & [-20.0; 11.9] & [17.9; 40.0] \\
CygX-N12 & 15.2 	& 2.5 & $12.9\pm0.1$ 	& $1.0\pm0.1$  & 1.9 & $12.2\pm0.2$ 	& $4.0\pm0.2$ 		& 9.1(30.1)$^{**}$	&  25.1 & [-10.0; 12.2] & [18.2; 25.0] \\
CygX-N40 & -3.5	& 1.9 & $-3.1\pm0.1$	& $1.2\pm0.1$  & 0.5 & $-4.5\pm0.8$ 	& $7.2\pm0.9$  	& 3.9(19.4)$^{**}$	& 12.6  & [-35.0; -6.5]  & [-0.5; 30.0] \\
CygX-N48 & -3.5 	& 2.7 & $-4.2\pm0.1$	& $2.0\pm0.1$ 	& 0.3 & $-10.5\pm2.0$ 	& $14.6\pm 2.1$ 	& 22.9			& 28.7  & [-35.0; -6.5]  & [-0.5; 27.0] \\
CygX-N53 & -4.4 	&  -    &         -		 	&    -			&  -    &  -                     	&     - 	  		& 34.4			& 48.5  & [-35.0; -7.4]  & [-1.4; 26.0] \\
CygX-N63 & -4.5	& 0.8 & $-6.0\pm0.3$	& $2.5\pm0.3$	& 0.4 & $-5.3\pm1.4$ 	& $14.9\pm1.7$ 	& 20.7			& 15.2  & [-50.0; -7.5]  & [-1.5; 50.0] \\
\hline
\end{tabular}
$^{*}$ The total flux was estimated using the integrated intensities over the entire velocity range (as plotted in Fig.~\ref{fig:sio_average_spectra}), and they are the same for the single-dish (SD) and PdBI data.\\
$^{**}$ The values inside brackets show the PdBI flux retrieved from a $50''\times50''$ region instead of $30''\times30''$.
\label{tab:velocities}
\end{table*}

The maps were cleaned using the natural weighting, and the resulting synthesised beam and rms in the continuum are summarised in Table~\ref{obs_summary}, together with the field names and centres of phase. The pixel size is of $0.62''$. The cleaning components were searched across the whole area of the primary beam. No support for cleaning was used to avoid introducing any bias into the resulting emission maps.


\section{Results}
\label{sec:results}

\subsection{Profile of the SiO emission}
\label{sec:profile}

\citet[][]{2007A&A...476.1243M} presented single-dish pointed observations of SiO (2-1) emission with the IRAM 30m telescope (beam size of $\sim29''$) towards the sample of IR-quiet cores we study here. Figure~\ref{fig:sio_average_spectra} shows the SiO average profiles from our PdBI observations as dark grey shadowed spectra (averaged within the central $30''\times30''$) overlaid with the dotted blue spectra from the single-dish  observations. 

There is a good correspondence between the two profiles for 3 out of 6 sources (CygX-N3, N48 and N63). For these three MDCs, the PdBI recovers most of the flux detected with the single-dish ($95\%$, $80\%$, and $>100\%$ for CygX-N3, N48, and N63, respectively; see Table~\ref{tab:velocities}). The fact that we recover slightly more emission with the PdBI with respect to the IRAM 30m telescope in N63 can be explained by the uncertainties on the calibration of the IRAM 30m data ($\sim 10\%$), by some loss of power towards the edges of the single-dish primary beam, and by the fact that the area we used to extract the average PdBI profiles is a rectangular area of $30''$ width, instead of a circular beam shape. In CygX-N53, even though the shape of the profile is similar, we miss some of the systemic velocity emission with the PdBI. Nevertheless, we still recover $\sim 70\%$ of the single-dish flux with the PdBI. This indicates that most of the SiO emission is compact on these four sources, and not filtered out with the PdBI.  

\begin{figure*}[!t]
	\centering
	{\renewcommand{\baselinestretch}{1.1}
	\includegraphics[width=0.33\textwidth]{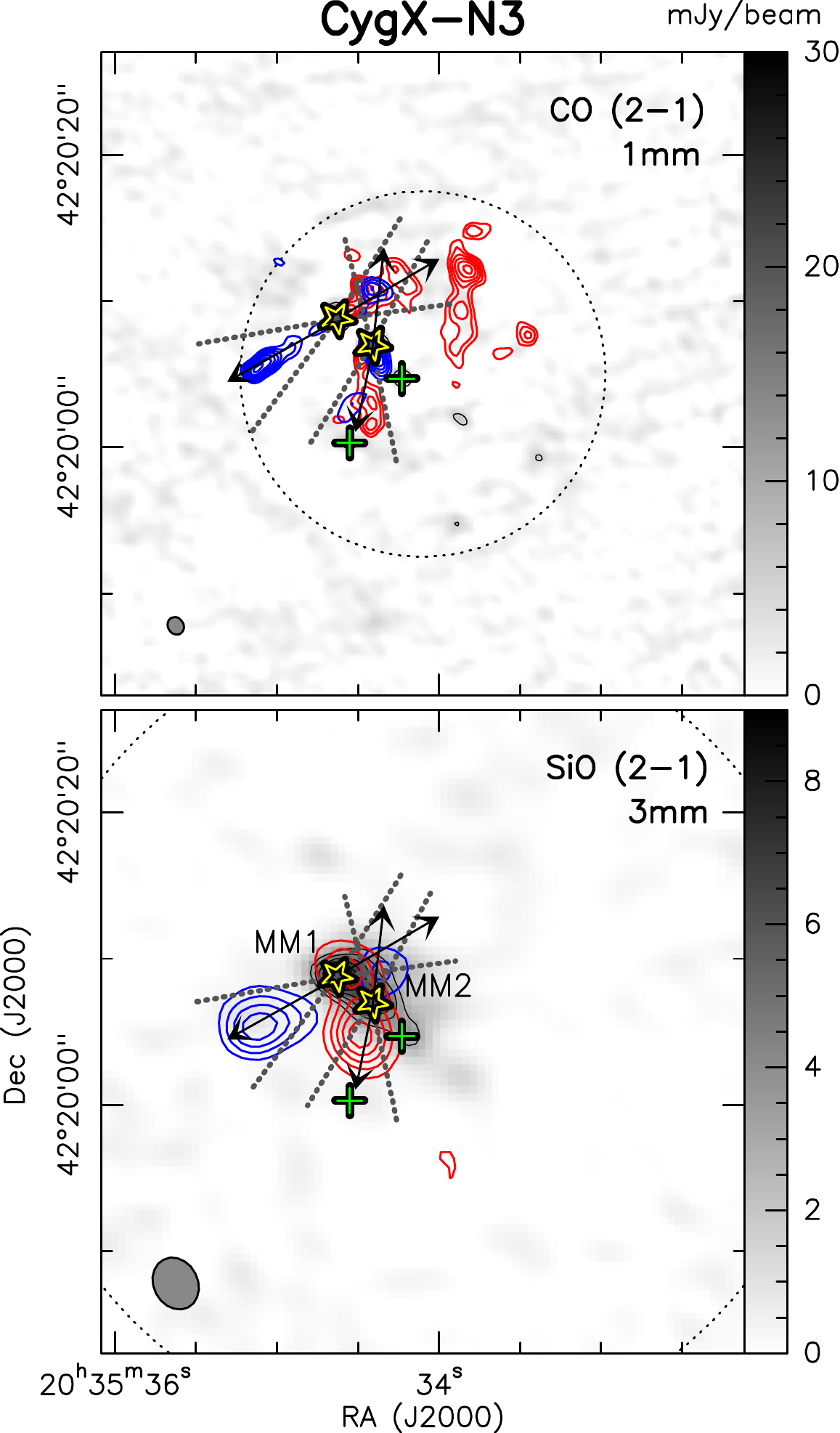}
	\hfill
	\includegraphics[width=0.33\textwidth]{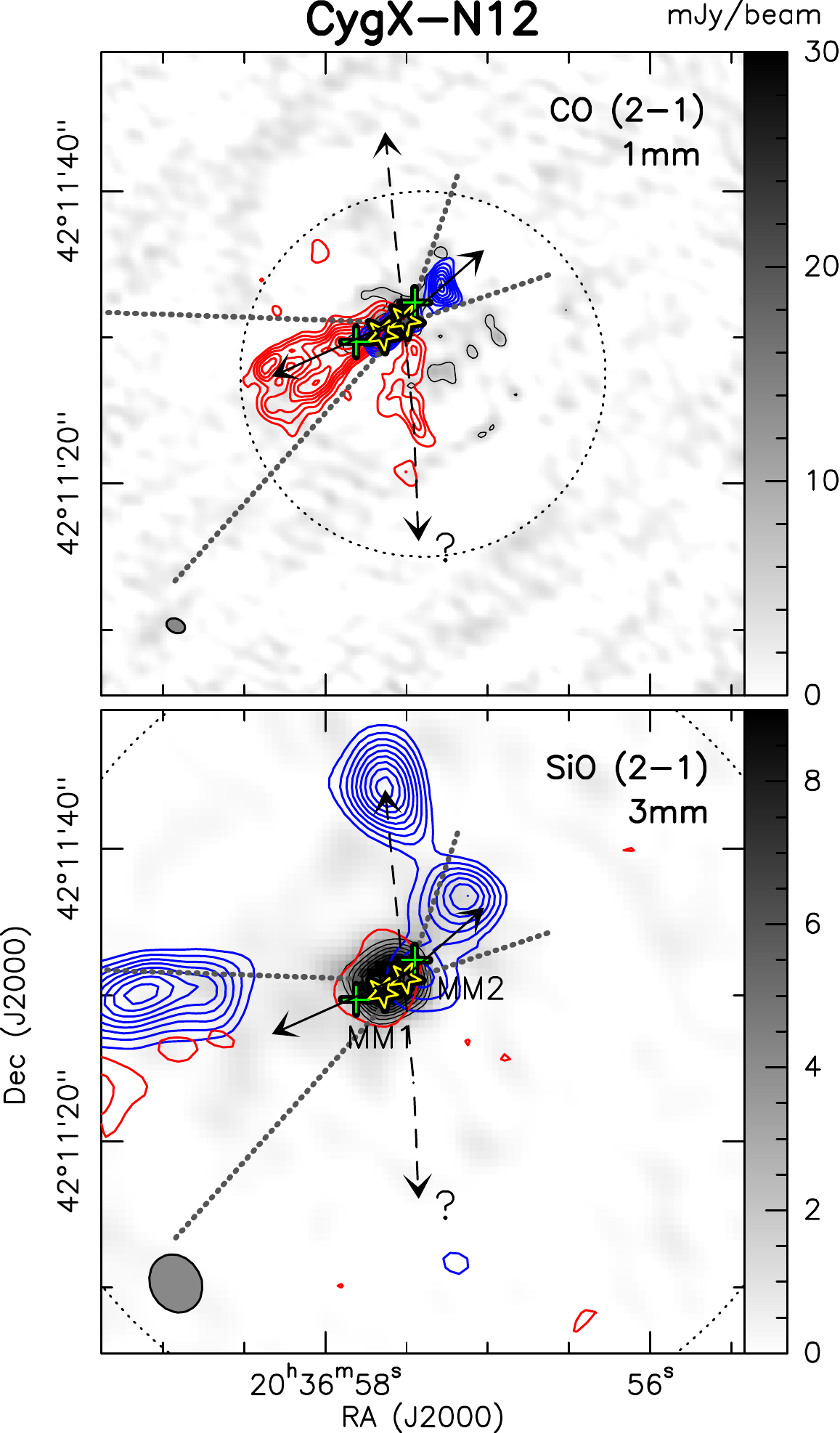}
	\hfill
	\includegraphics[width=0.33\textwidth]{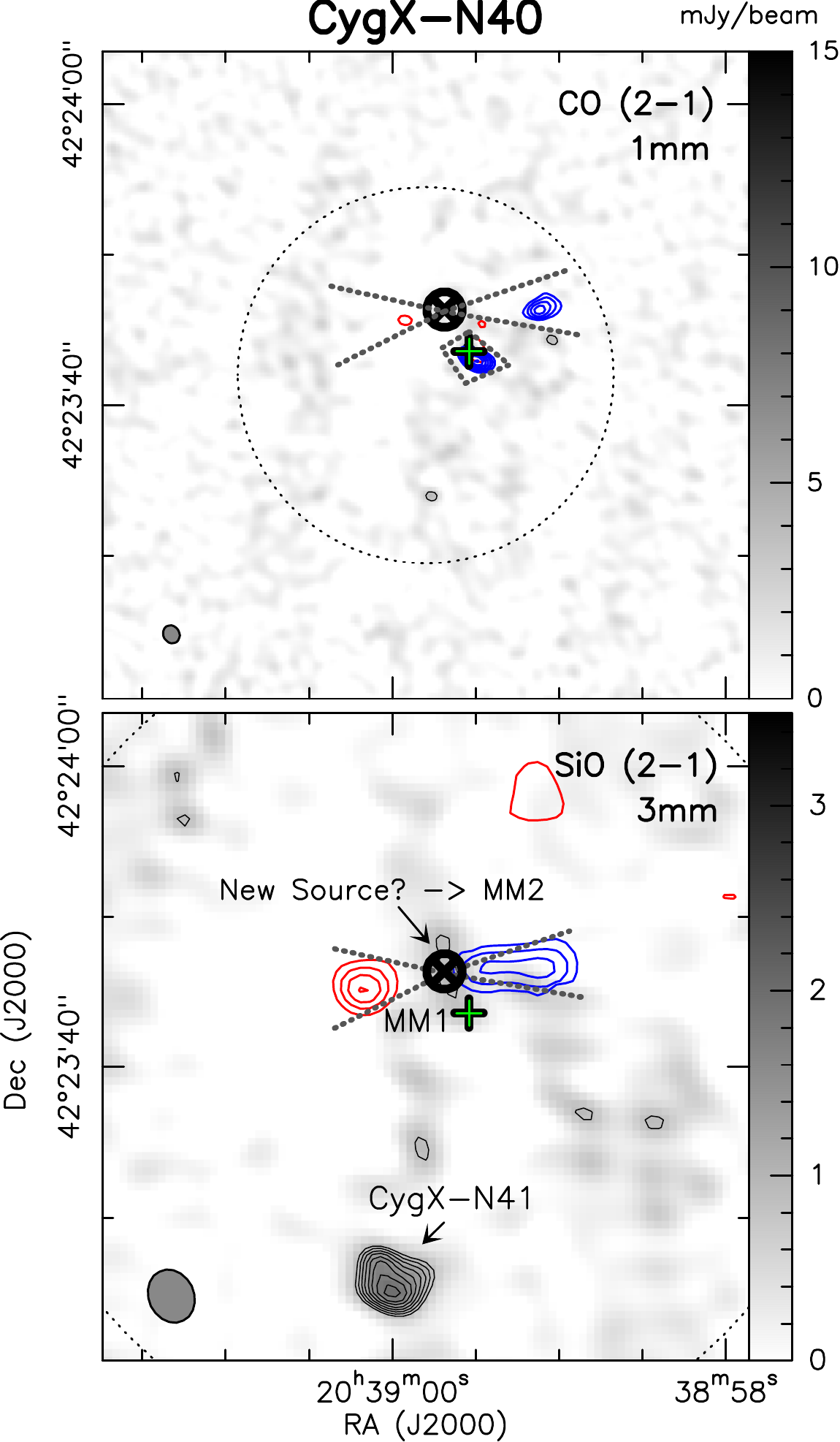}
	\caption[]{\small{Comparison of the high-velocity CO and SiO emission in CygX-N3 (left), N12 (centre), and N40 (right). The top panels show the high-velocity CO emission in contours \citep[with the velocity ranges as in][]{2013A&A...558A.125D} over the 1mm continuum emission in greyscale. The primary beam of the CO and 1mm observations is shown as a dashed circle, and the synthetic beam is in the lower-left corner. Contours are stepped by 0.25\,Jy\,beam$^{-1}$\,km\,s$^{-1}$, starting at 2.0\,(2.75)\,Jy\,beam$^{-1}$\,km\,s$^{-1}$ for N3 blue (red), 1.5\,Jy\,beam$^{-1}$\,km\,s$^{-1}$ for N12, and 1.75\,(1.25)\,Jy\,beam$^{-1}$\,km\,s$^{-1}$ for N40 blue (red). The lower panels show the high-velocity SiO emission in contours (with velocity ranges as in Table~\ref{tab:velocities}) over the 3mm continuum emission in greyscale for the same sources. The primary beams are shown as dashed circles and the synthetic beams are plotted in the lower-left corners. Contours are stepped by 0.1\,Jy\,beam$^{-1}$\,km\,s$^{-1}$, starting at 0.15\,Jy\,beam$^{-1}$\,km\,s$^{-1}$ for N3, 0.2\,(0.05)\,Jy\,beam$^{-1}$\,km\,s$^{-1}$ for N12 blue (red), and 0.2\,Jy\,beam$^{-1}$\,km\,s$^{-1}$ for N40. The most massive protostars are marked with yellow stars (and labelled in the lower panels), and the least massive fragments from \citet[][]{2010A&A...524A..18B} are shown as green crosses. The arrows are the directions of the outflows inferred from \citet[][]{2013A&A...558A.125D}, and the dashed lines show the respective outflow cones.}}
	\label{fig:sio_co_highv}}
\end{figure*}

The two most striking exceptions are CygX-N12 and N40, where a significant fraction of the single-dish flux ($\sim 65-70\%$) is missing with the PdBI, when only the central $30''$ emission is taken, either because some flux is filtered out in the interferometric observations or because the single-dish observations suffer from contamination from strong emission sitting in the immediacies of the single-dish primary beam. In order to discriminate between these possibilities, we have integrated the PdBI spectra for these two regions considering a larger area of $50''\times50''$ (light grey shadowed spectra in the panels of these two sources). Doing so for CygX-N12 recovers not only the flux of the single-dish observations, but also the line shape. We conclude that this is due to the contamination from the region east of the single-dish primary beam, with strong outflow emission (see central panels of Fig.~\ref{fig:sio_co_highv}). Doing this for CygX-N40, despite a considerable improvement, still does not recover the peak flux detected with the single-dish at systemic velocities, suggesting that there is a considerable amount of extended SiO emission at systemic velocities which is filtered by the PdBI (with a need for short-spacings). On the other hand, it recovers some line-wings that were not seen with the single-dish.

\begin{figure*}[!t]
	\centering
	{\renewcommand{\baselinestretch}{1.1}
	\includegraphics[width=0.33\textwidth]{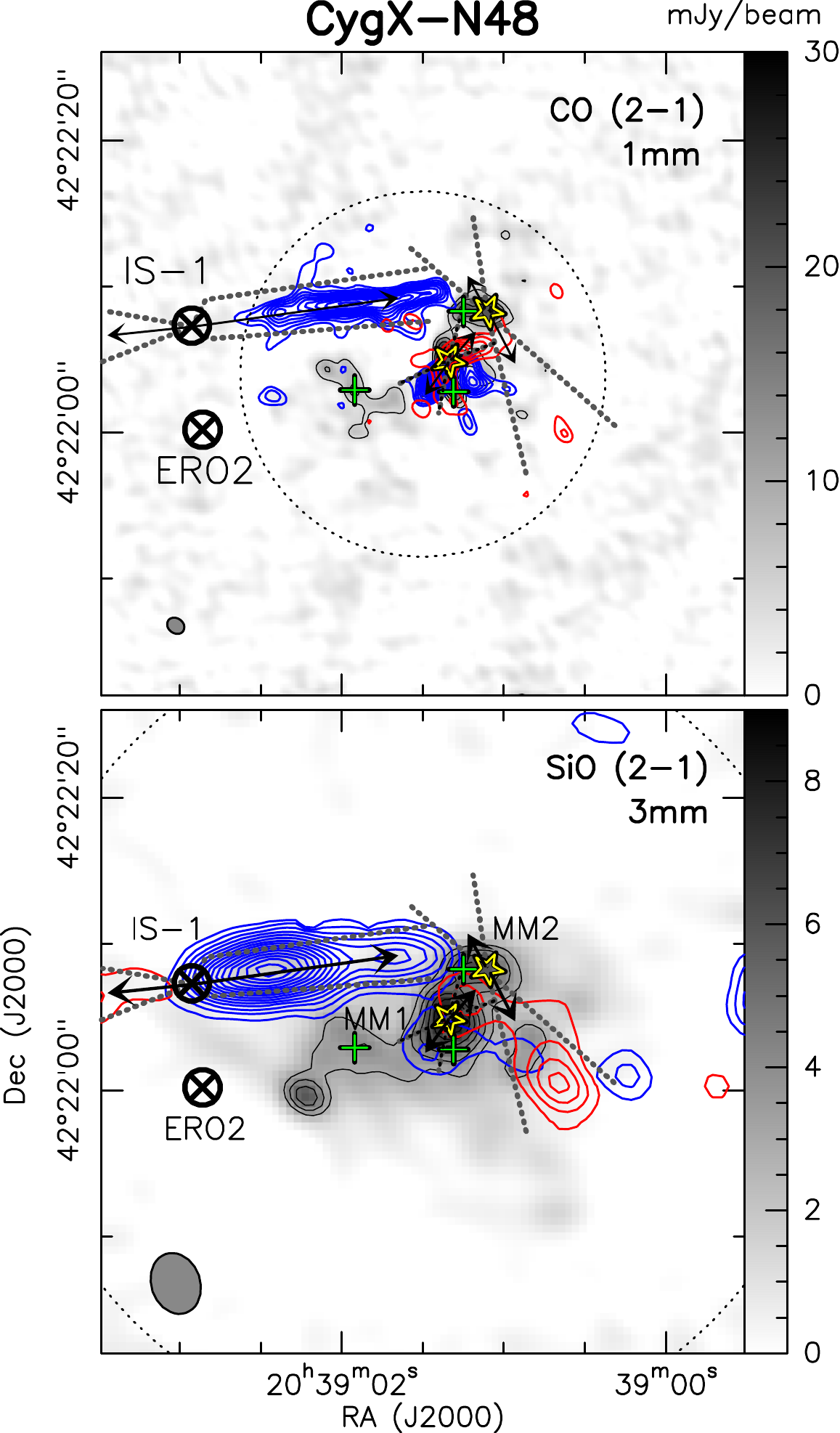}
	\hfill
	\includegraphics[width=0.33\textwidth]{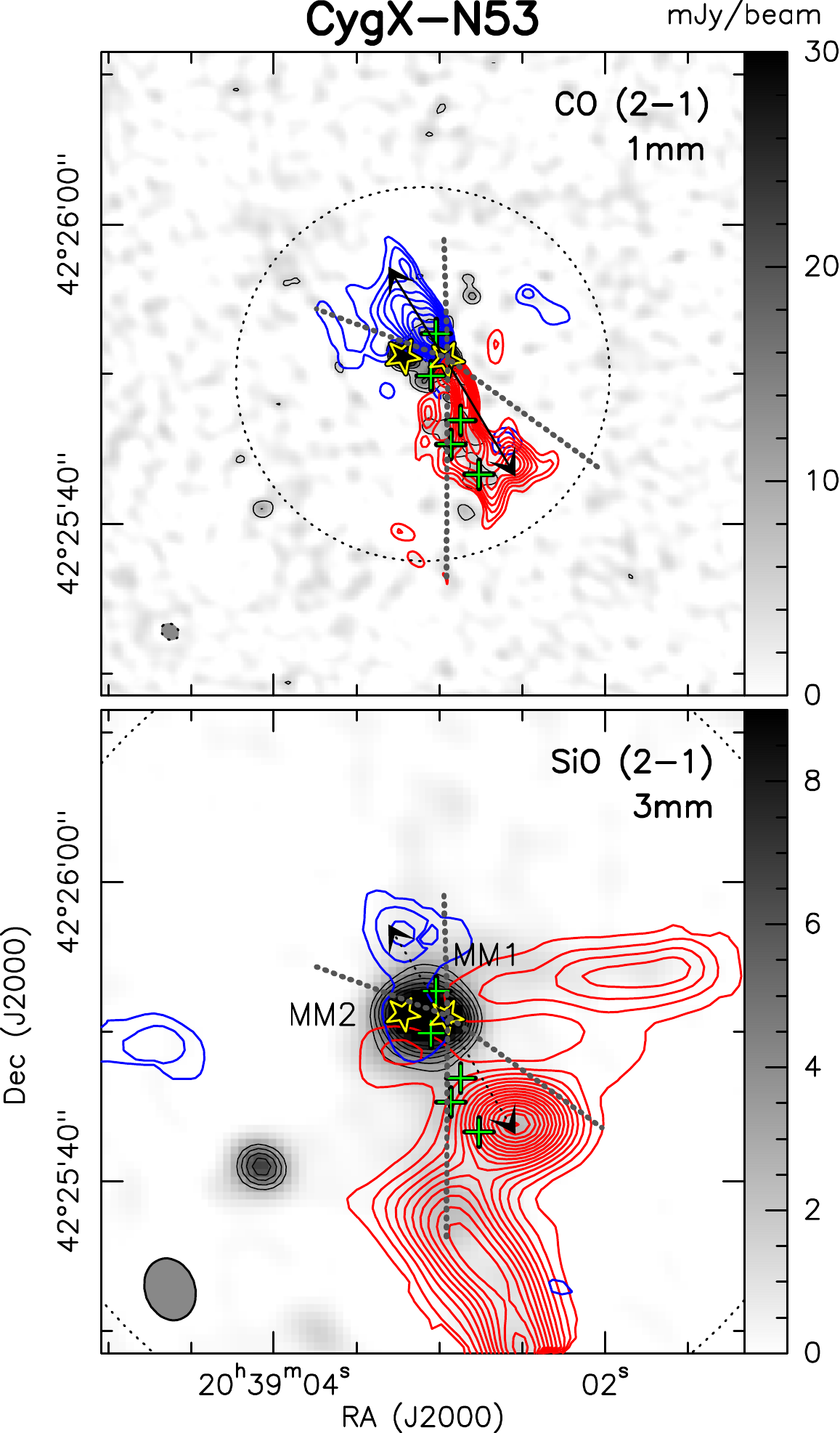}
	\hfill
	\includegraphics[width=0.33\textwidth]{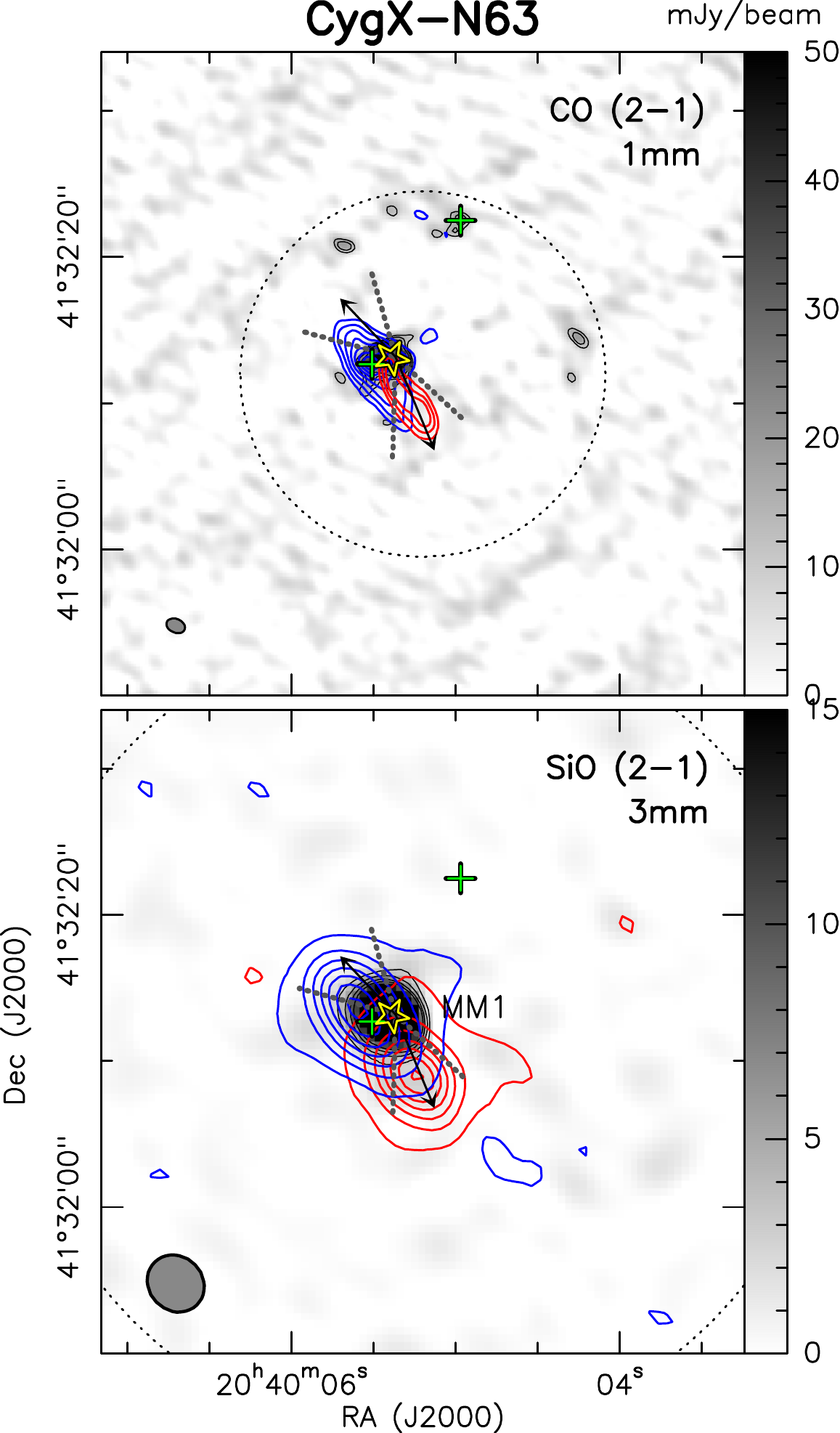}
	\caption[]{\small{Same as Fig.~\ref{fig:sio_co_highv} for CygX-N48 (left), N53 (centre), and N63 (right). In N48 we mark with circled crosses two Spitzer sources: ERO2 \citep[][]{2004ApJS..154..333M} and IS-1, which is a resolved source detected in the IRAC bands at 3 and 4~$\mu$m, and unresolved at 8$\mu$m and onwards. IS-1 is likely responsible for the outflow directed east-west. Contour key for CO emission, top panels: N48 (left) contour steps of 0.5\,Jy\,beam$^{-1}$\,km\,s$^{-1}$ starting at 4\,(3)\,Jy\,beam$^{-1}$\,km\,s$^{-1}$ for blue (red); N53 (centre) contour steps of 1.5\,Jy\,beam$^{-1}$\,km\,s$^{-1}$ starting at 8.5\,Jy\,beam$^{-1}$\,km\,s$^{-1}$; N63 (right) contour steps of 1.5\,Jy\,beam$^{-1}$\,km\,s$^{-1}$ starting at 5.0\,Jy\,beam$^{-1}$\,km\,s$^{-1}$. Contours of SiO emission, lower panels: N48 (left) contour steps of 0.1\,Jy\,beam$^{-1}$\,km\,s$^{-1}$, starting at 0.2\,(0.15)\,Jy\,beam$^{-1}$\,km\,s$^{-1}$ for blue (red); N53 (centre) contour steps of 0.15\,(0.25)\,Jy\,beam$^{-1}$\,km\,s$^{-1}$, starting at 0.2\,(0.15)\,Jy\,beam$^{-1}$\,km\,s$^{-1}$ for blue (red); N63 (right) contour steps of 0.5\,Jy\,beam$^{-1}$\,km\,s$^{-1}$, starting at 0.2\,Jy\,beam$^{-1}$\,km\,s$^{-1}$, for both blue- and redshifted emission.}}
	\label{fig:sio_co_highv2}}
\end{figure*}

Based on the single-dish SiO emission, \citet[][]{2007A&A...476.1243M} discussed the existence of two components of SiO emission. 
Our observations confirm that the average spectra of most sources can be accurately represented by the emission of two Gaussian components (Fig.~\ref{fig:sio_average_spectra}), with the exception of CygX-N53 which shows a clear asymmetric profile, with broad redshifted emission (reaching a terminal velocity of $\sim30$\,km\,s$^{-1}$). In Table~\ref{tab:velocities} we show the systemic velocities, $\varv_{0}$, of the different regions as measured by \citet[][]{2010A&A...524A..18B} using N$_{2}$H$^{+}$ pointed observations, also performed with the IRAM 30m telescope (and therefore within an equivalent beam size). The uncertainties on these velocities are of 0.1\,km\,s$^{-1}$. These are shown as dashed vertical lines in Fig.~\ref{fig:sio_average_spectra}. For the MDCs where two Gaussian components fit well the observed average spectra, we present, in Table~\ref{tab:velocities}, the peak intensity, $I_{\rm peak}$; the peak velocity, $\varv_{\rm{peak}}$; and the velocity dispersion, $\sigma_{\varv}$, for both the narrow and the broad components. The peak of the narrow component of the SiO emission is only slightly different from the systemic velocities (less than 1\,km\,s$^{-1}$ difference), which is not significant considering the linewidths of the emission and the uncertainties on the peak velocity. This is with the exception of N12, where there is an important shift between the two (of 2\,km\,s$^{-1}$). We will discuss possible explanations for this shift later in the article (Sect.~\ref{sec:narrow_sio_outflows}).

While the broad emission whose average width at the base reaches $\gtrsim$30\,km\,s$^{-1}$ suggests unequivocally outflow shocks, \citet[][]{2007A&A...476.1243M} considered the possibility of the narrower component originating from shocks inside a hot core and/or lower velocity outflows.The only way to distinguish between a high-velocity or low-velocity shock origin is by carefully inspecting the spectra along the maps, and investigate the spatial distribution of the systemic-velocity and broad emissions (see Sects.~\ref{sec:outflow_id} and \ref{sec:narrow_sio}). The spectra throughout the maps vary significantly from the average spectra, and examples of the different line profiles found in each MDC can be found in Appendix~\ref{ap:SiO_spectra}, Fig.~\ref{fig:sio_indiv_spectra}.

\subsection{Distribution of high-velocity SiO versus CO}
\label{sec:outflow_id}

To investigate how effectively SiO is tracing protostellar outflows, we compare the spatial distribution of the high-velocity SiO emission with that of CO \citep[][]{2013A&A...558A.125D}. Table~\ref{tab:velocities} shows the velocity ranges used to estimate the SiO integrated intensities and luminosities at high-velocities (high-$\varv$ range), which is in essence the emission offset from the systemic velocity of the cloud by more than $\pm$3\,km\,s$^{-1}$ \citep[corresponding to roughly 3$\sigma_v$ from the fits to the N$_{2}$H$^{+}$ (1-0) spectra by][]{2010A&A...524A..18B}. Figures~\ref{fig:sio_co_highv} and \ref{fig:sio_co_highv2} show the high-velocity CO and SiO emission in the top and bottom panels, respectively. In these figures we also show the outflow cones (dotted lines) and directions (arrows). Although with a different resolution ($\sim1''$) and field of view (see dashed circles in Figs.~\ref{fig:sio_co_highv} and \ref{fig:sio_co_highv2}), the CO is used as a reference to identify the different outflow lobes and respective driving sources.

From a fast inspection of these figures we can see a number of similarities in the distribution of the high-velocity SiO and high-velocity CO outflow emission. In particular, for CygX-N3 (left column of Fig.~\ref{fig:sio_co_highv}) and CygX-N63 (right column of Fig.~\ref{fig:sio_co_highv2}), the correspondence is quite clear both spatially and in terms of relative intensity of the blue- and redshifted outflow lobes.

In other regions, the correspondence in terms of spatial extent and intensities is poorer. This is the case, for instance, of CygX-N53 (central column of Fig.~\ref{fig:sio_co_highv2}) where SiO is almost only redshifted while CO is roughly symmetric. This is puzzling. Perhaps this is due to the powerful red lobe encountering more material which is efficiently shocked to form and entrain SiO. On the other hand, the blueshifted SiO traces only vaguely the blueshifted emission of CO. In addition, the redshifted emission seen towards the west could be due to an additional outflow, since there is a blueshifted counterpart in the same direction towards the east, at the edge of the primary beam. However, although it may be that the corresponding CO emission is outside our field of view, the absence of such outflow in CO makes it hard to understand its origin. It could be that it arises from CygX-N53 MM2, but this is not the outflow we consider for this source. \citet[][]{2013A&A...558A.125D} suggest a tentative detection of a very compact outflow using CO emission for which we estimate the respective SiO luminosities here (and hence these are only to be taken as an upper limit). For CygX-N48 (left column of Fig.~\ref{fig:sio_co_highv2}), there is also some discrepancy between the intensities of the two molecules, namely on the outflow powered by MM2. In this region, only 25\% of the high-velocity SiO luminosity arises from the outflows of MM1 and MM2, and most high-velocity outflow emission in this field is due to the outflow from IS-1 (IS-1 stands for Infrared Source 1, and corresponds to a resolved source detected in the IRAC bands at 3 and 4~$\mu$m, and unresolved at 8$\mu$m and onwards, at RA $20^h39^m02.927^s$ and Dec $42^{\circ}22' 07.32''$; using Herschel, Hennemann et al. (in prep) define this source as a more evolved object). This demonstrates the need to identify individual outflows to perform good evaluations of individual outflow power (unresolved observations of this region would have assigned the emission from IS-1 outflow to the stronger millimetre peak, MM1). 

\begin{figure*}[!t]
	\centering
	{\renewcommand{\baselinestretch}{1.1}
	\hspace{-0.1cm}
	\includegraphics[width=0.324\textwidth]{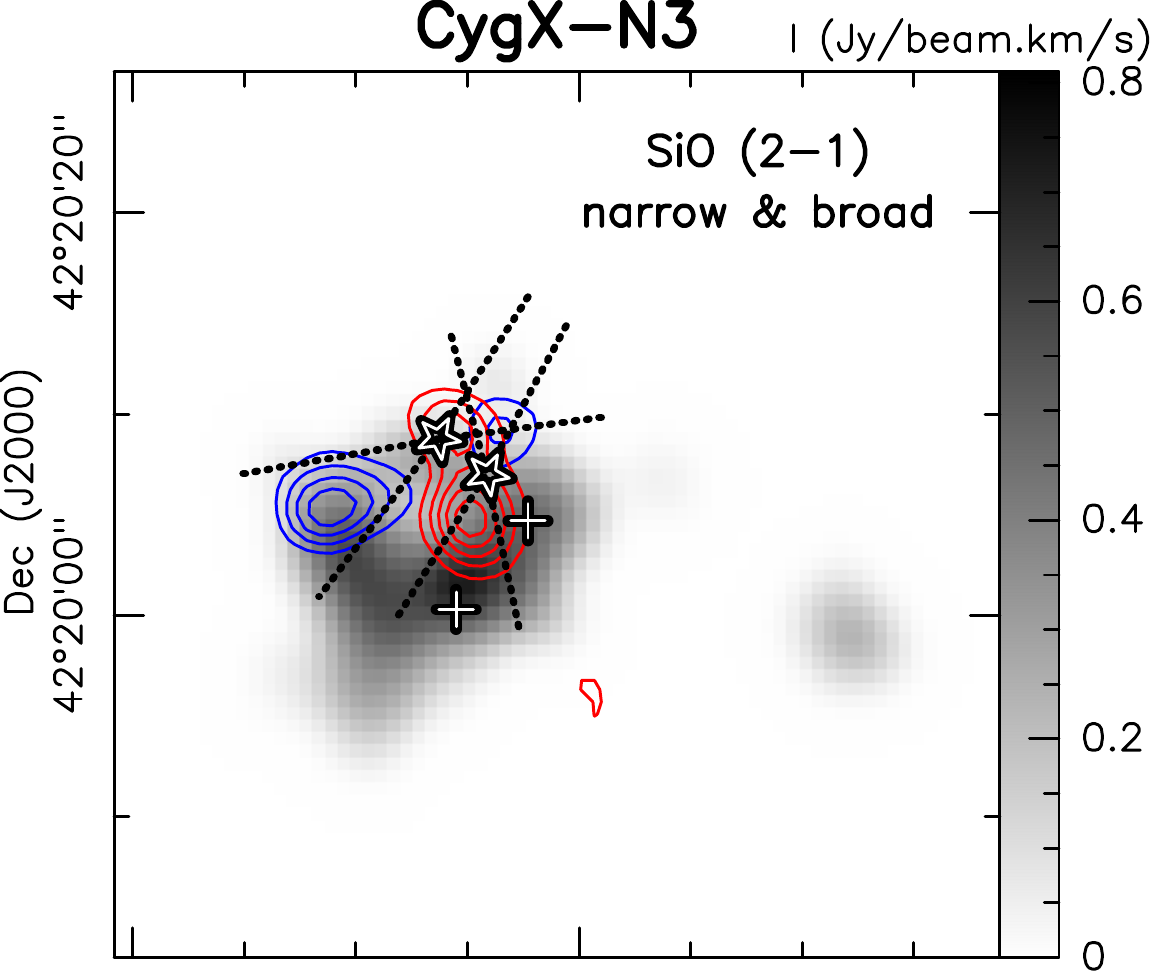}
	\hspace{0.3cm}
	\includegraphics[width=0.312\textwidth]{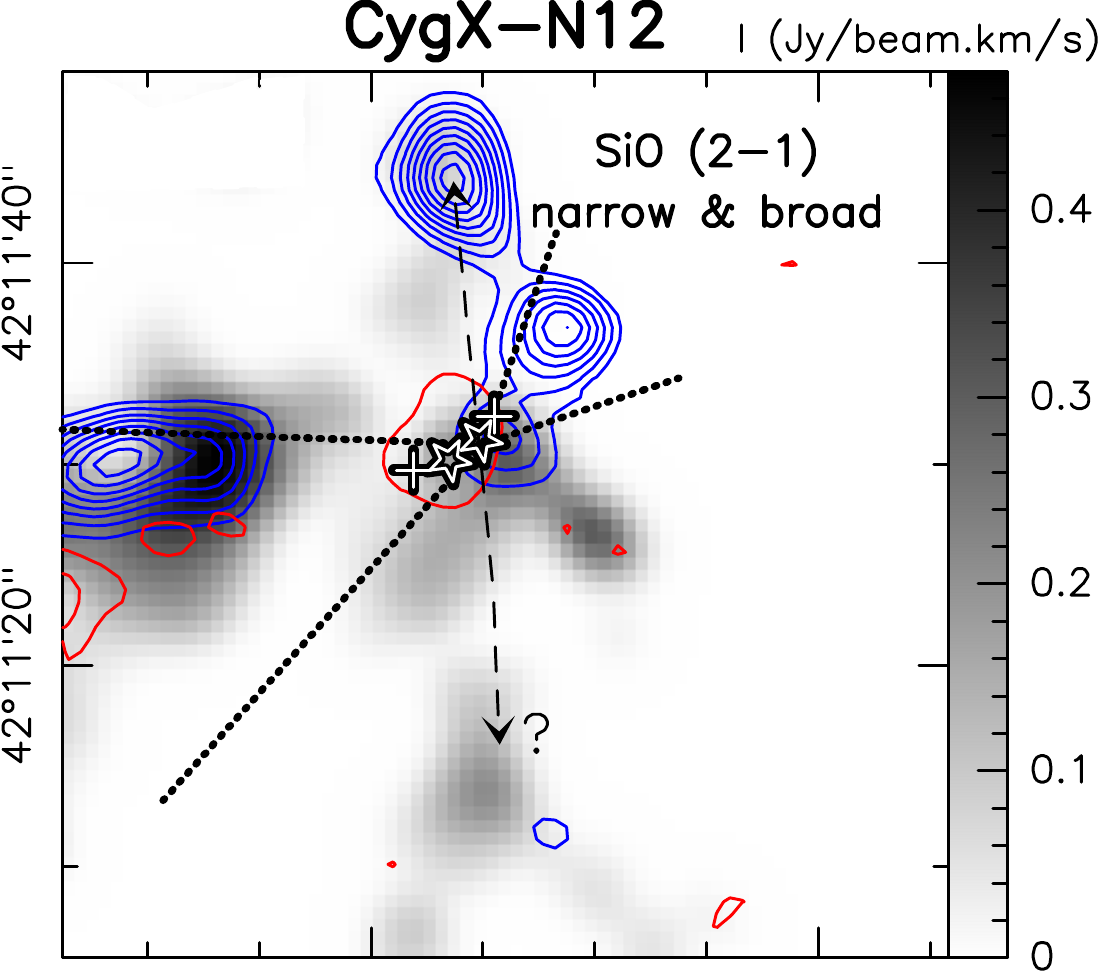}
	\hspace{0.3cm}
	\includegraphics[width=0.313\textwidth]{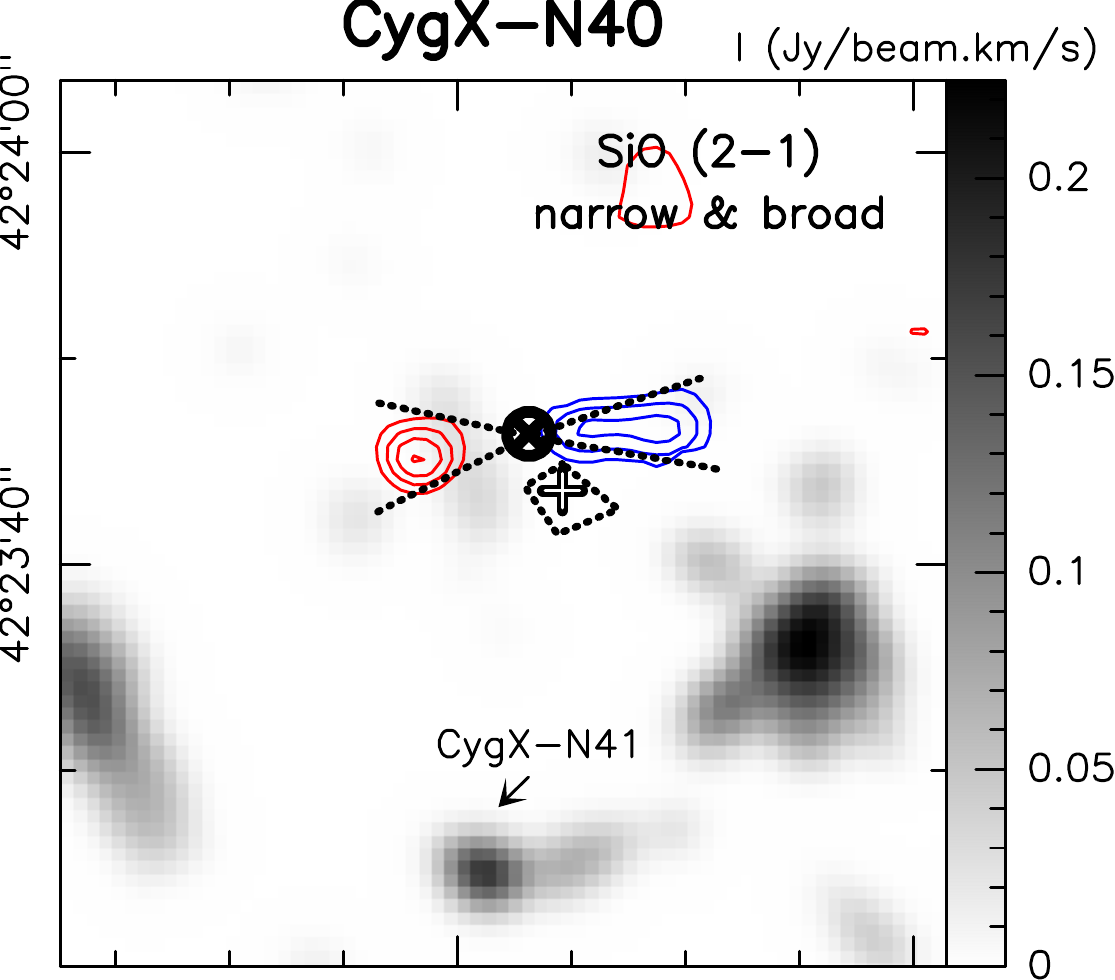}
	\includegraphics[width=0.33\textwidth]{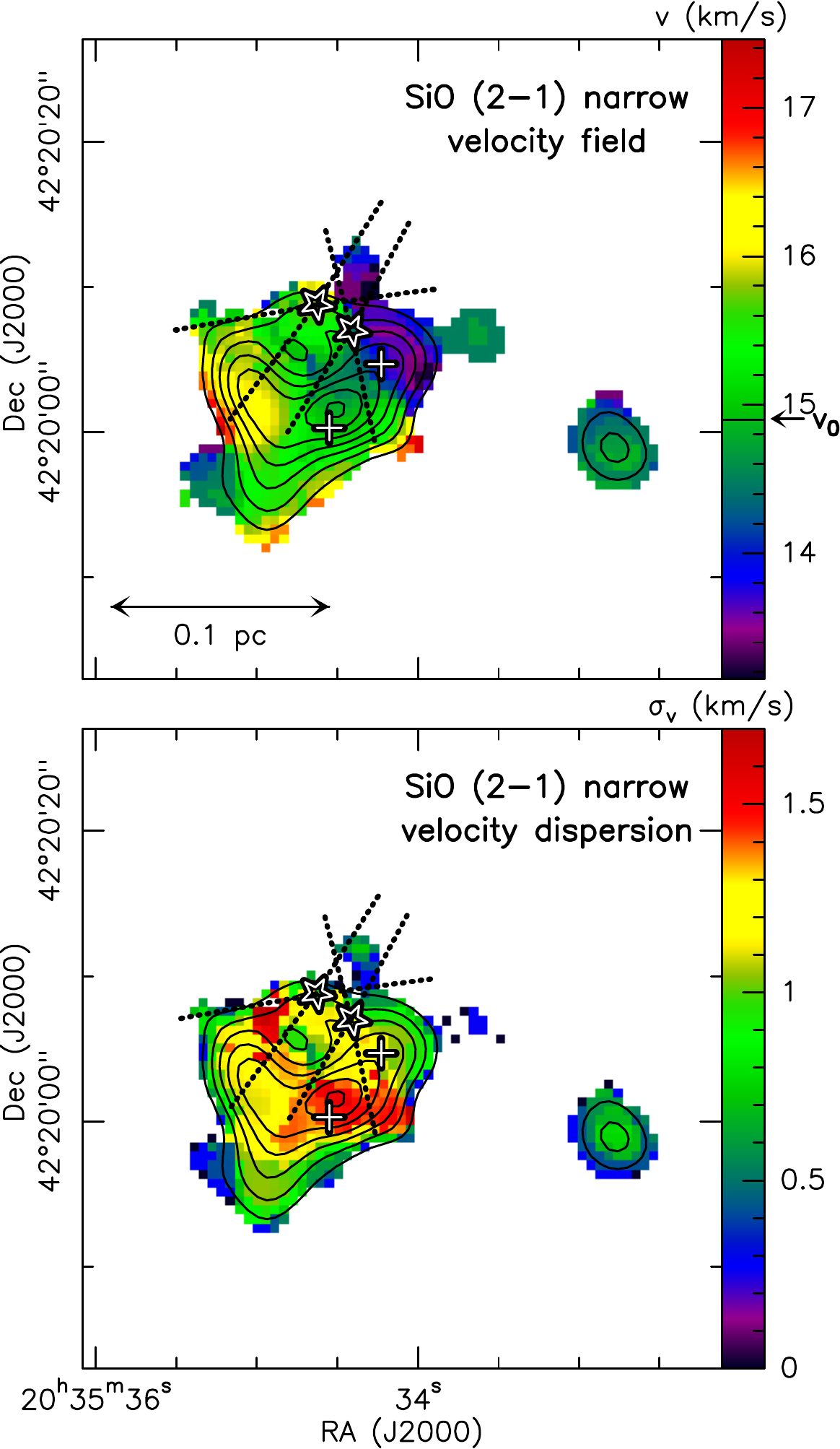}
	\hfill
	\includegraphics[width=0.315\textwidth]{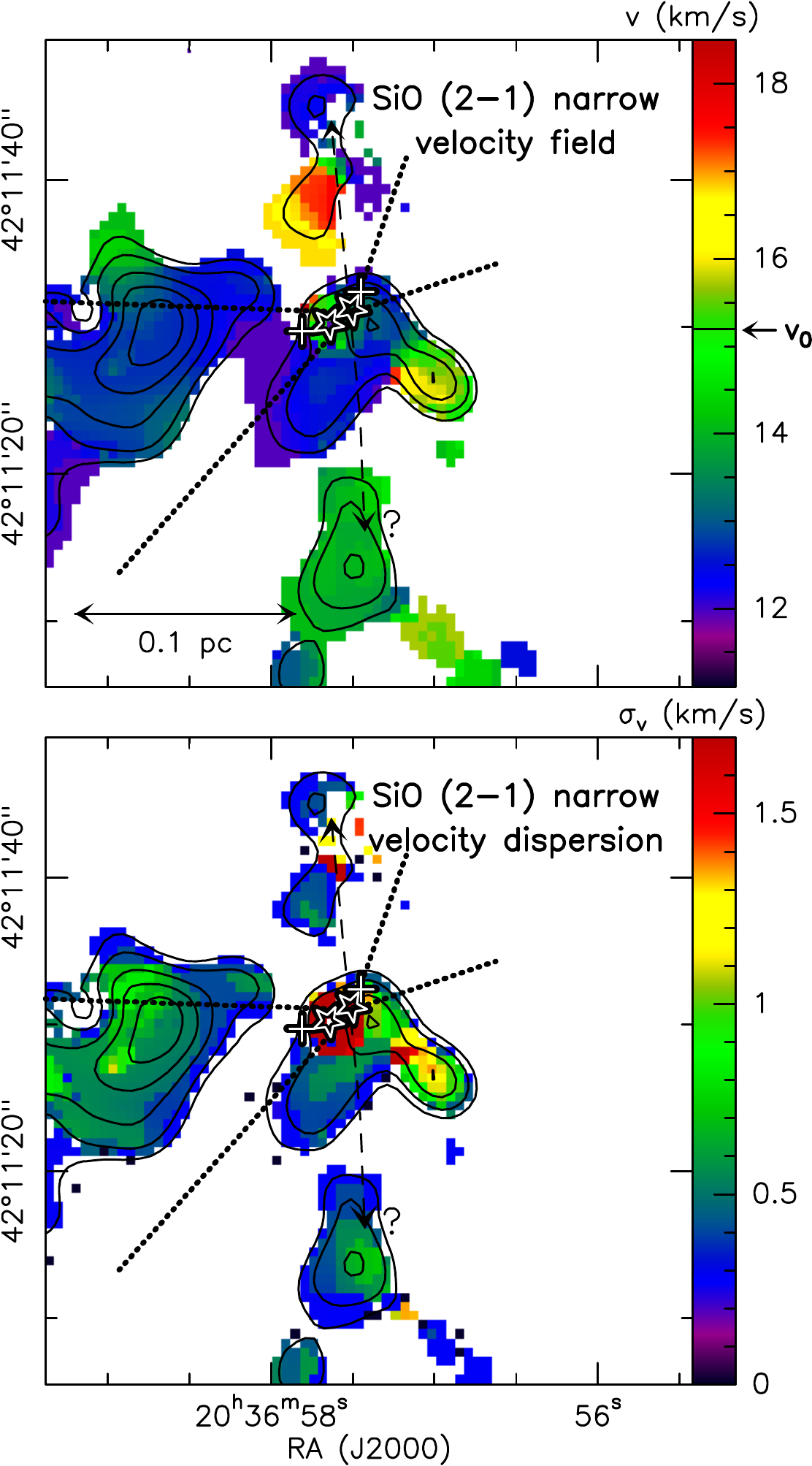}
	\hfill
	\includegraphics[width=0.316\textwidth]{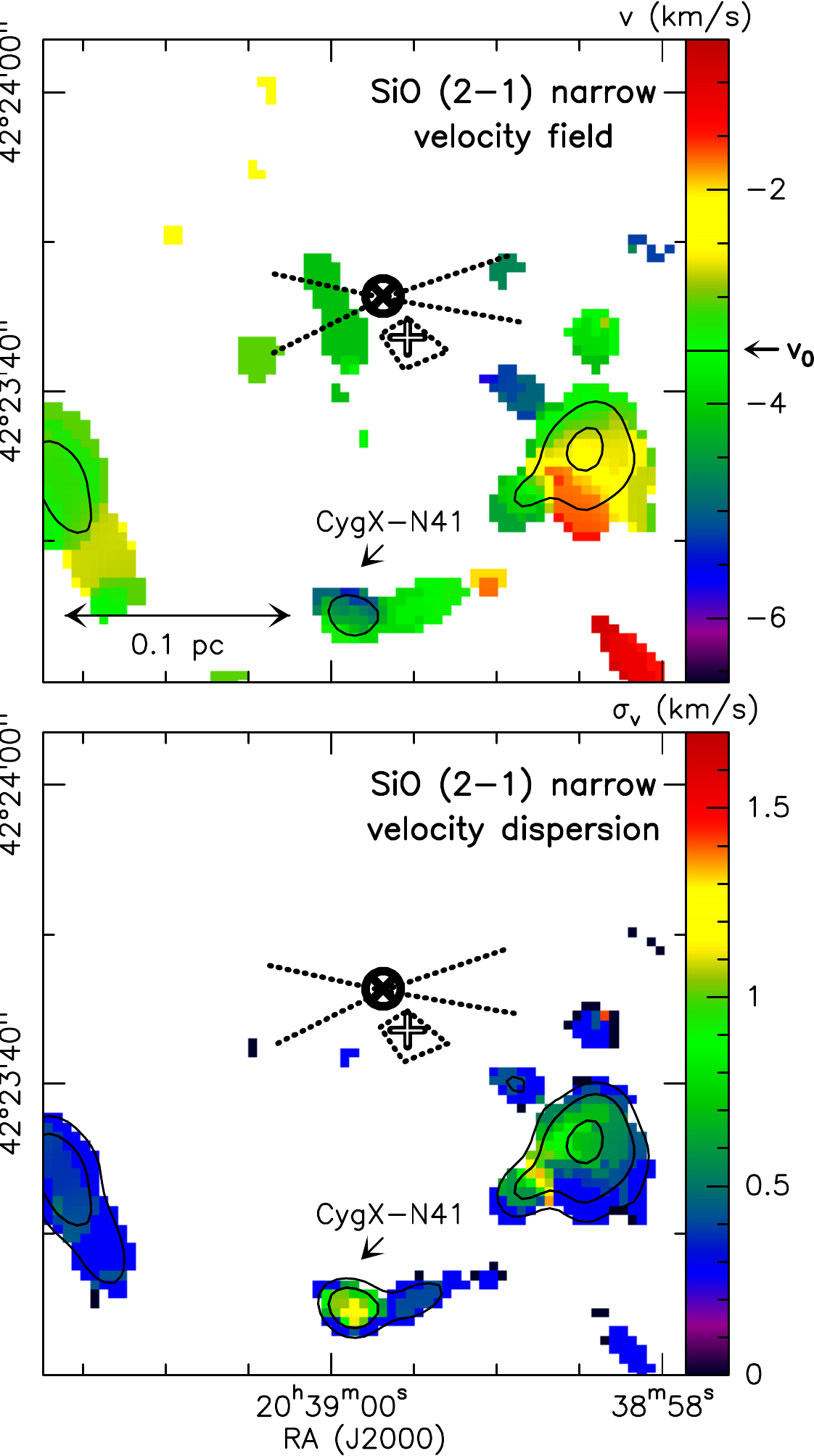}
	\caption[]{\small{Integrated intensity of the SiO emission at ambient velocities with the contribution from the broad outflow emission subtracted, for CygX-N3 (left), N12 (centre), and N40 (right). The top row panels show the narrow emission as greyscale, with contours from the high-velocity (broad) SiO emission from Fig.~\ref{fig:sio_co_highv} overlaid. The two lower rows show the integrated narrow SiO emission in contours (with steps of 0.1\,Jy\,beam$^{-1}$\,km\,s$^{-1}$, starting at 0.1\,Jy\,beam$^{-1}$\,km\,s$^{-1}$), with the colour scale showing the respective moment maps: the velocity field (first moment map) in the middle row, and the velocity dispersion (second moment map) in the last row. Annotations are as in Fig.~\ref{fig:sio_co_highv}.}}
	\label{fig:sio_narrow}}
\end{figure*}

Finally, other regions have high-velocity SiO emission at slightly different positions to where we had previously detected the CO outflows. For instance, in CygX-N12 (central column of Fig.~\ref{fig:sio_co_highv}), because of the larger field of view of the SiO observations, any outflow at the edge of the SiO maps was not covered by CO.  Close to the sources, the high-velocity emission from CO and SiO is similar (redshifted), and the larger extent of the SiO emission to the east could indicate that the red CO outflow becomes blue farther away, which suggests that this is an outflow cone close to the plane of the sky. Although it could be that the SiO blue lobe is simply part of another outflow (and unrelated with the CO red outflow lobe), its small velocity range ($\sim 10$\,km\,s$^{-1}$) and large spatial extent ($\sim 20''$, i.e. $\sim 0.13$\,pc) support the hypothesis of an outflow close to the plane of the sky. There is also some blue high-velocity SiO emission to the north which could be due to a second outflow not detected in CO (and which would follow the opposite direction of the south-directed CO lobe, marked with a dashed line). Using CO only, in \citet[][]{2013A&A...558A.125D}, we had interpreted this region as having two overlapping outflows, both E-W directed.  With SiO, this picture could perhaps be revised with the existence of a N-S outflow, but because the emission does not trace back clearly to any of the two sources, we have refrained from doing so. Instead, and for the remainder of the paper, we will take into account only the outflow directed E-W. In CygX-N40 (right column of Fig.~\ref{fig:sio_co_highv}), the outflow emission we had detected with CO in \citet[][]{2013A&A...558A.125D} close to MM1 has no SiO counterpart (perhaps it is too weak to be detected within out noise levels), while the two weak high-velocity lobes to the north-east and north-west of MM1 are detected with both molecules. This, along with the existing weak 3mm peak, lead us to think that there is a second low-mass source ($M_{\rm env}\lesssim1$\,M$_{\odot}$), to the north of MM1, and which could be responsible for the additional E-W CO and SiO outflow. We will refer to this source as CygX-N40 MM2, whose position and estimate of the $F_{\rm CO}$ \citep[as done for the other sources in][]{2013A&A...558A.125D} is shown in Table.~\ref{tab:luminosities}. In this region, only 10$\%$ of the total PdBI SiO luminosity is close to the central sources (MM2 in particular), 90\% of which is high-velocity emission.

\subsection{Distribution of the narrow SiO emission}
\label{sec:narrow_sio}

\begin{figure*}[!t]
	\centering
	{\renewcommand{\baselinestretch}{1.1}
	\hspace{-0.1cm}	
	\includegraphics[width=0.324\textwidth]{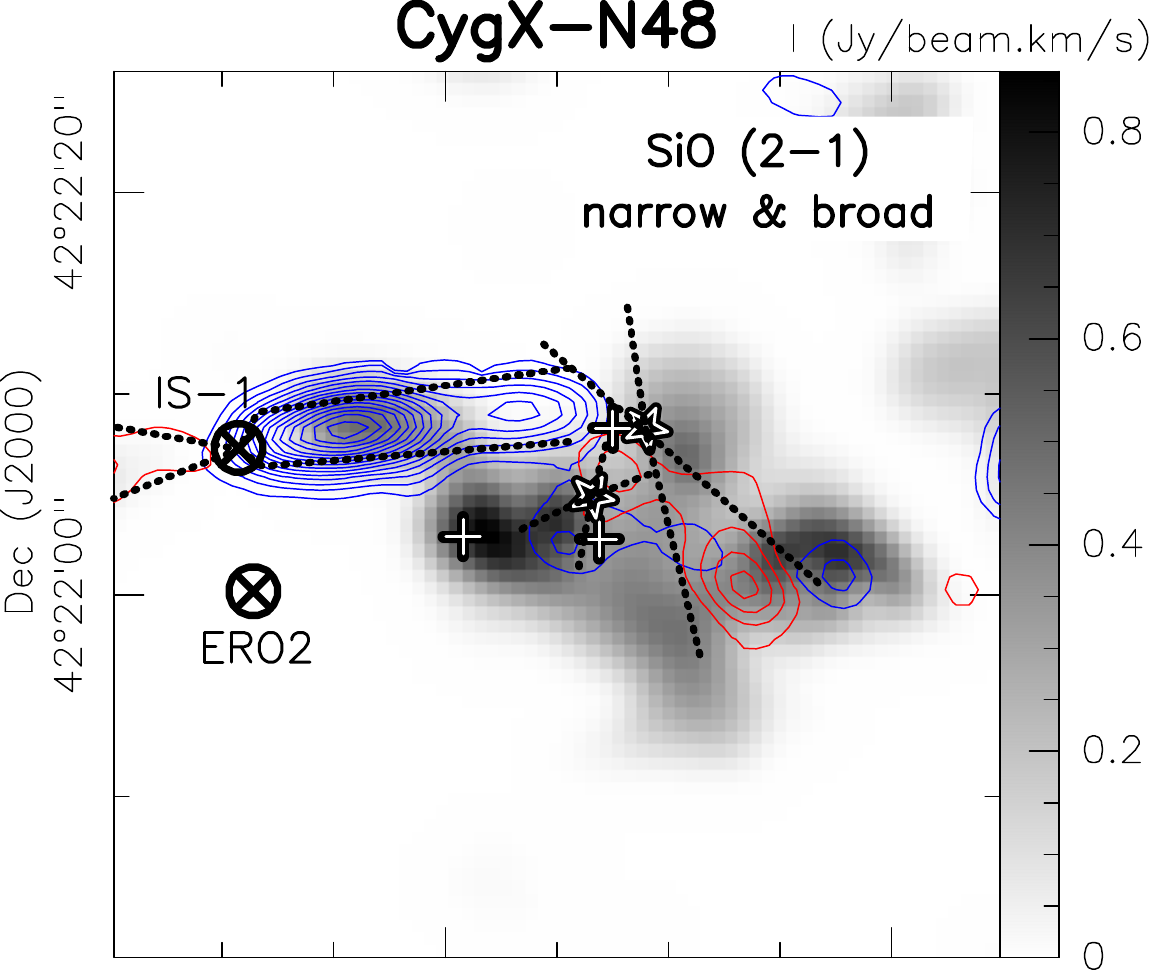}
	\hspace{0.31cm}
	\includegraphics[width=0.312\textwidth]{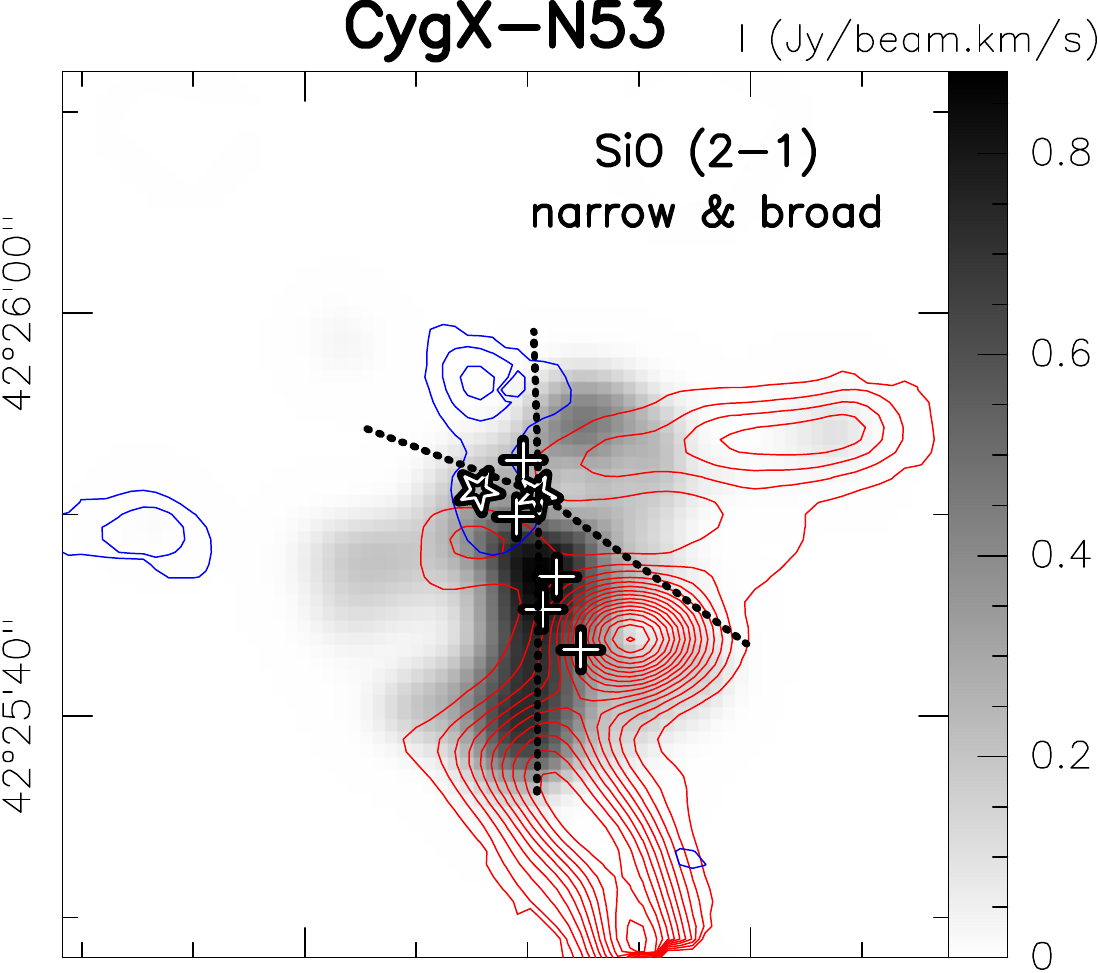}
	\hspace{0.28cm}
	\includegraphics[width=0.313\textwidth]{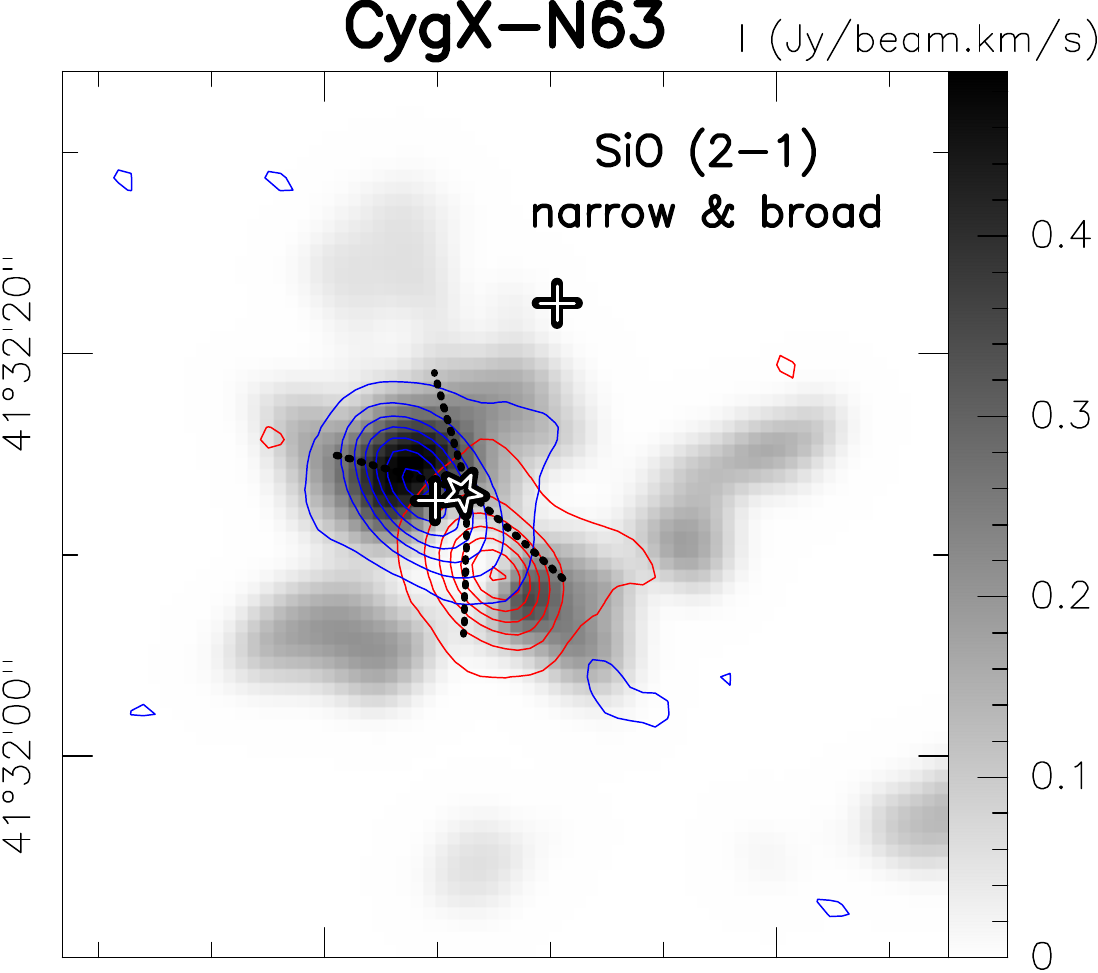}\\
	\includegraphics[width=0.33\textwidth]{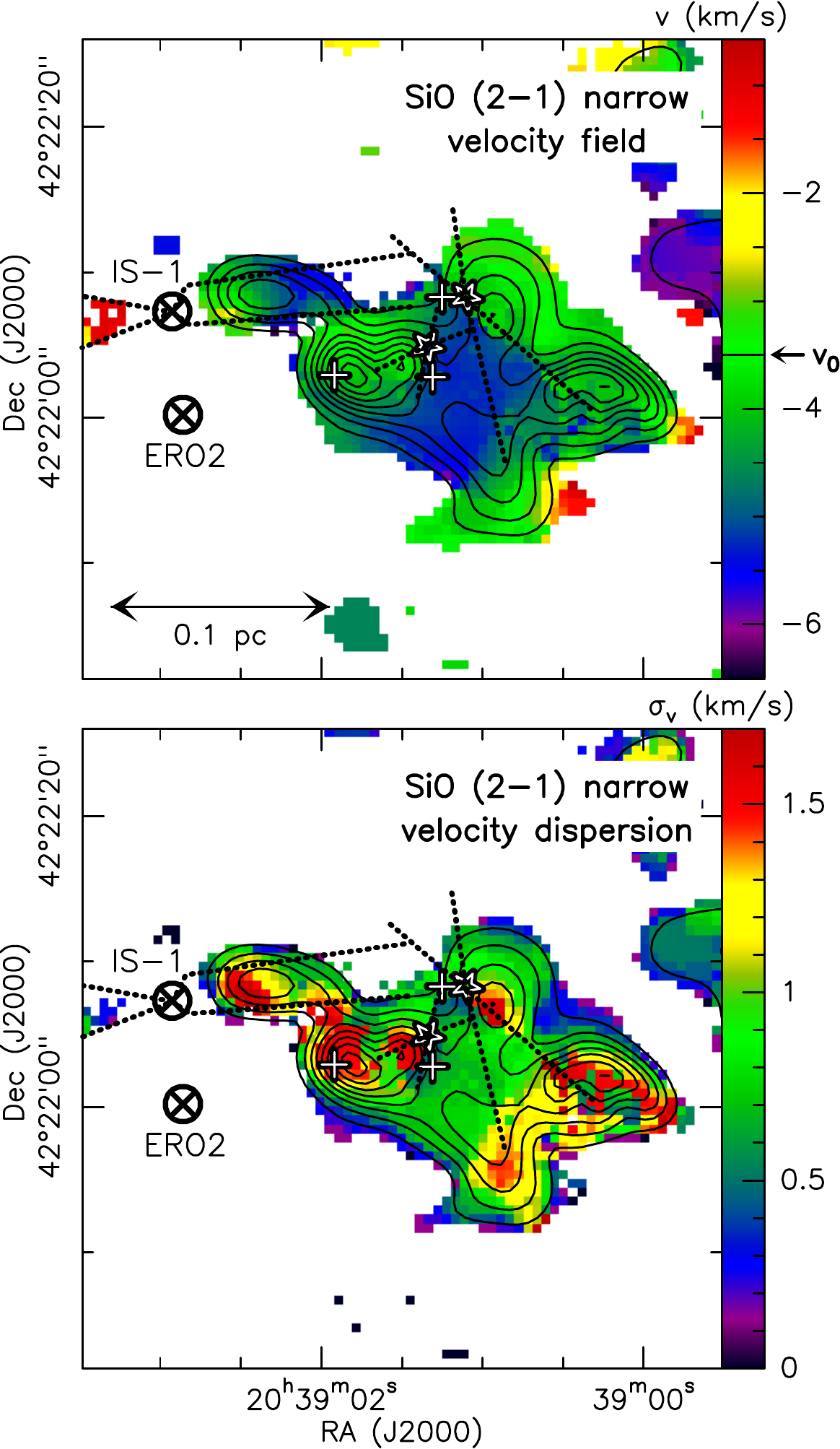}
	\hfill
	\includegraphics[width=0.316\textwidth]{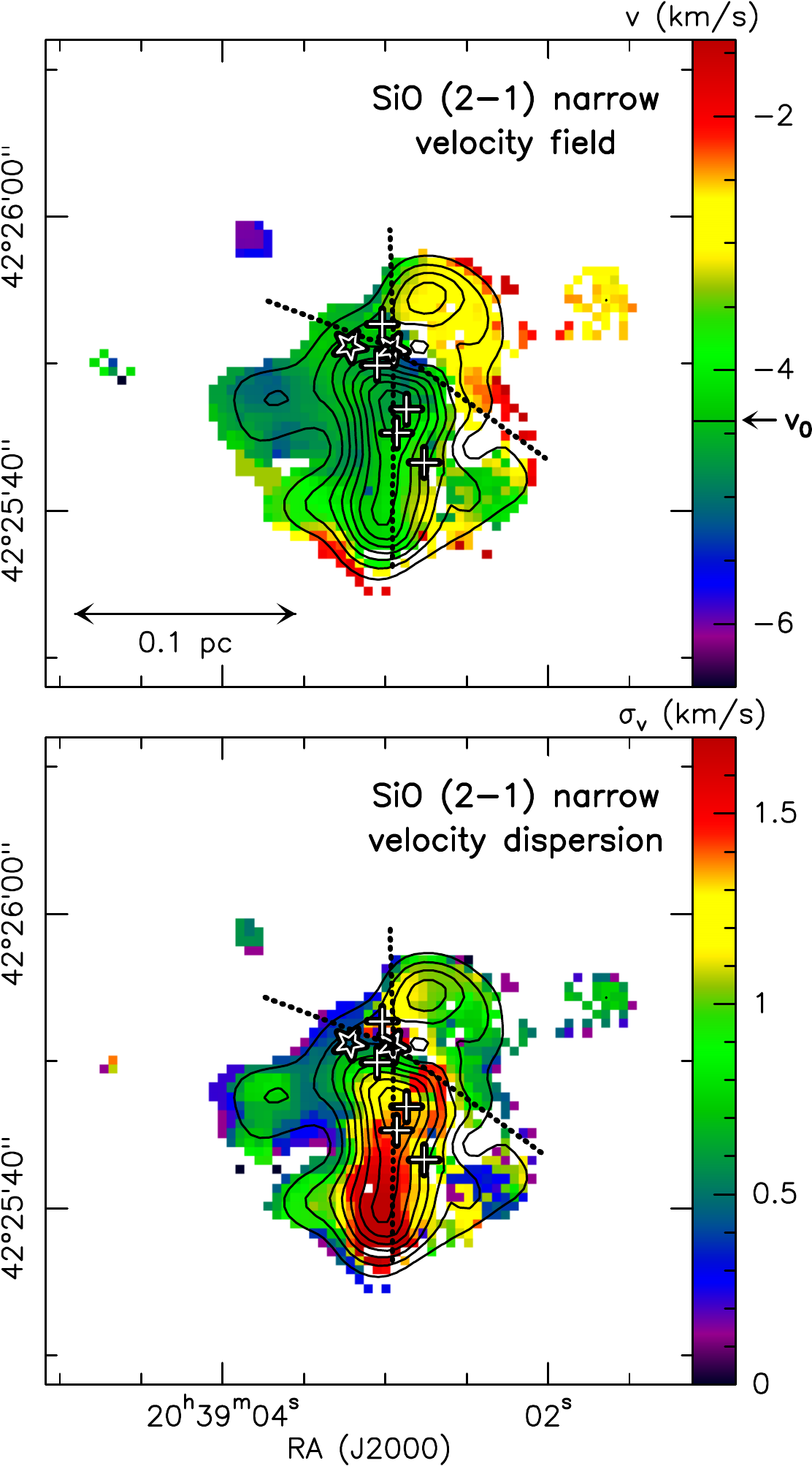}
	\hfill
	\includegraphics[width=0.316\textwidth]{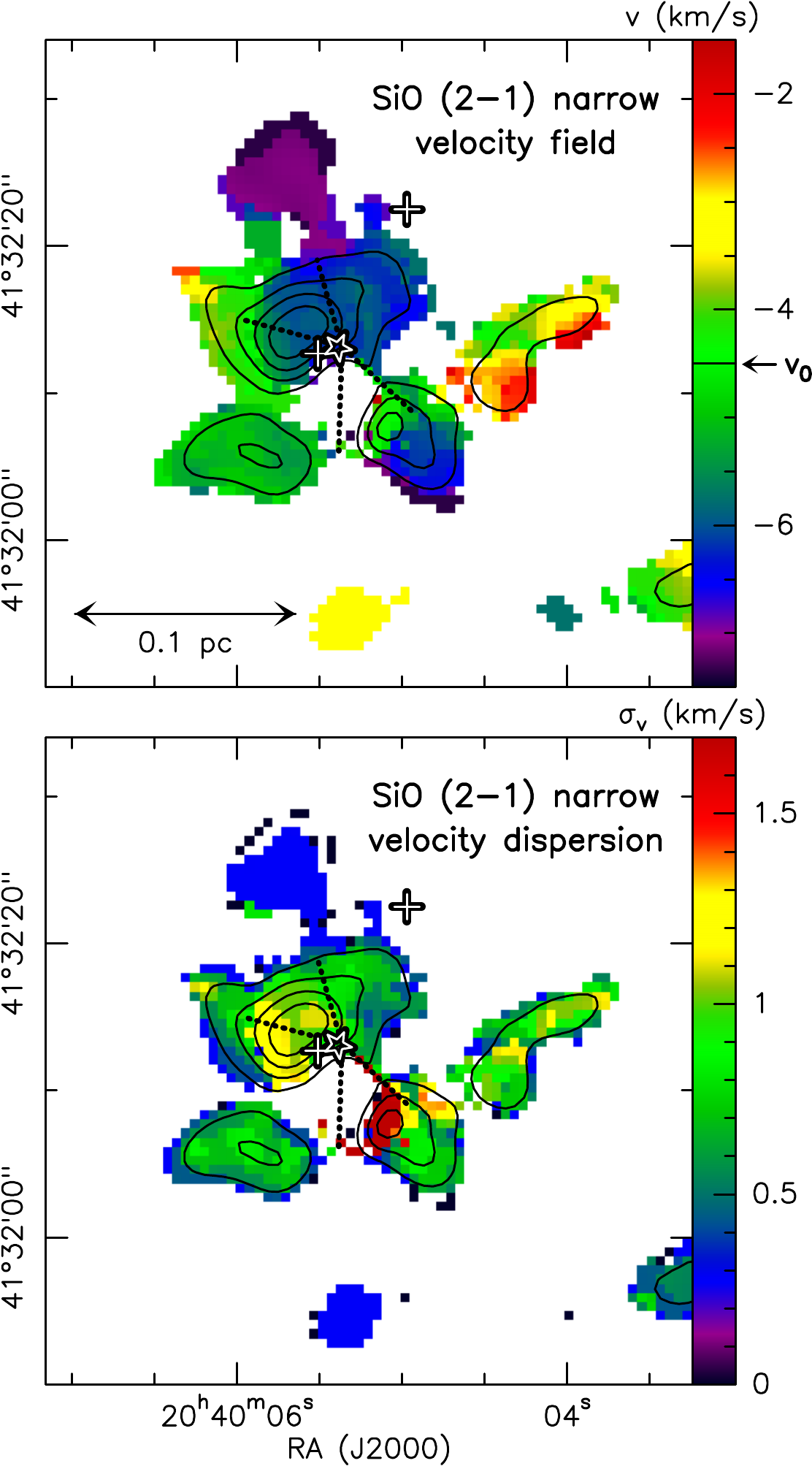}
	\caption[]{\small{Same as Fig.~\ref{fig:sio_narrow}, for CygX-N48 (left), N53 (centre), and N63 (right). Top panels have the narrow SiO emission in greyscale and the high-velocity (broad) SiO emission in contours as in Fig.~\ref{fig:sio_co_highv2}. The two lower rows show the integrated narrow SiO emission in contours (with steps of 0.1\,Jy\,beam$^{-1}$\,km\,s$^{-1}$, starting at 0.1\,Jy\,beam$^{-1}$\,km\,s$^{-1}$), with the respective moment maps in colour scale: the velocity field (first moment map) in the middle row, and the velocity dispersion (second moment map) in the last row. Annotations are as in Fig.~\ref{fig:sio_co_highv2}.}}
	\label{fig:sio_narrow2}}
\end{figure*}

We have investigated the distribution of the SiO emission close to the systemic velocities of the cloud, and whose main contribution is adequately fitted by a narrow Gaussian (see spectra in Fig.~\ref{fig:sio_average_spectra}). Since sometimes the narrow SiO emission is along the same line of sight as some high-velocity gas, separating the contribution of the outflows from the emission at systemic velocities becomes critical. In an exercise to try and remove this contribution, we have created datacubes of the narrow emission of each field, by subtracting the contribution from a broad outflow profile whenever there was significant high-velocity emission. To do so, we selected all the pixels where the integrated high-velocity SiO emission was above the rms noise level, and fitted a double Gaussian profile to the spectrum at each pixel. The fitted broad Gaussian component was then subtracted from the original spectra. By doing so, we do not make an a priori assumption on the profile of the narrow component. Instead, we remove the broad contribution and analyse what is left from this process (see Fig.~\ref{fig:example_fits_spectra} in Appendix ~\ref{ap:SiO_spectra} for an example spectrum before and after the subtraction of the broad component). Figures~\ref{fig:sio_narrow} and \ref{fig:sio_narrow2} show the spatial distribution of the moment maps (integrated intensities, velocity field, and velocity dispersion) of the SiO emission at ambient cloud velocities as a result of this exercise. The three moments were calculated using the central 6\,km\,s$^{-1}$ velocity range, centred on the $\varv_{0}$ value given in Table~\ref{tab:velocities}. Since the narrow component (as measured from the average spectra) has a $\sigma_\varv \lesssim 1.5$\,km\,s$^{-1}$, this velocity range is enough to contain more than $95\%$ of the narrow emission. Naturally, this method is only correct if/when the broad outflow emission is Gaussian, and when both narrow and broad components are well distinguished, but this is not always the case. Therefore, the resulting narrow emission maps may still show some residual broad emission (see the velocity dispersion maps of Figs.~\ref{fig:sio_narrow} and \ref{fig:sio_narrow2}).
For the purpose of these figures, we degraded slightly the spatial resolution of the integrated intensity maps (i.e. the zeroth-moment maps), by convolving with a Gaussian kernel of two arcseconds, merely so as to smooth the appearance of the contours, namely around individual pixels where the broad component was not successfully removed\footnote{This occurred in pixels where either the emission was not purely Gaussian, or the broad emission was too weak to pass our initial selection criteria for performing the broad Gaussian fitting.}. We did not do this spatial convolution for the velocity field and velocity dispersion maps, and the latter can be used to spot the positions where this method was not successful (pixels that still have large velocity dispersions, with $\sigma_\varv > 1.5$\,km\,s$^{-1}$).
 
In CygX-N3 (left column of Fig.~\ref{fig:sio_narrow}), based on the location of the CO outflows, we have determined that about 60\% of the total PdBI SiO luminosity is not constrained to the location of known outflows (as defined by the existence of SiO high-velocity emission). There is, therefore, a significant fraction of PdBI SiO emission that is not clearly associated with any outflow, and which has narrow line profiles ($\sigma_{\varv} \sim 1.2$\,km\,s$^{-1}$, i.e. FWHM $\sim 2.8$\,km\,s$^{-1}$). The velocity field of this narrow component shows a large velocity gradient across the sources (of 2\,km\,s$^{-1}$ in 0.05\,pc), with a shear-like structure aligned with the continuum filamentary structure. Interestingly, the velocity gradient seen in SiO is opposite to that of the H$^{13}$CO$^{+}$ flows seen by \citet[][also with a velocity gradient of $\sim2$\,km\,s$^{-1}$ in 0.05\,pc, but in the reverse direction; see their Fig.~11]{2011A&A...527A.135C}. 

\begin{table*}[!t]
\caption{Source properties and estimated SiO luminosities}
\begin{tabular}{l | c c c c c | c r  | c r}
\hline 
\hline 
 \multirow{4}{*} {Source}   	&	 \multirow{4}{1cm} {\hfil RA (J2000)} & \multirow{4}{1cm} {\hfil Dec J(2000)} & \multirow{4}{0.6cm} {$M_{\rm env}$ (M$_{\odot}$)}  & \multirow{4}{0.55cm} { $L_{\rm bol}$ (L$_{\odot}$)} & \multirow{4}{2.1cm} {\hfil $F_{\rm{CO}} \times 10^{-5} $  (M$_{\odot}$km\,s$^{-1}$yr$^{-1}$) } &  \multicolumn{4}{c}{L$_{\mathrm{SiO}}$ ($10^{8}$ K\,km\,s$^{-1}$pc$^{2}$)}   \\
		&	 		&  			&	 				&	  				&			& \multicolumn{2}{c |}{\multirow{2}{*}{Outflow emission}} & \multicolumn{2}{c } {\multirow{2}{1.5cm}{Entire field} }\\
     		&	 		& 			& 	 				&	 				&			&  \multirow{2}{*}{Total}			& 		\multirow{2}{1.2cm}{High-$\varv$ } 			& \multirow{2}{*}{Total} & \multirow{2}{1.2cm}{Narrow} \\		
     		&	 		& 			& 	 				&	 				&			& 		& 			& \\
\hline
N3-MM1	& 20:35:34.63	& 42:20:08.8	&	12.5	$\pm$ 3.7		&	106	$\pm$	60 	&	131		&    0.75		&    0.37 	(50\%) 	& \multirow{2}{*} {3.85} & \multirow{2}{*} {1.86 (48\%)} \\
N3-MM2	& 20:35:34.41	& 42:20:07.0	&	13.8	$\pm$ 5.6		&	121	$\pm$	50 	&	72		&    0.79		&    0.29 	(37\%)	& & \\
\hline
N12-MM1	& 20:36:57.65	& 42:11:30.2	&	17.7	$\pm$ 6.9		&	485	$\pm$	130	&	$>$36	& \multirow{2}{*} {2.65$^{(a)}$}  & \multirow{2}{*} {1.39 $^{(a)}$ (52\%)} & \multirow{2}{*} {4.61} & \multirow{2}{*} {2.07 (45\%)} \\
N12-MM2	& 20:36:57.51	& 42:11:31.2	&	16.4	$\pm$ 5.8 	&	195	$\pm$	75	&	$>$12	&   			& 	 	 	      	& & \\
\hline
N40-MM1	& 20:38:59:54	& 42:23:43.6	&	1.9	$\pm$ 0.5		&	-				&	7		&   -		  	 & -	 			&  \multirow{2}{*} {3.29} & \multirow{2}{*} {1.77 (54\%)} \\
N40-MM2	& 20:38:59:69	& 42:23:46.3	&	$1.0\pm$ 0.5		&	-				&	28		&   0.33	  	 & 0.29	(88\%) 	&  & \\
\hline
N48-MM1	& 20:39:01.34	& 42:22:04.9	&	17.0	$\pm$ 6.4		&	102	$\pm$	70	&	135		&   0.99		&  0.89  	(90\%) 	& \multirow{2}{*} {11.84$^{(b)}$ }& \multirow{2}{*} {4.6 (38\%)} \\
N48-MM2	& 20 39 01.10	& 42 22 08.3	&	8.1 	$\pm$  3.0 	&	85 	$\pm$ 60		&	45		&   2.00		&  0.51  	(26\%) 	& & \\
\hline
N53-MM1	& 20:39:02.96	& 42:25:51.0	&	34.2	$\pm$ 11.1	&	199	$\pm$	70	&	412		&   10.34  		&  7.58 	(73\%)	& \multirow{2}{*} {11.57} & \multirow{2}{*} {2.06 (18\%)} \\
N53-MM2	& 20:39:03.22	& 42:25:51.2	&	21.4	$\pm$ 6.1		&	144	$\pm$	50	&	$<$121	&   $<$0.23    	&  $<$0.11 (48\%)	& & \\
\hline
N63-MM1	& 20:40:05.39	& 42:32:13.1	&	44.3	$\pm$ 11.9	&	339	$\pm$	50	&	291		&  4.89	   	&  3.49  	(71\%) 	& 5.93 &  1.08 (18\%) \\
\hline
\end{tabular}\\
\flushleft
{\footnotesize
$^{(a)}$ With the resolution of our SiO maps and the configuration of the outflows in N12, we cannot distinguish the outflow from the two millimetre sources. The values presented are for the combination of the two.\\ 
$^{(b)}$ The total SiO luminosity in N48 includes the high-velocity outflow from IS-1 and a high velocity blue and red emission to the west, each of which account for 20\% of the total SiO luminosity. }
\label{tab:luminosities}
\end{table*}

In CygX-N12 (central column of Fig.~\ref{fig:sio_narrow}), $45\%$ of the PdBI SiO luminosity arises from narrow low-velocity gas ($\sigma_{\varv} \sim 0.8$\,km\,s$^{-1}$, i.e. FWHM $\sim 1.8$\,km\,s$^{-1}$). This includes the narrow blueshifted emission towards the southern edge of the blueshifted outflow lobe (see blue emission in the velocity field panel of Fig.~\ref{fig:sio_narrow}, or narrow spectra from Fig.~\ref{fig:sio_indiv_spectra}), which accounts for the missed emission from the inner $30''$ but recovered in the $50''$ area. In this particular case, the narrow SiO emission is not spatially overlapping with an outflow and yet it is clearly associated with one, perhaps indicating that we are seeing the post-shock gas at the edges of an outflow cavity. The remaining ambient gas has a spatial distribution slightly extended along the N-S direction around the central sources, with a rather complex velocity field which makes it hard to compare with the (also complex) velocity field from H$^{13}$CO$^{+}$ \citep[][]{2011A&A...527A.135C}. 

For CygX-N40 (right column of Fig.~\ref{fig:sio_narrow}), a significant amount of the total PdBI emission arises from systemic velocities (54$\%$), in the form of extended narrow emission, with a velocity dispersion reaching as low as $\sigma_{\varv} \sim 0.3$\,km\,s$^{-1}$ (i.e. FWHM $\sim 0.7$\,km\,s$^{-1}$). Some of the SiO emission is associated with DR21(OH)N1 \citep[or CygX-N41 from][]{2007A&A...476.1243M}, seen as a 3mm continuum peak to the south of MM1. The existence of negative side-lobes in SiO between MM1 and the arc-shaped emission around MM1 and MM2 means that despite recovering a good fraction of the emission observed with single-dish, the emission is quite extended and filtered out by the PdBI. Therefore, the fidelity of the imaging may not be very good (and the arc structure may not be real). Nevertheless, the existence of such emission is significant. It is not associated with the central cores, nor with outflows, and its origin is to be elucidated.

For CygX-N48 (left column of Fig.~\ref{fig:sio_narrow2}), even though most of the PdBI SiO luminosity is associated with low-velocity gas ($\sim$ 70\%), there is a significant spatial overlap of the narrow SiO emission with high-velocity SiO, suggesting that a fraction of the low-velocity SiO emission could be associated with the outflows of the region (e.g. the emission associated with the IS-1, MM1, and/or MM2 outflows). Nevertheless, there is some narrow SiO emission which lies outside outflow areas ($\sim 20\%$). The origin of this emission could be associated with the low-velocity shocks detected in this region by \citet[][]{2011ApJ...740L...5C} in N$_{2}$H$^{+}$ (see Sect.~\ref{sec:othernarrow} below).

In CygX-N53 (central column of Fig.~\ref{fig:sio_narrow2}) there is only a relatively small amount of PdBI SiO emission at systemic velocities ($\sim$35\%). This emission shows a SE-NW elongation that has a velocity gradient of 2\,km\,s$^{-1}$ in 0.1pc, with a direction reminiscent of the rotation of the MDC and the envelope \citep[as seen in H$^{13}$CO$^{+}$; see Fig.\,5 of][]{2011A&A...527A.135C}. The narrow SiO emission to the south, however, could be tracing the lower-velocities of the redshifted outflowing gas.

Finally, in CygX-N63 (right column of Fig.~\ref{fig:sio_narrow2}), the narrow emission is broken up in two areas. One to the north of (and including) MM1, which shows a velocity gradient across MM1 (E-W), with a direction consistent with that seen in H$^{13}$CO$^{+}$ \citep[][]{2011A&A...527A.135C}. This emission is coincident with the blueshifted outflow lobe, and the sharp velocity gradient of $\sim2$\,km\,s$^{-1}$ in 0.05pc is coincident with the eastern edge of the lobe, perpendicular to the direction of the outflow. This could indicate that the SiO is tracing the cavity walls of the outflow. Farther south, there is another E-W extended emission that has narrow linewidths ($\sigma_{\varv} \sim 0.8$\,km\,s$^{-1}$, i.e. FWHM $\sim 1.8$\,km\,s$^{-1}$). While the central peak could be interpreted as being part of the redshifted outflow walls, the origin of the extensions to the east and west are less clear.

\subsection{Narrow SiO column densities}
\label{sec:coldens}

To estimate the order of magnitude of the column densities of the SiO narrow component, we have used the IRAM 30m spectra of SiO  \citep[from][]{2007A&A...476.1243M}. Since these spectra include a contribution from broad outflow emission, we performed double Gaussian fittings to extract the narrow component. The H$_{2}$ volume densities in the MDCs are of the order of $\sim10^{6} - 10^{7}$\,cm$^{-3}$ \citep[$\sim10^{6}$\,cm$^{-3}$ being the average MDC densities, while $10^{7}$\,cm$^{-3}$ is more representative of the average density of the protostellar envelopes; see e.g.][]{2014arXiv1404.4843L}, which means that we are above the SiO critical densities, and hence we can assume LTE to estimate the column densities \citep[as e.g. in][]{2013ApJ...775...88N}. Assuming a temperature of $20-40$\,K, we estimate that the observed beam-averaged SiO integrated intensities for the narrow component correspond to SiO column densities of the order of $0.1-4 \times 10^{12}$\,cm$^{-2}$, where the optical depths are largely below one. We note, however, that this corresponds to the beam-averaged SiO column densities, within the IRAM 30m beam. From the PdBI, we can see that the emission is more compact than the 30$''$ beam, and therefore these column densities are merely a lower limit for the effective column densities. The PdBI emission having linear dimensions typically of the order of $\sim 15''$ (i.e. $\sim 0.1$\,pc) implies that the beam filling factors are easily low enough to bring the column densities to values closer to $\sim 10^{13}$\,cm$^{-2}$. Since the H$_{2}$ column densities in the MDCs range between $6\times10^{23}$\,cm$^{-2}$ and $2\times10^{24}$\,cm$^{-2}$, we retrieve SiO abundances of the order of $0.5-1.0 \times 10^{-11}$ for the narrow SiO emission \citep[which are similar to the abundance values of narrow SiO emission found in IRDCs by, e.g.][]{2010MNRAS.406..187J,2013ApJ...773..123S}.

\section{Analysis}
\label{sec:analyse}

\subsection{SiO outflow luminosity}

\begin{figure}[!t]
	\centering
	{\renewcommand{\baselinestretch}{1.1}
	\includegraphics[width=0.45\textwidth]{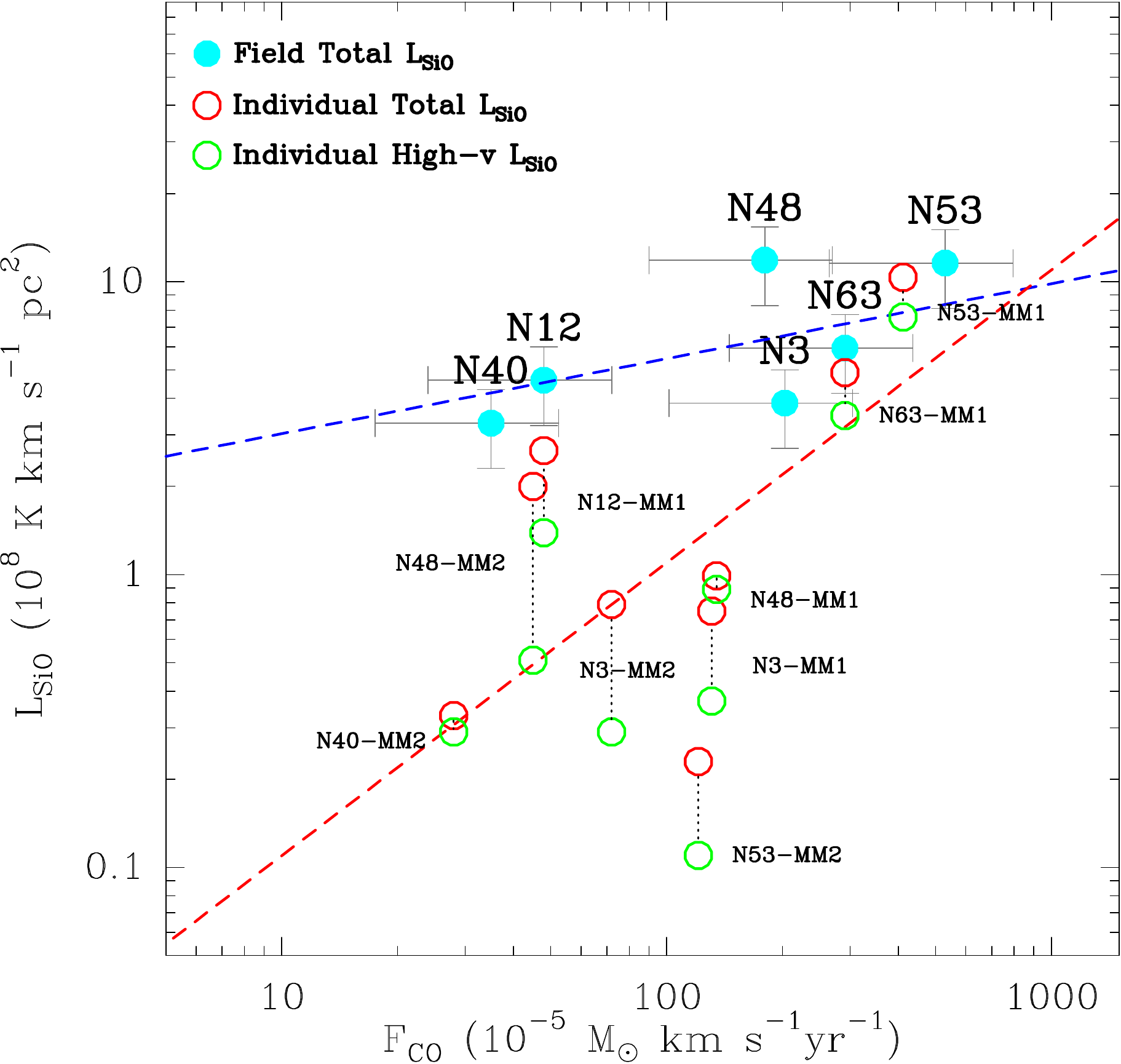}
	\caption[]{\small{Observed $L_{\rm SiO}$ against the corresponding $F_{\rm{CO}}$ \citep[from][]{2013A&A...558A.125D}. The light blue filled circles correspond to the total $L_{\rm SiO}$ and total $F_{\rm{CO}}$ for each of the six fields, and the blue dashed line is the respective linear fit in a log-scale. The open circles are for the resolved individual outflows (integration spatially constrained to location of the individual outflows), with the $F_{\rm{CO}}$ estimated in \citet[][]{2013A&A...558A.125D} (Table~\ref{tab:luminosities}, Col. 6). The red open circles have $L_{\rm SiO}$ estimated using the entire velocity range (Table~\ref{tab:luminosities}, Col. 7), and the respective linear fit to the log-values is shown as a dashed red line, which translates into a linear correlation of $L_{\rm SiO}  \propto F_{\rm{CO}}$. The green open circles show the $L_{\rm SiO}$ only from the outflow high-velocity wing emission (Table~\ref{tab:luminosities}, Col. 8).}}
	\label{fig:lsio_fco}}
\end{figure} 

\begin{figure}[!t]
	\centering
	{\renewcommand{\baselinestretch}{1.1}
	\includegraphics[width=0.45\textwidth]{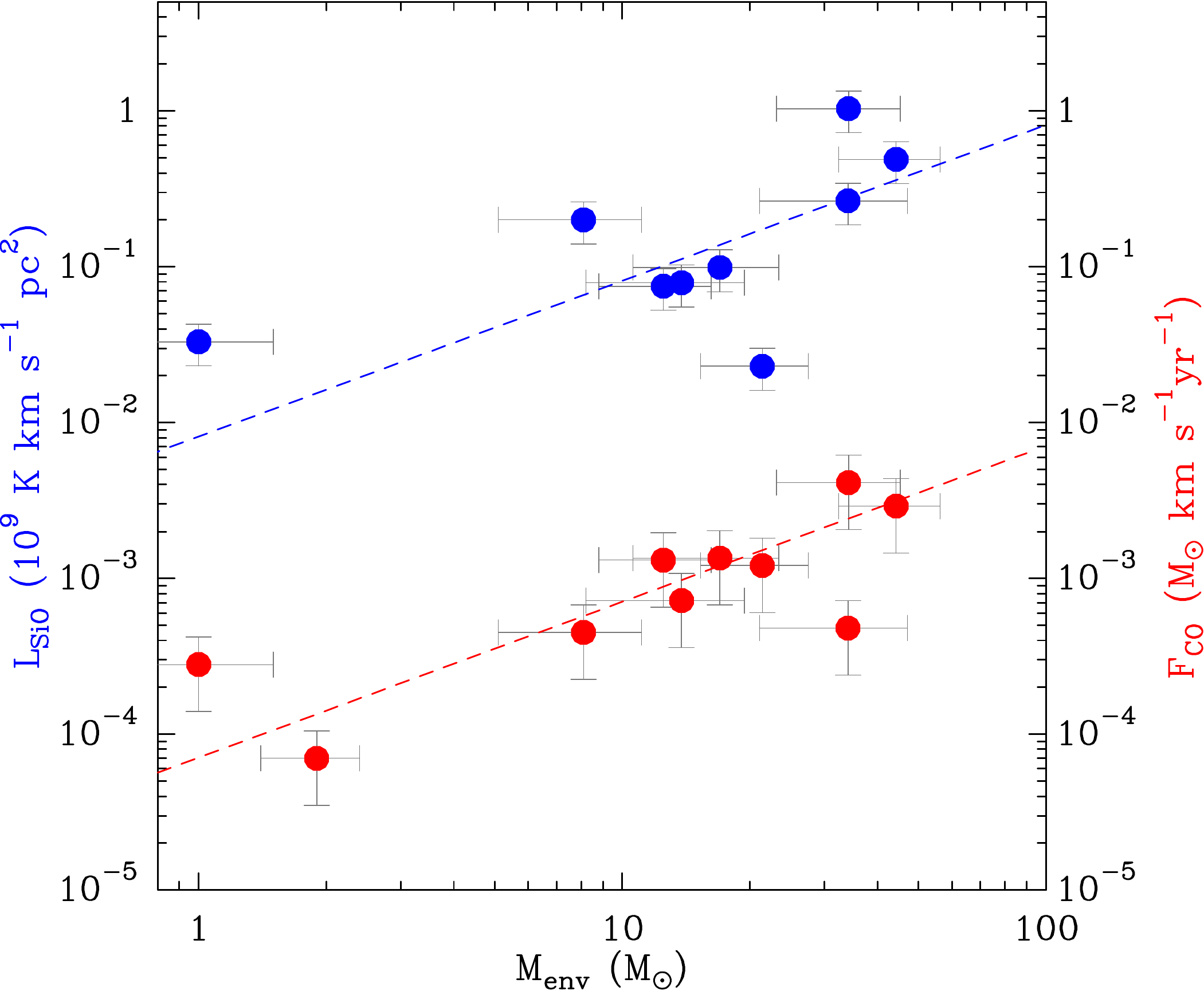}
	\vspace{-0.1cm}	
	\caption[]{\small{Observed  $L_{\rm SiO}$ (same as the red open circles in Fig.~\ref{fig:lsio_fco}) and $F_{\rm{CO}}$, in blue and red circles, respectively, with respect to the envelope masses ($M_{\rm env}$) for individual sources. The second red point from the left corresponds to CygX-N40 MM1, for which we have an $F_{\rm{CO}}$ measurement, but no $L_{\rm{SiO}}$. The dashed red line shows the linear correlation of $F_{\rm{CO}} = 7.5 \times 10^{-5} M_{\rm env}$, originally found by \citet[][]{1996A&A...311..858B} for low-mass protostars, and confirmed in \citet[][]{2013A&A...558A.125D} to extend to the high-mass regime. The dashed blue line shows the linear correlation of $L_{\rm SiO}  \simeq 5\times 10^{6} M_{\rm{env}}$ that we find for the blue points. This figure accentuates the fact that  $L_{\rm SiO}$ of individual outflows follows the same trend as $F_{\rm{CO}}$ (both being $\propto$\,$M_{\rm env}$), even though it introduces more scatter with respect to the linear relation found for $F_{\rm{CO}}$.}}
	\label{fig:lsio_menv_fco}}
\end{figure}

To investigate the relation between the SiO luminosity,  $L_{\rm SiO}$, and the respective CO outflow momentum flux, $F_{\rm{CO}}$, we have estimated the $L_{\rm SiO}$ associated with a given outflow,  whenever possible. In practice, this involved the usage of polygons to limit the areas where to estimate the SiO flux. These areas were determined based on the extent and morphology of the SiO emission, but with the help of the CO emission presented in \citet[][]{2013A&A...558A.125D}, as the CO observations allow a more accurate identification of the driving sources of the different outflow lobes. We estimated this $L_{\rm SiO}$ for each source, first using the total velocity range, and then by restraining the calculations to the high-velocity SiO emission. 

Table~\ref{tab:luminosities} summarises the results from this exercise, for the sample of sources for which we have been able to individually estimate the SiO luminosity. Columns 1 to 6 indicate the source names and positions, the envelope mass and bolometric luminosities retrieved from SED fittings, and the CO outflow momentum flux \citep[the last three are from][]{2013A&A...558A.125D}. Columns 7 and 8 show the SiO luminosities associated with individual outflows, and the second-to-last column shows the total SiO luminosity in the entire PdBI field (equivalent to the integrated SiO emission that would be observed over the same field by a single-dish telescope).  The last column shows the SiO luminosity in the entire PdBI field that arises from narrow emission (i.e. calculated from the integrated intensity maps shown in Figs.~\ref{fig:sio_narrow} and \ref{fig:sio_narrow2}).

The first result from this estimate demonstrates that there is no 1:1 relation between $L_{\rm SiO}$ and $F_{\rm{CO}}$. Nevertheless, the two stronger outflows in the sample (from the two most massive sources, CygX-N53 MM1 and CygX-N63 MM1) are also those showing the most luminous SiO emission. This is illustrated well in Fig.~\ref{fig:lsio_fco}, where we can see that the two stronger outflows (with $F_{\rm{CO}} > 2\times10^{-3}$ M$_{\odot}$\,km\,s$^{-1}$\,yr$^{-1}$) also have strong SiO luminosities (both when taking the entire fields, and the individual sources). However, for the remaining fields, a correlation is less evident. When measuring the SiO luminosities on entire fields, which in essence mimics observations of SiO that do not resolve individual outflows, we see no correlation of SiO luminosity with the corresponding outflow power (the dashed blue line  in Fig.~\ref{fig:lsio_fco} shows the tentative fit to the light blue circles, with a weak dependency of $L_{\rm SiO} \propto F_{\rm{CO}}^{\,\,\alpha \lesssim 0.3}$). When we are able to distinguish individual outflows, then the SiO luminosities do appear to correlate linearly with the outflow momentum fluxes measured from CO lines (see dashed red line in Fig.~\ref{fig:lsio_fco}, which corresponds to $L_{\rm SiO}  \propto F_{\rm{CO}}$). This result suggests that SiO luminosities associated with individual outflows could statistically be used as a proxy for the outflow power. However, as shown in Table 3, there is a significant scatter of more than one order of magnitude for the SiO luminosities, for similar values of $F_{\rm{CO}}$. The larger scatter in $L_{\rm SiO}$ is also visible in Fig.~\ref{fig:lsio_menv_fco}, where we show the relation between both $L_{\rm SiO}$ (in blue) and $F_{\rm{CO}}$ (in red) with the $M_{\rm env}$, for the protostars of our sample. This figure shows that both $L_{\rm SiO}$ and $F_{\rm{CO}}$ scale linearly with envelope mass, when individual outflows can be resolved, in spite of the larger apparent scatter from the $L_{\rm SiO}$. We caution that these results are based on our small sample, and need to be confirmed with larger samples.

\subsection{Nature of the narrow SiO emission}

The fraction of SiO emission at low velocities is highly variable (Table~\ref{tab:luminosities}), and while in regions with strong outflows (e.g. CygX-N53 and CygX-N63) the SiO high-velocity emission accounts for 80\% of the full SiO luminosity, other sources have the inverse situation, where most of the SiO emission arises from narrow emission at systemic velocities. Such emission could be due to shocks other than outflows. To help clarify the nature of the narrow SiO emission, we summarise the results from Sect.~\ref{sec:narrow_sio} in Table~\ref{tab:narrow_properties}, by describing the emission in the regions according to four distinct cases: (1) the existence of spatially overlapping broad and narrow emission; (2) the existence of broad emission with little/no narrow emission associated; (3) the existence of narrow emission adjacent to (but not coincident with) broad emission; and (4) the existence of narrow emission with absolutely no broad emission in the vicinity. We will look in more detail into each case in the following sections.

\begin{figure}[!t]
	\centering
	{\renewcommand{\baselinestretch}{1.1}
	\includegraphics[width=0.45\textwidth]{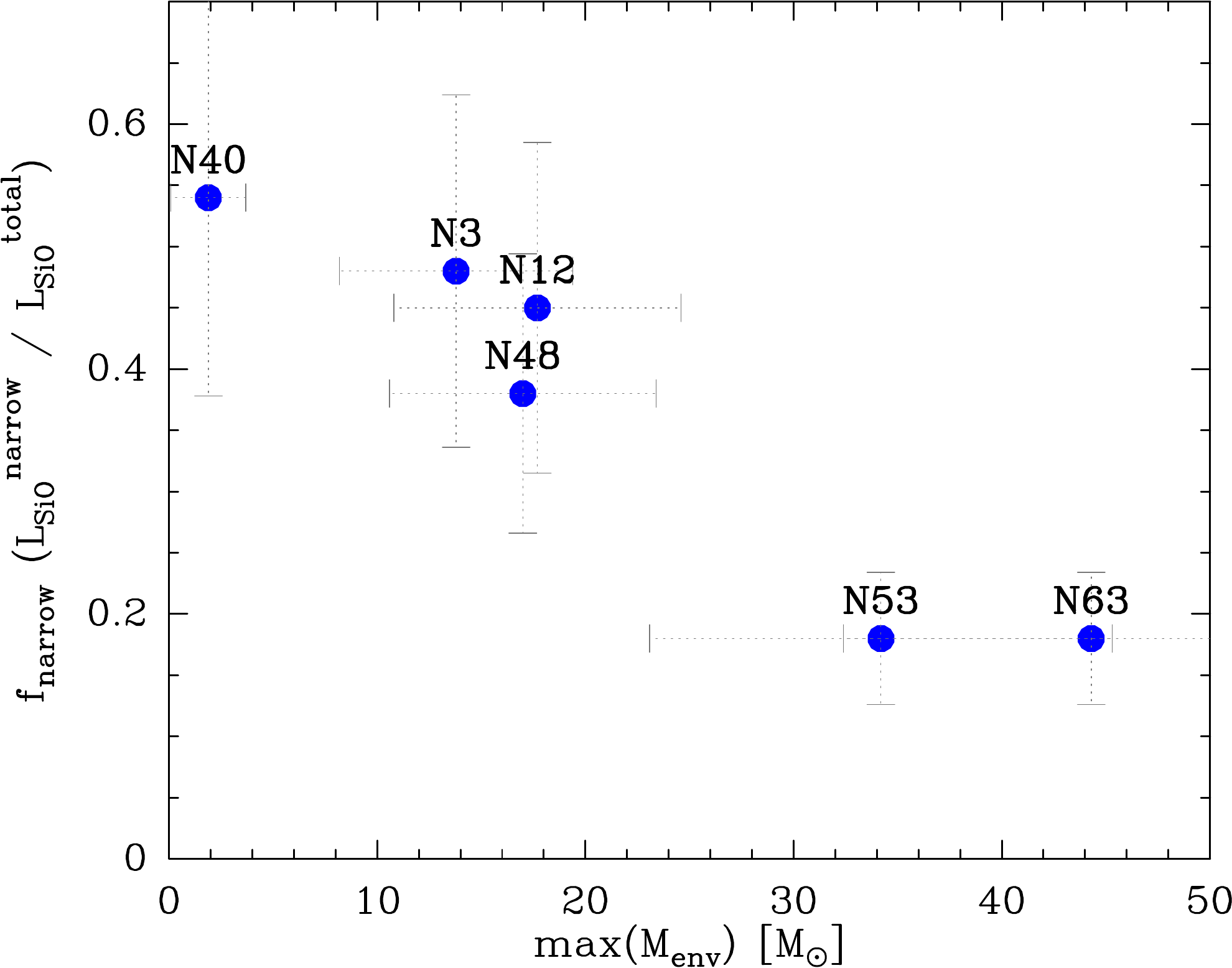}
	\caption[]{\small{ Fraction of the total SiO luminosity that arises from the narrow component, f$_{\rm narrow}$, as a function of the mass of the most massive protostellar envelope which is currently forming in each MDC. This plot accentuates the trend of a higher fraction of the emission in the form of narrow emission for the MDCs which have not yet gathered most of their mass in the envelopes of the forming protostars.}}
	\label{fig:lsio_maxmass}}
\end{figure}

\subsubsection{Narrow SiO emission from outflows}
\label{sec:narrow_sio_outflows}

Strong high-velocity outflows (as those powered by the two most massive sources CygX-N53 and CygX-N63), have naturally a strong impact on the ambient gas, and can easily form and entrain SiO to high velocities. There are, however, several factors that come into play when SiO is formed in such a high-velocity shock, and these factors can change the observed line profile. 

The typical profile of SiO produced in high-velocity shocks has a highly asymmetric profile, including a broad and a narrow component \citep[from 1D models, e.g.][]{2008A&A...482..809G,2013A&A...556A..69A}. This typical profile explains case (1) of Table~\ref{tab:narrow_properties}, where broad and narrow emission are spatially overlapping. While such models predict the narrow profiles to be shifted by tens of km\,s$^{-1}$ from the systemic velocities, projection effects can contribute to significantly decrease the observed velocity shifts of the shocked gas with respect to the ambient velocities. On the other hand, projection effects can also have another consequence, which is the fact that the profile observed is a result of the superposition of several ideal 1D profiles. One of the results of this can be the symmetrisation of the asymmetric profiles, i.e. the narrow component will become less obvious and the emission we see is mostly a broad (near Gaussian) profile. This explains the profiles of case (2) of Table~\ref{tab:narrow_properties}.

If, however, the composition of the pre-shock gas is such that the silicon is solely in the grain cores, the high-velocity shocks will produce SiO with narrower profiles \citep[][]{2008A&A...482..809G,2013A&A...556A..69A}. Therefore, the existence of narrow emission offset from the systemic velocities, with no broad emission associated, might be the signature of a fraction of SiO as emanating from the grain cores. Another way to have narrow SiO emission associated with high-velocity shocks, is as a result of the mixing of the shocked gas with the colder ambient gas, which can still form SiO but with lower velocity dispersion \citep[e.g.][]{1998ApJ...504L.109L,1999A&A...343..585C}. Even though the timescales for the deceleration of the shocked gas or alternatively the oxidisation of SiO or its depletion back into the grain ice mantles are short \citep[both of the order of $\sim10^4$\,yr, e.g.][]{1992A&A...254..315M,1999A&A...343..585C}, the high-mass protostars in these MDCs are Class 0 equivalents, and therefore also relatively young. Furthermore, the outflows being a continuous process, they have the potential to continuously replenish the decaying decelerated SiO in outskirts of outflows. In this case, we expect to see narrow SiO emission at systemic velocities, in the vicinity of high-velocity shocks, for example, along outflow walls. Alternatively, and as suggested by \citet[][]{2004ApJ...603L..49J}, the presence of narrow SiO at systemic velocities could also be explained by recent sputtering of the grain mantles containing a small fraction of Si/SiO and therefore could be indicative of young shocks (or simply tracing of the most recently shocked material). All three options could explain configuration (3) of Table~\ref{tab:narrow_properties}.

Finally, another way that we could achieve narrow profiles from a high-velocity shock would be with a strong UV field from the protostar, illuminating either the pre- or post-shock region, effectively decreasing the thickness of the SiO emitting-region, and subsequently narrowing its line profiles (Gusdorf et al. in prep.). However, our sample of protostars (some of which host hot cores, e.g. CygX-N53 MM1 and CygX-N63 MM1) are Class 0 equivalents \citep[with $L_{\rm bol} \sim 100-500 L_{\odot}$, see][and Table~\ref{tab:luminosities}]{2013A&A...558A.125D}, and therefore the radiation from the forming protostars is not yet strong enough to ionise the gas outside the inner protostellar envelopes/hot-core regions (which are unresolved by our observations), ruling out this hypothesis.

In our study, we do find narrow SiO line profiles associated with high-velocity outflow shocks. On one hand, we have narrow SiO emission at systemic velocities close to high-velocity outflow lobes, which could be due to the mixing of shocked gas and ambient gas along outflow walls. The most obvious candidates for this specific case in our sample are CygX-N40, next to the red outflow lobe; CygX-N53, along the red outflow cone, and to the west of the blue outflow lobe; and CygX-N63, around the blue outflow lobe, and at the tip of the red outflow. In CygX-N48 this process can also be in play, for instance along the edges of the red outflow from MM2 and next to the outflow from MM1. On the other hand, we also detected narrow SiO emission offset from systemic velocities, which could be an indication of a specific environment where the Si is in grain cores. This is the case of CygX-N12, where there are narrow line profiles (FWHM $\sim$\,1\,km\,s$^{-1}$) associated with the east-bound outflow (see Fig.~\ref{fig:sio_indiv_spectra}) and slightly blueshifted from the systemic velocities. 

\begin{table}[!t]
\caption{Distribution of the SiO (2-1) emission.}
\begin{tabular}{l | l }
\hline 
\hline 
Field  		&	Configuration \\
\hline
CygX-N3		& (1), (3), (4) \\ 
CygX-N12		& (1), (3), (4), marginal (2) in the NW of the cone \\
CygX-N40		& (2), (4), marginal (3) near the red lobe \\ 
CygX-N48		& (1), (3), marginal (4) between southbound outflows \\ 
CygX-N53		& (1), (2), (3), marginal (4) to the SE of MM2 \\ 
CygX-N63		& (1), (3), (4) \\ 
\hline
\end{tabular}
\\
(1) spatial overlap between broad and narrow emission.\\
(2) broad emission without narrow emission.\\
(3) narrow emission with nearby broad emission.\\
(4) narrow emission with no broad emission in the vicinity.
\label{tab:narrow_properties}
\end{table}

\subsubsection{Other narrow emission}
\label{sec:othernarrow}

Despite the existence of SiO narrow emission associated with outflows, we also detect some narrow SiO emission, at systemic velocities, that does not have any broad outflow emission in the immediacies. This is our configuration (4). 

This is most striking for CygX-N40 where most of the SiO emission is narrow, extended, and offset (by $\gtrsim 0.1$\,pc) from the forming protostars, hence definitely not associated with outflows ($\lesssim 0.05$\,pc) nor any hot-core ($< 0.005$\,pc). For this specific region, we have also investigated whether this extended SiO emission could be due to external radiation, but this seems unlikely as we have found no evidence of higher dust temperatures surrounding the N40 core \citep[see e.g.][]{2012A&A...543L...3H}. It is perhaps worth noting that this region sits close to the convergence point between the DR21 ridge and another filament thought to be feeding material into the main ridge \citep[][]{2010A&A...520A..49S}. The shocks from this infall of material into the MDC could explain the existence of extended SiO emission \citep[in line with what had been interpreted for other regions with extended narrow SiO emission with FWHM of $0.8-2$\,km\,s$^{-1}$, e.g. by][]{2010MNRAS.406..187J,2013ApJ...775...88N,2013ApJ...773..123S}.

In CygX-N3 and N48, even though the narrow low-velocity emission has a significant overlap/adjacency to outflow regions, the SiO narrow emission has a similar spatial distribution to the interface regions of the small-scale converging flows observed here \citep[][]{2011A&A...527A.135C,2011ApJ...740L...5C}, suggesting that the SiO could also have been formed from the low-velocity shocks of such flows. Interestingly, in both cases, the SiO narrow emission has a velocity gradient which does not mimic the velocity shears seen in high-density tracers \citep[H$^{13}$CO$^{+}$ and N$_{2}$H$^{+}$,][]{2011A&A...527A.135C,2011ApJ...740L...5C}. The mismatched velocity field could be explained by the fact that the SiO is not formed on the flows themselves, but as part of the post-shock material. If this post-shock material is originally at systemic velocities, and is dense enough to retain its original velocity field despite the external flow, the SiO would retain the original velocity of the (denser) local gas. 

\begin{figure}[!t]
	\centering
	{\renewcommand{\baselinestretch}{1.1}
	\includegraphics[width=0.45\textwidth]{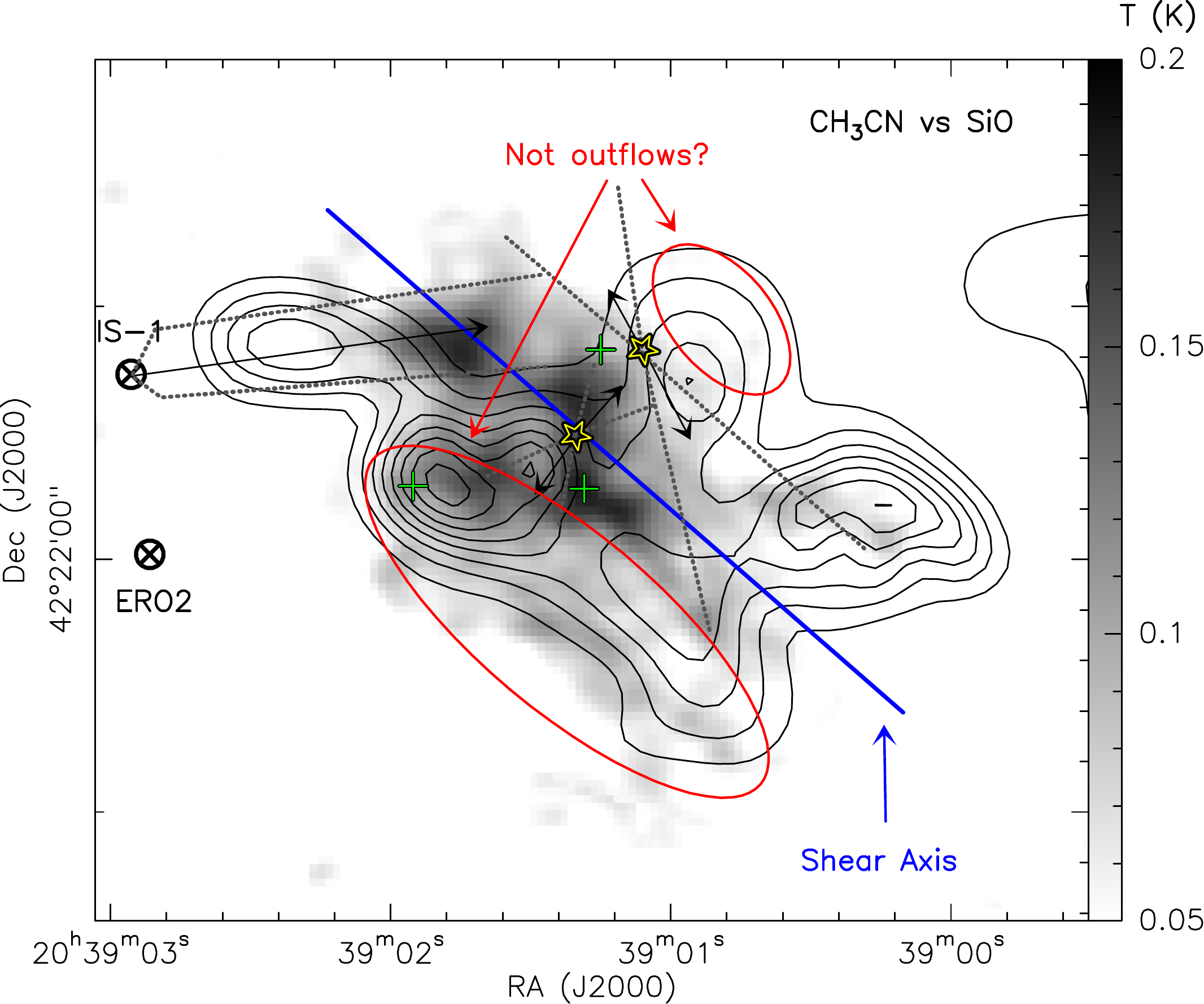}
	\caption[]{\small{Map of the CygX-N48 integrated intensity of CH$_{3}$CN from \citep[][]{2011ApJ...740L...5C} in greyscale, overlaid with the SiO systemic velocity emission from this paper in contours (for both molecules the velocity integration range was from -6.5 to -0.5\,km\,s$^{-1}$). The red ellipses show the regions with narrow SiO emission adjacent to the clear velocity shear found in N$_{2}$H$^{+}$, shown here as a blue line. Even though there are strong and compact SiO peaks coincident with high-velocity emission as seen in Fig.~\ref{fig:sio_narrow2}, the weaker and more extended SiO systemic velocity emission could be associated with the lower-velocity shocks from this velocity shear.}}
	\label{fig:sio_ch3cn}}
\end{figure}

In Cyg-X N48, \citet[][]{2011ApJ...740L...5C} detected CH$_{3}$CN as well, and suggest that it is tracing the gas shocked by such converging low-velocity flows. In Fig.~\ref{fig:sio_ch3cn} we show how the narrow SiO emission (in contours) correlates with this CH$_{3}$CN emission. Some of the CH$_{3}$CN emission could be associated with high-velocity outflows as they peak in regions adjacent to outflow lobes (e.g. the peak next to the outflow from IS-1). Nevertheless, it is interesting to note that the interface between the two converging flows (marked with a blue line) coincides with the most massive protostar of this MDC, and could be responsible for the high star formation activity in this region. This interface has both CH$_{3}$CN and narrow SiO emission, supporting the idea that they could both potentially arise not only from the several high-velocity outflows, but also from the shocks from small-scale converging flows. 

In CygX-N12, CygX-N53, and CygX-N63, apart from the narrow emission that is associated with outflows (either blueshifted or in outflow cavity walls), there are small extensions with narrow emission at systemic velocities where no outflows co-exist. This SiO narrow emission extensions in CygX-N53, N63, and N12 could, similarly to N3 and N48, be pinpointing the convergence zones of the global collapse of the MDC material (or perhaps even the collapse of larger scale cloud structures) onto the protostellar envelopes. 

\section{Discussion and conclusions}
\label{concl}

\subsection{SiO as a probe of outflow shocks}

Our study confirms that SiO is a good indicator of outflows in massive star-forming regions. However, we advise caution when using the SiO luminosity as a proxy of outflow power when individual outflows cannot be resolved. Doing so in our sample of sources would have induced an overestimation of the outflow power which could reach more than an order of magnitude (e.g. in CygX-N3, N40, and N48). This is mostly due either to the existence of a narrow component of SiO emission, often extended, which may not be due to outflow shocks. Nevertheless, the stronger the outflows, the more the outflow SiO luminosity will dominate over any other contributing SiO emission, resulting in a tighter correlation with outflow power.

These problems are attenuated when we are able to separate the individual outflows, where we tentatively recover a linear relation between $F_{\rm CO}$ and the total $L_{\rm SiO}$, but because SiO needs to be formed and entrained by the outflows (unlike CO that needs only to be entrained), the SiO outflow emission depends not only on the efficiency of the outflows in forming and entraining SiO (which is likely linked to the outflow power and, as such, the evolutionary stage of the protostar), but it also depends on the amount of material encountered and shocked by the outflows along their propagating axis as they pierce the cloud. The scatter is therefore more significant for the relation of SiO luminosity with envelope mass, than it is when using the CO outflow momentum fluxes.

\subsection{Origin of the narrow SiO emission}

From our sample of sources, we have realised that MDCs forming the least massive stars have their SiO luminosity dominated by a narrow component of SiO at systemic velocities (see Fig.~\ref{fig:lsio_maxmass}). This narrow component, however, does seem to exist in all cases. While for the most massive cores it seems mostly related to outflow walls or compact narrow emission close to the protostellar envelopes (at systemic velocities), the least massive sources have a tendency to show more extended SiO narrow emission. 
Such narrow SiO emission could have various origins, and in the following we shall explore some of the possibilities.

One hypothesis is that it arises from the mixing of the gas due to the interaction of different outflows from a population of low-mass stars undetected in the PdBI continuum images. We note, however, that the sensitivity in the 1mm continuum observations is equivalent to $\sim 1$\,M$_{\odot}$, and that all the potential low-mass protostars with envelope masses above $\sim 1$\,M$_{\odot}$ are listed by \citet[][]{2010A&A...524A..18B} and shown as crosses in our figures. The positions of these candidate low-mass protostars are not associated with any CO high-velocity outflow (which we would expect to detect as we did in CygX-N40), and they also do not lie near most of the narrow emission detected. Judging by the number of low-mass fragments detected we do not expect a significant underlying population of very low-mass protostars ($< 1$\,M$_{\odot}$). Even though we cannot rule out this hypothesis completely, we consider that this scenario is thus unlikely.

Another hypothesis is that the shocks that may create the narrow SiO emission where no nearby high-velocity shocks exist could be related to converging flows shocking with the cloud's material. In our sample, these shocks could be due to the small-scale converging flows reported by \citet[][]{2011A&A...527A.135C,2011ApJ...740L...5C} or due to the large scale collapse of the DR21 ridge as suggested by \citet[][]{2010A&A...520A..49S}. This is in line with what had been suggested for other high-mass star-forming regions by e.g. \citet[][]{2010MNRAS.406..187J,2013ApJ...775...88N}. Perhaps seeing a dominant SiO narrow emission is actually a sign that there is a significant and ongoing infall of material from beyond the MDC scale ($\gtrsim$\,0.1-0.3\,pc) to the envelope scale ($\lesssim$\,0.02\,pc), which could lead to a further increase of the masses of existing individual cores or to the formation of new ones. The most striking case of our sample is undoubtedly CygX-N40 where the inner envelope is not yet massive at all, but there are SiO shocks everywhere around it. We note that in spite of comprising nearly $\sim 100 $\,M$_{\odot}$ inside less than 0.15\,pc \citep[][]{2007A&A...476.1243M}, Cygx-N40 has a rather shallow density profile, with no high-mass protostars yet formed in the centre. Perhaps N40 is at an earlier stage than the other sources, where the material of the MDC is yet to become centrally condensed and form one or more high-mass protostellar cores, and the SiO emission is the evidence of the shocks of the infall of large-scale flows and the collapsing MDC itself.

One of the problems with having SiO at low velocities not associated with outflows is to understand if and how low-velocity shocks could form SiO. As mentioned by \citet[][]{2013ApJ...775...88N}, SiO can be efficiently formed by low-velocity shocks ($<$\,10\,km\,s$^{-1}$) provided that there is already some Si in the gas phase or in ice mantles. Having Si surviving in the gas phase in the cold and dense environment of an MDC is, however, not very likely since Si is expected to either deplete back onto the grain ice mantles, or to be oxidised in SiO$_{2}$, itself being subsequently depleted back onto the grain ice mantles, both processes leading to its disappearance from the gas phase in about $\sim10^4$\,yr  \citep[e.g.][]{1992A&A...254..315M,1999A&A...343..585C,2008A&A...482..809G}. Nevertheless, it is possible to admit that the converging flows are feeding the MDCs with pre-shock material originating from the diffuse medium, which contains Si in the gas phase due to the exposition to the radiation field of nearby OB stars, or even that in some cases there is some SiO released into the gas phase due to shocks from earlier high-velocity outflows. This SiO could then easily be maintained in the gas phase by lower velocity shocks. If this is the case, shocks with velocities below 5\,km\,s$^{-1}$ could start producing enough SiO to be detected. We retrieve SiO column densities of the order of $0.5-1.0 \times 10^{13}$\,cm$^{-2}$ (see Sect.~\ref{sec:coldens}) for the narrow SiO emission \citep[similar to what was found by][]{2013ApJ...775...88N}, which is consistent with the possibility that SiO has been formed with low-velocity shocks.
The low-velocity shears detected in the Cygnus-X MDCs have a line of sight velocity difference around 2-3\,km\,s$^{-1}$  \citep[][]{2011A&A...527A.135C,2011ApJ...740L...5C}. Although this lies in the very lower limits to produce any SiO at all, these are lower limits for the actual velocity shears because projection effects are most certainly in play. Furthermore, and as mentioned in Sect.~\ref{sec:othernarrow}, since SiO molecules form in the post-shock gas, the actual velocities of SiO and the respective velocity dispersions can be different from those of the shocks, supporting the idea that the low-velocity converging flows in the region \citep[][]{2011A&A...527A.135C,2011ApJ...740L...5C} could be responsible for the widespread SiO narrow emission, despite the different observed velocities.

We therefore propose an evolutionary picture for the existence of SiO at systemic velocities and with a narrow line profile, not associated with outflows. In this scenario, the least centrally condensed MDCs are interpreted as being in an earlier stage of the collapse and the SiO would be tracing shocks from the global collapse of cloud onto the MDC. This would be the case for CygX-N40, which is a MDC with only a couple of low-mass protostars ($< 2$\,M$_{\odot}$) at its centre, a very weak outflow, and where 90\% of the SiO emission arises from the outskirts. This could also be the case of, for instance, IRDC G028.23-00.19 \citep[][]{2013ApJ...773..123S}, but higher resolution observations would be necessary to confirm this. As the MDC collapses, the SiO emission becomes more confined to the close surroundings of denser parts, tracing the post-shock material from the shocks of the infalling MDC directly against the dense central cores. This is in essence the equivalent of the shocks from the small-scale low-velocity converging flows seen in CygX-N3, N12, and N48. At later stages, the MDC infall is less prominent and a more significant fraction of the MDC mass is already gathered in the central massive protostars. At this stage, the SiO luminosity is largely dominated by the powerful outflows, and there is only a weak narrow component which would point to the narrow SiO emission arising from the shocks from the very last remnants of the collapse of the MDC against the protostellar envelopes - this could be the case of CygX-N53 and N63. In this scenario, we would not expect to see a great amount of extended narrow SiO emission at systemic velocities in low-mass star-forming regions (apart from the narrow SiO emission intrinsically originating from high-velocity shocks), because the simple gravitational collapse of the clumps is unlikely to create shocks with enough speed to produce SiO (infall velocities being typically $\sim1$\,km\,s$^{-1}$), and these regions are generally less dynamic and less dense environments, less prone to shocks in the ISM capable of forming SiO. This is consistent with the fact that the SiO emission detected so far in low-mass star-forming regions is associated with high-velocity protostellar outflows, either as broad SiO emission, or as narrow emission with a velocity peak offset from systemic velocities \citep[e.g.][]{1998A&A...333..287G,2007A&A...462..163N,2011A&A...532A..53G,2013A&A...558A..94G}.

Even though this dynamical view would predict a smaller amount of widespread SiO narrow emission at systemic velocities in low-mass star-forming regions, it is sill puzzling to understand why we do not seem to observe any. If low-velocity shocks are able to form SiO (whether from outflows or other turbulent shocks in the ISM), then they should also be present in the low-mass case. One could think of two ways to explain the lack of detection of this component: either the composition of the gas is different in massive star-forming regions (e.g. with a greater amount of Si on grain mantles or in the gas phase), making it easier to form SiO with low-velocity shocks, or there is a density threshold where the lower velocity shocks become efficient at forming SiO only if the volume densities are sufficiently high. 
Hence it would be interesting to test the minimum densities needed to be able to form SiO through low-velocity shocks. This will be addressed in upcoming papers from Gusdorf et al. and Louvet et al. (in prep.). Observationally, progress can be made by observing higher energy transitions of SiO to help constrain all the characteristics of the shocks (velocities, magnetic field, pre-shock densities, chemistry, UV field, etc.).


\begin{acknowledgements}
      
ADC acknowledges funding from the European Research Council for the FP7 ERC starting grant project LOCALSTAR. Part of the work by ADC was accomplished under funding by the project PROBeS, and NS is supported by the project STARFICH, both funded by the French National Research Agency (ANR). TCs acknowledges financial support for the ERC Advanced Grant GLOSTAR under contract no. 247078. IRAM is supported by INSU/CNRS (France), MPG (Germany), and IGN (Spain). The data reduction and analysis made use of the GILDAS software (http://www.iram.fr/IRAMFR/GILDAS).  
     
\end{acknowledgements}

\bibliographystyle{aa}	
\bibliography{references}		

\begin{thebibliography}{40}
\expandafter\ifx\csname natexlab\endcsname\relax\def\natexlab#1{#1}\fi

\bibitem[{{Anderl} {et~al.}(2013){Anderl}, {Guillet}, {Pineau des For{\^e}ts},
  \& {Flower}}]{2013A&A...556A..69A}
{Anderl}, S., {Guillet}, V., {Pineau des For{\^e}ts}, G., \& {Flower}, D.~R.
  2013, \aap, 556, A69

\bibitem[{{Beuther} {et~al.}(2002){Beuther}, {Schilke}, {Gueth}, {McCaughrean},
  {Andersen}, {Sridharan}, \& {Menten}}]{2002A&A...387..931B}
{Beuther}, H., {Schilke}, P., {Gueth}, F., {et~al.} 2002, \aap, 387, 931

\bibitem[{{Bontemps} {et~al.}(1996){Bontemps}, {Andre}, {Terebey}, \&
  {Cabrit}}]{1996A&A...311..858B}
{Bontemps}, S., {Andre}, P., {Terebey}, S., \& {Cabrit}, S. 1996, \aap, 311,
  858

\bibitem[{{Bontemps} {et~al.}(2010){Bontemps}, {Motte}, {Csengeri}, \&
  {Schneider}}]{2010A&A...524A..18B}
{Bontemps}, S., {Motte}, F., {Csengeri}, T., \& {Schneider}, N. 2010, \aap,
  524, A18

\bibitem[{{Caselli} {et~al.}(1997){Caselli}, {Hartquist}, \&
  {Havnes}}]{1997A&A...322..296C}
{Caselli}, P., {Hartquist}, T.~W., \& {Havnes}, O. 1997, \aap, 322, 296

\bibitem[{{Codella} {et~al.}(1999){Codella}, {Bachiller}, \&
  {Reipurth}}]{1999A&A...343..585C}
{Codella}, C., {Bachiller}, R., \& {Reipurth}, B. 1999, \aap, 343, 585

\bibitem[{{Csengeri} {et~al.}(2011{\natexlab{a}}){Csengeri}, {Bontemps},
  {Schneider}, {Motte}, \& {Dib}}]{2011A&A...527A.135C}
{Csengeri}, T., {Bontemps}, S., {Schneider}, N., {Motte}, F., \& {Dib}, S.
  2011{\natexlab{a}}, \aap, 527, A135

\bibitem[{{Csengeri} {et~al.}(2011{\natexlab{b}}){Csengeri}, {Bontemps},
  {Schneider}, {Motte}, {Gueth}, \& {Hora}}]{2011ApJ...740L...5C}
{Csengeri}, T., {Bontemps}, S., {Schneider}, N., {et~al.} 2011{\natexlab{b}},
  \apjl, 740, L5

\bibitem[{{Duarte-Cabral} {et~al.}(2013){Duarte-Cabral}, {Bontemps}, {Motte},
  {Hennemann}, {Schneider}, \& {Andr{\'e}}}]{2013A&A...558A.125D}
{Duarte-Cabral}, A., {Bontemps}, S., {Motte}, F., {et~al.} 2013, \aap, 558,
  A125

\bibitem[{{Godard} \& {Cernicharo}(2013)}]{2013A&A...550A...8G}
{Godard}, B. \& {Cernicharo}, J. 2013, \aap, 550, A8

\bibitem[{{G{\'o}mez-Ruiz} {et~al.}(2013){G{\'o}mez-Ruiz}, {Hirano}, {Leurini},
  \& {Liu}}]{2013A&A...558A..94G}
{G{\'o}mez-Ruiz}, A.~I., {Hirano}, N., {Leurini}, S., \& {Liu}, S.-Y. 2013,
  \aap, 558, A94

\bibitem[{{Gueth} {et~al.}(1998){Gueth}, {Guilloteau}, \&
  {Bachiller}}]{1998A&A...333..287G}
{Gueth}, F., {Guilloteau}, S., \& {Bachiller}, R. 1998, \aap, 333, 287

\bibitem[{{Guillet} {et~al.}(2009){Guillet}, {Jones}, \& {Pineau Des
  For{\^e}ts}}]{2009A&A...497..145G}
{Guillet}, V., {Jones}, A.~P., \& {Pineau Des For{\^e}ts}, G. 2009, \aap, 497,
  145

\bibitem[{{Guillet} {et~al.}(2007){Guillet}, {Pineau Des For{\^e}ts}, \&
  {Jones}}]{2007A&amp;A...476..263G}
{Guillet}, V., {Pineau Des For{\^e}ts}, G., \& {Jones}, A.~P. 2007, \aap, 476,
  263

\bibitem[{{Guillet} {et~al.}(2011){Guillet}, {Pineau Des For{\^e}ts}, \&
  {Jones}}]{2011A&A...527A.123G}
{Guillet}, V., {Pineau Des For{\^e}ts}, G., \& {Jones}, A.~P. 2011, \aap, 527,
  A123

\bibitem[{{Gusdorf} {et~al.}(2008{\natexlab{a}}){Gusdorf}, {Cabrit}, {Flower},
  \& {Pineau Des For{\^e}ts}}]{2008A&A...482..809G}
{Gusdorf}, A., {Cabrit}, S., {Flower}, D.~R., \& {Pineau Des For{\^e}ts}, G.
  2008{\natexlab{a}}, \aap, 482, 809

\bibitem[{{Gusdorf} {et~al.}(2011){Gusdorf}, {Giannini}, {Flower}, {Parise},
  {G{\"u}sten}, \& {Kristensen}}]{2011A&A...532A..53G}
{Gusdorf}, A., {Giannini}, T., {Flower}, D.~R., {et~al.} 2011, \aap, 532, A53

\bibitem[{{Gusdorf} {et~al.}(2008{\natexlab{b}}){Gusdorf}, {Pineau Des
  For{\^e}ts}, {Cabrit}, \& {Flower}}]{2008A&A...490..695G}
{Gusdorf}, A., {Pineau Des For{\^e}ts}, G., {Cabrit}, S., \& {Flower}, D.~R.
  2008{\natexlab{b}}, \aap, 490, 695

\bibitem[{{Hennemann} {et~al.}(2012){Hennemann}, {Motte}, {Schneider},
  {Didelon}, {Hill}, {Arzoumanian}, {Bontemps}, {Csengeri}, {Andr{\'e}},
  {Konyves}, {Louvet}, {Marston}, {Men'shchikov}, {Minier}, {Nguyen Luong},
  {Palmeirim}, {Peretto}, {Sauvage}, {Zavagno}, {Anderson}, {Bernard}, {Di
  Francesco}, {Elia}, {Li}, {Martin}, {Molinari}, {Pezzuto}, {Russeil}, {Rygl},
  {Schisano}, {Spinoglio}, {Sousbie}, {Ward-Thompson}, \&
  {White}}]{2012A&A...543L...3H}
{Hennemann}, M., {Motte}, F., {Schneider}, N., {et~al.} 2012, \aap, 543, L3

\bibitem[{{Jim{\'e}nez-Serra} {et~al.}(2008){Jim{\'e}nez-Serra}, {Caselli},
  {Mart{\'{\i}}n-Pintado}, \& {Hartquist}}]{2008A&A...482..549J}
{Jim{\'e}nez-Serra}, I., {Caselli}, P., {Mart{\'{\i}}n-Pintado}, J., \&
  {Hartquist}, T.~W. 2008, \aap, 482, 549

\bibitem[{{Jim{\'e}nez-Serra} {et~al.}(2010){Jim{\'e}nez-Serra}, {Caselli},
  {Tan}, {Hernandez}, {Fontani}, {Butler}, \& {van Loo}}]{2010MNRAS.406..187J}
{Jim{\'e}nez-Serra}, I., {Caselli}, P., {Tan}, J.~C., {et~al.} 2010, \mnras,
  406, 187

\bibitem[{{Jim{\'e}nez-Serra} {et~al.}(2004){Jim{\'e}nez-Serra},
  {Mart{\'{\i}}n-Pintado}, {Rodr{\'{\i}}guez-Franco}, \&
  {Marcelino}}]{2004ApJ...603L..49J}
{Jim{\'e}nez-Serra}, I., {Mart{\'{\i}}n-Pintado}, J.,
  {Rodr{\'{\i}}guez-Franco}, A., \& {Marcelino}, N. 2004, \apjl, 603, L49

\bibitem[{{Kauffmann} {et~al.}(2013){Kauffmann}, {Pillai}, \&
  {Zhang}}]{2013ApJ...765L..35K}
{Kauffmann}, J., {Pillai}, T., \& {Zhang}, Q. 2013, \apjl, 765, L35

\bibitem[{{Le Picard} {et~al.}(2001){Le Picard}, {Canosa}, {Pineau des
  For{\^e}ts}, {Rebrion-Rowe}, \& {Rowe}}]{2001A&A...372.1064L}
{Le Picard}, S.~D., {Canosa}, A., {Pineau des For{\^e}ts}, G., {Rebrion-Rowe},
  C., \& {Rowe}, B.~R. 2001, \aap, 372, 1064

\bibitem[{{Lefloch} {et~al.}(1998){Lefloch}, {Castets}, {Cernicharo}, \&
  {Loinard}}]{1998ApJ...504L.109L}
{Lefloch}, B., {Castets}, A., {Cernicharo}, J., \& {Loinard}, L. 1998, \apjl,
  504, L109

\bibitem[{{L{\'o}pez-Sepulcre} {et~al.}(2011){L{\'o}pez-Sepulcre}, {Walmsley},
  {Cesaroni}, {Codella}, {Schuller}, {Bronfman}, {Carey}, {Menten}, {Molinari},
  \& {Noriega-Crespo}}]{2011A&A...526L...2L}
{L{\'o}pez-Sepulcre}, A., {Walmsley}, C.~M., {Cesaroni}, R., {et~al.} 2011,
  \aap, 526, L2

\bibitem[{{Louvet} {et~al.}(2014){Louvet}, {Motte}, {Hennebelle}, {Bonnell},
  {Bontemps}, {Gusdorf}, {Hill}, {Gueth}, {Peretto}, {Duarte-Cabral},
  {Stephan}, {Schilke}, {Csengeri}, {Nguyen Luong}, \&
  {Lis}}]{2014arXiv1404.4843L}
{Louvet}, F., {Motte}, F., {Hennebelle}, P., {et~al.} 2014, ArXiv e-prints

\bibitem[{{Marston} {et~al.}(2004){Marston}, {Reach}, {Noriega-Crespo}, {Rho},
  {Smith}, {Melnick}, {Fazio}, {Rieke}, {Carey}, {Rebull}, {Muzerolle},
  {Egami}, {Watson}, {Pipher}, {Latter}, \&
  {Stapelfeldt}}]{2004ApJS..154..333M}
{Marston}, A.~P., {Reach}, W.~T., {Noriega-Crespo}, A., {et~al.} 2004, \apjs,
  154, 333

\bibitem[{{Martin-Pintado} {et~al.}(1992){Martin-Pintado}, {Bachiller}, \&
  {Fuente}}]{1992A&A...254..315M}
{Martin-Pintado}, J., {Bachiller}, R., \& {Fuente}, A. 1992, \aap, 254, 315

\bibitem[{{Motte} {et~al.}(2007){Motte}, {Bontemps}, {Schilke}, {Schneider},
  {Menten}, \& {Brogui{\`e}re}}]{2007A&A...476.1243M}
{Motte}, F., {Bontemps}, S., {Schilke}, P., {et~al.} 2007, \aap, 476, 1243

\bibitem[{{Nguy{\^e}n-Lu'o'ng} {et~al.}(2013){Nguy{\^e}n-Lu'o'ng}, {Motte},
  {Carlhoff}, {Louvet}, {Lesaffre}, {Schilke}, {Hill}, {Hennemann}, {Gusdorf},
  {Didelon}, {Schneider}, {Bontemps}, {Duarte-Cabral}, {Menten}, {Martin},
  {Wyrowski}, {Bendo}, {Roussel}, {Bernard}, {Bronfman}, {Henning}, {Kramer},
  \& {Heitsch}}]{2013ApJ...775...88N}
{Nguy{\^e}n-Lu'o'ng}, Q., {Motte}, F., {Carlhoff}, P., {et~al.} 2013, \apj,
  775, 88

\bibitem[{{Nisini} {et~al.}(2007){Nisini}, {Codella}, {Giannini}, {Santiago
  Garcia}, {Richer}, {Bachiller}, \& {Tafalla}}]{2007A&A...462..163N}
{Nisini}, B., {Codella}, C., {Giannini}, T., {et~al.} 2007, \aap, 462, 163

\bibitem[{{Qiu} {et~al.}(2007){Qiu}, {Zhang}, {Beuther}, \&
  {Yang}}]{2007ApJ...654..361Q}
{Qiu}, K., {Zhang}, Q., {Beuther}, H., \& {Yang}, J. 2007, \apj, 654, 361

\bibitem[{{Rygl} {et~al.}(2012){Rygl}, {Brunthaler}, {Sanna}, {Menten}, {Reid},
  {van Langevelde}, {Honma}, {Torstensson}, \&
  {Fujisawa}}]{2012A&A...539A..79R}
{Rygl}, K.~L.~J., {Brunthaler}, A., {Sanna}, A., {et~al.} 2012, \aap, 539, A79

\bibitem[{{Sanhueza} {et~al.}(2013){Sanhueza}, {Jackson}, {Foster},
  {Jimenez-Serra}, {Dirienzo}, \& {Pillai}}]{2013ApJ...773..123S}
{Sanhueza}, P., {Jackson}, J.~M., {Foster}, J.~B., {et~al.} 2013, \apj, 773,
  123

\bibitem[{{Schilke} {et~al.}(2001){Schilke}, {Pineau des For{\^e}ts},
  {Walmsley}, \& {Mart{\'{\i}}n-Pintado}}]{2001A&A...372..291S}
{Schilke}, P., {Pineau des For{\^e}ts}, G., {Walmsley}, C.~M., \&
  {Mart{\'{\i}}n-Pintado}, J. 2001, \aap, 372, 291

\bibitem[{{Schilke} {et~al.}(1997){Schilke}, {Walmsley}, {Pineau des Forets},
  \& {Flower}}]{1997A&A...321..293S}
{Schilke}, P., {Walmsley}, C.~M., {Pineau des Forets}, G., \& {Flower}, D.~R.
  1997, \aap, 321, 293

\bibitem[{{Schneider} {et~al.}(2010){Schneider}, {Csengeri}, {Bontemps},
  {Motte}, {Simon}, {Hennebelle}, {Federrath}, \&
  {Klessen}}]{2010A&A...520A..49S}
{Schneider}, N., {Csengeri}, T., {Bontemps}, S., {et~al.} 2010, \aap, 520, A49+

\bibitem[{{Turner}(1991)}]{1991ApJ...376..573T}
{Turner}, B.~E. 1991, \apj, 376, 573

\bibitem[{{Walmsley} {et~al.}(1999){Walmsley}, {Pineau des For{\^e}ts}, \&
  {Flower}}]{1999A&A...342..542W}
{Walmsley}, C.~M., {Pineau des For{\^e}ts}, G., \& {Flower}, D.~R. 1999, \aap,
  342, 542

\end{thebibliography}

\begin{appendix}

\section{Examples of the SiO spectra}
\label{ap:SiO_spectra} 

\begin{figure}[!ht]
	\centering
	{\renewcommand{\baselinestretch}{1.1}
	\includegraphics[width=0.35\textwidth]{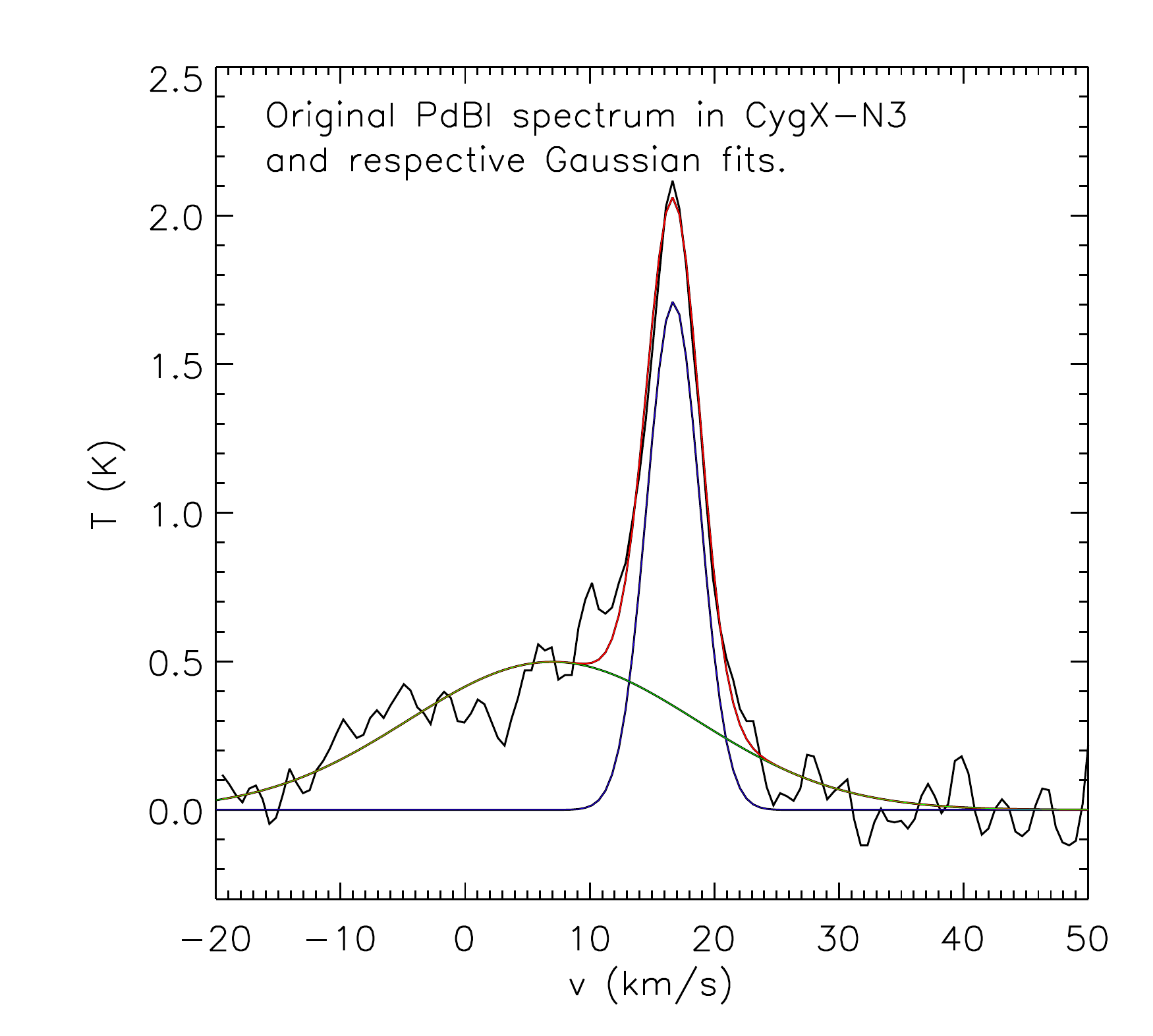}
	\includegraphics[width=0.35\textwidth]{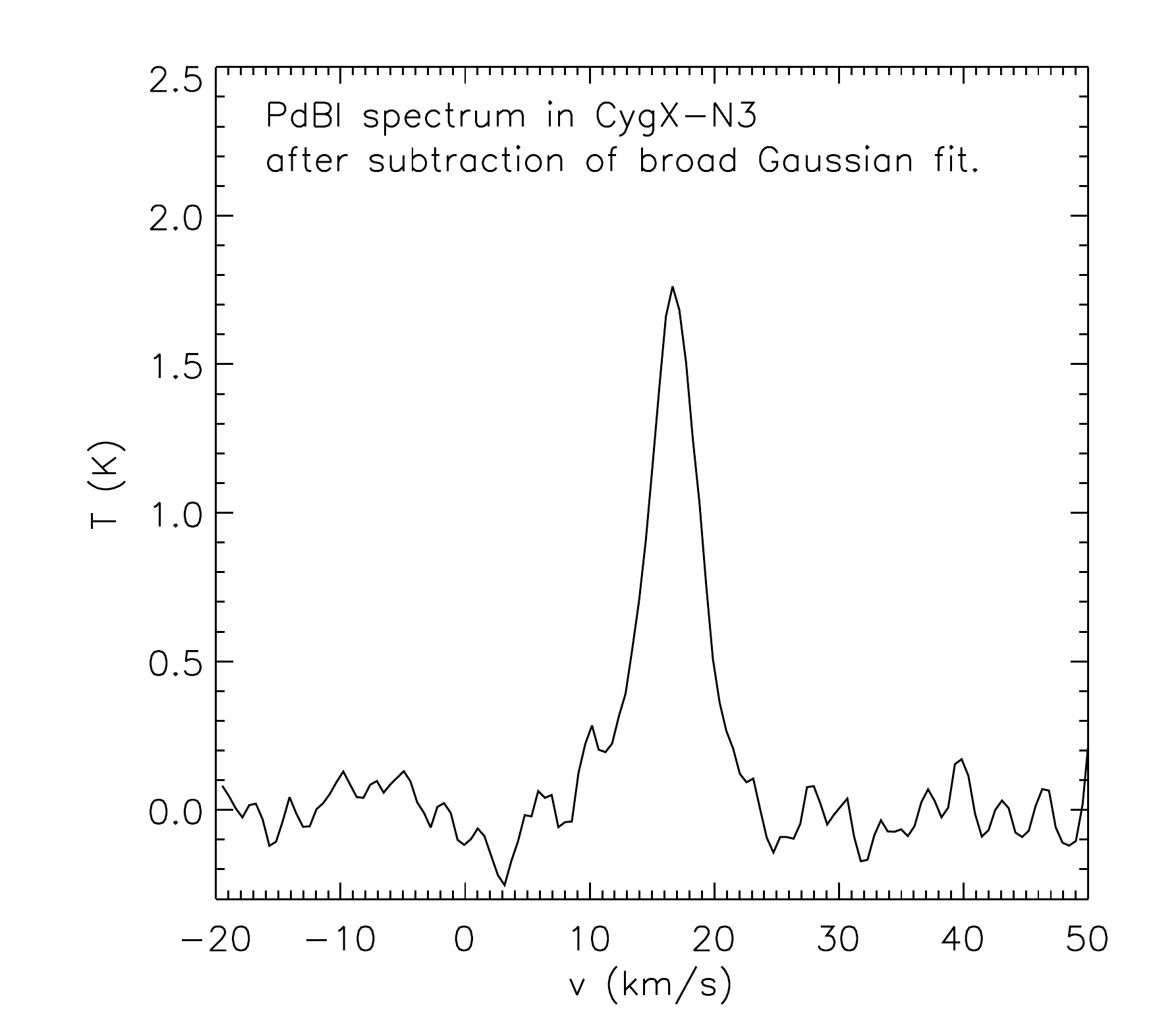}
	\caption[]{\small{Figure showing the procedure used to remove the broad high-velocity SiO emission from the datacubes. {\it Top:} Example of the line fitting of a spectrum where both broad and narrow emission exist (corresponding to the blue spectrum of CygX-N3 shown in Fig.~\ref{fig:sio_indiv_spectra}). The two Gaussian fits are shown in green (for the broad component) and blue (for the narrow component). The red line shows the sum of the two Gaussian components. {\it Bottom:} Same spectrum as the top panel, here with the respective broad Gaussian fit subtracted.}}
	\label{fig:example_fits_spectra}}
\end{figure}

\begin{figure*}[!ht]
	\centering
	{\renewcommand{\baselinestretch}{1.1}
	\includegraphics[width=0.27\textwidth]{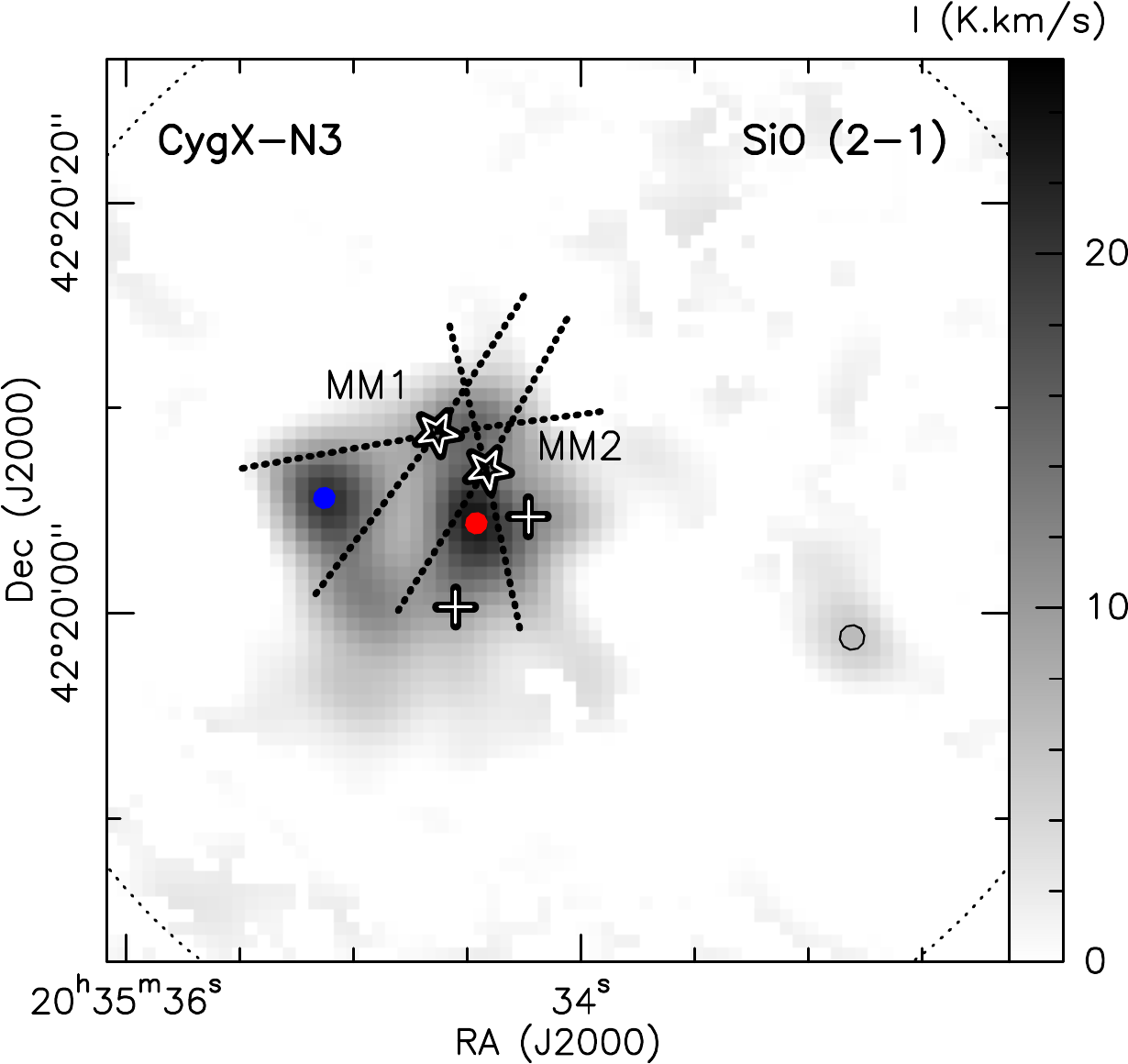}
	\hspace{-0.6cm}
	\includegraphics[width=0.245\textwidth]{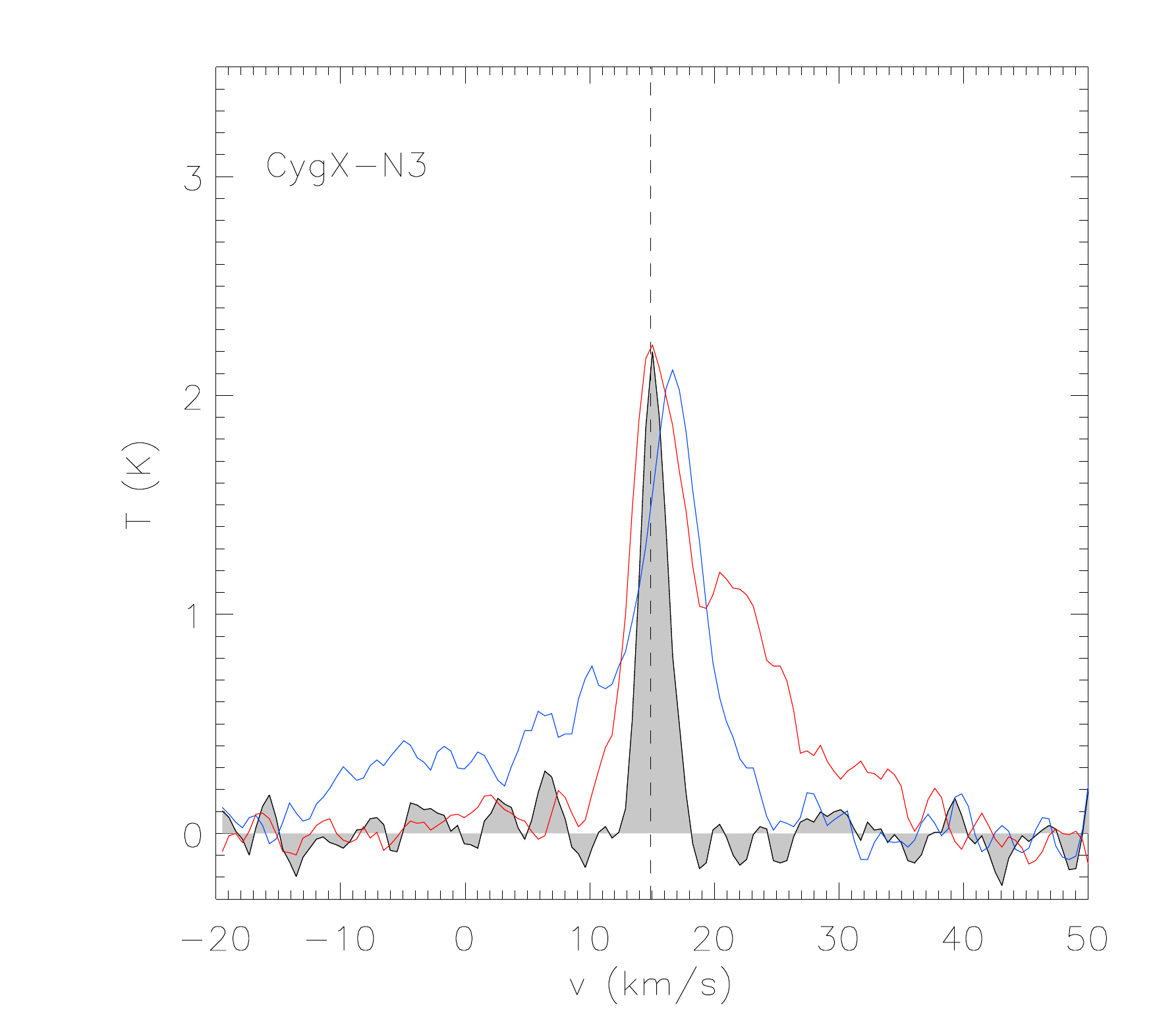}
	\hfill
	\includegraphics[width=0.27\textwidth]{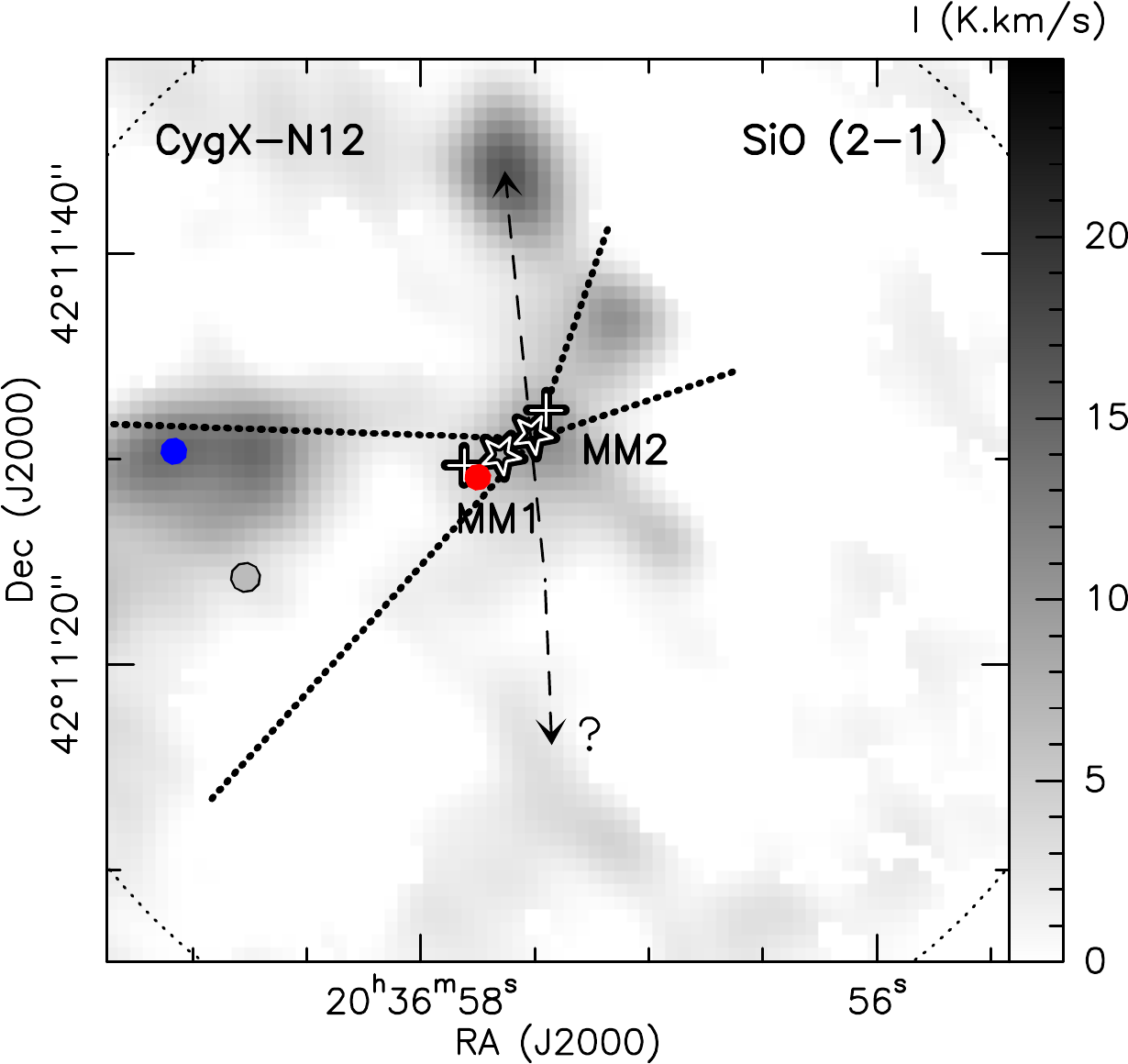}
	\hspace{-0.6cm}
	\includegraphics[width=0.245\textwidth]{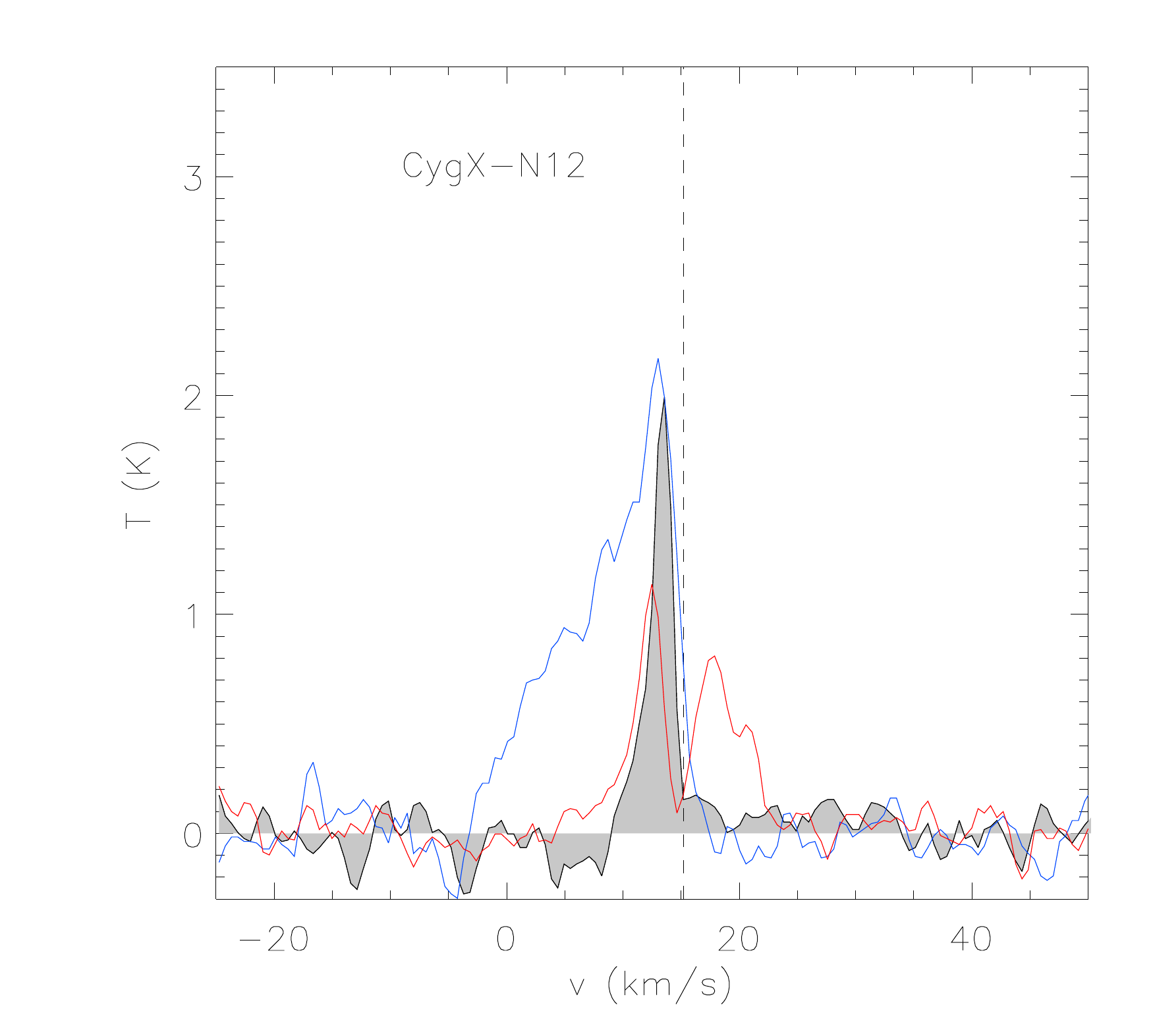}\\
	\includegraphics[width=0.27\textwidth]{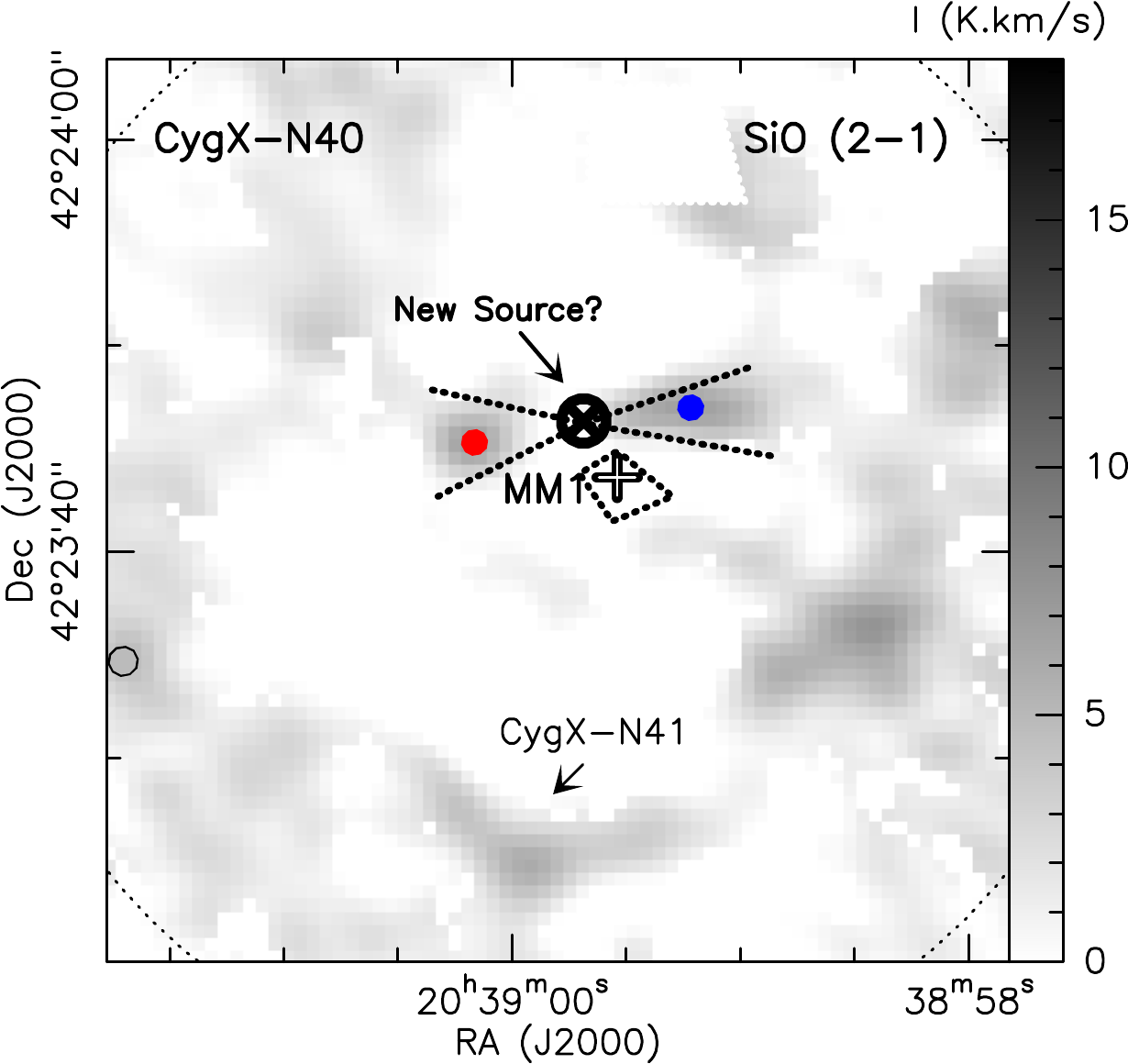}
	\hspace{-0.6cm}
	\includegraphics[width=0.245\textwidth]{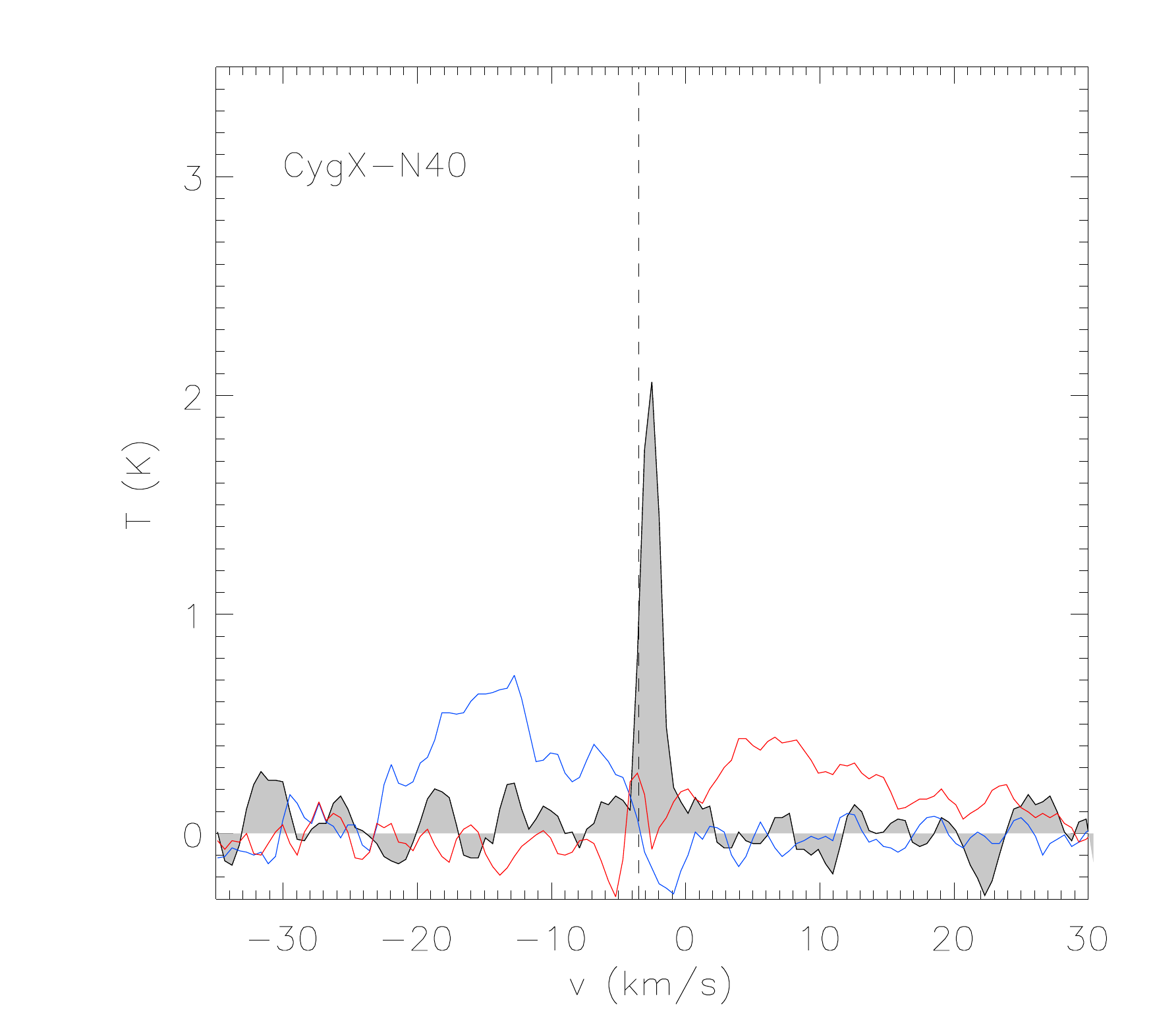}
	\hfill
	\includegraphics[width=0.27\textwidth]{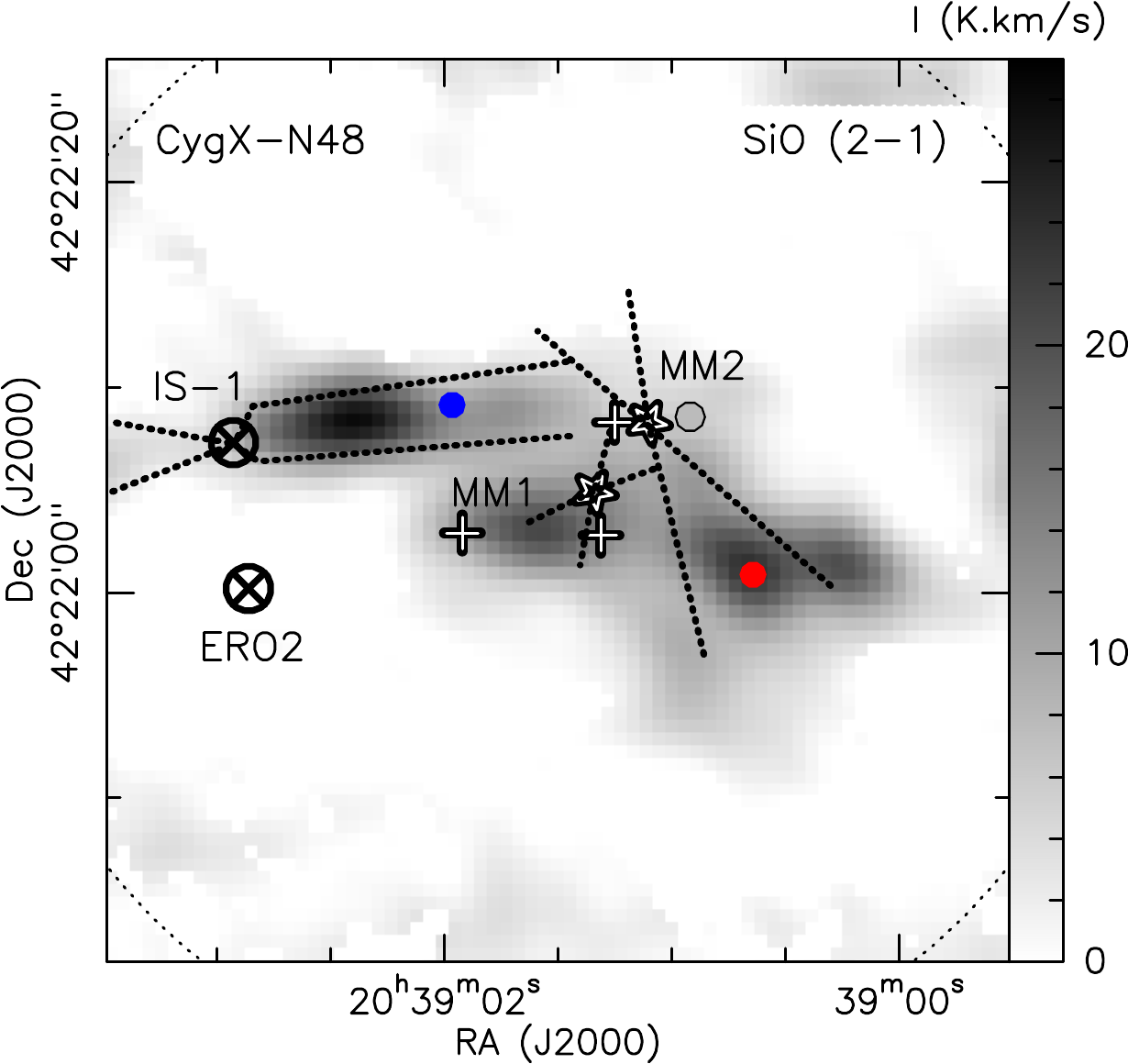}
	\hspace{-0.6cm}
	\includegraphics[width=0.245\textwidth]{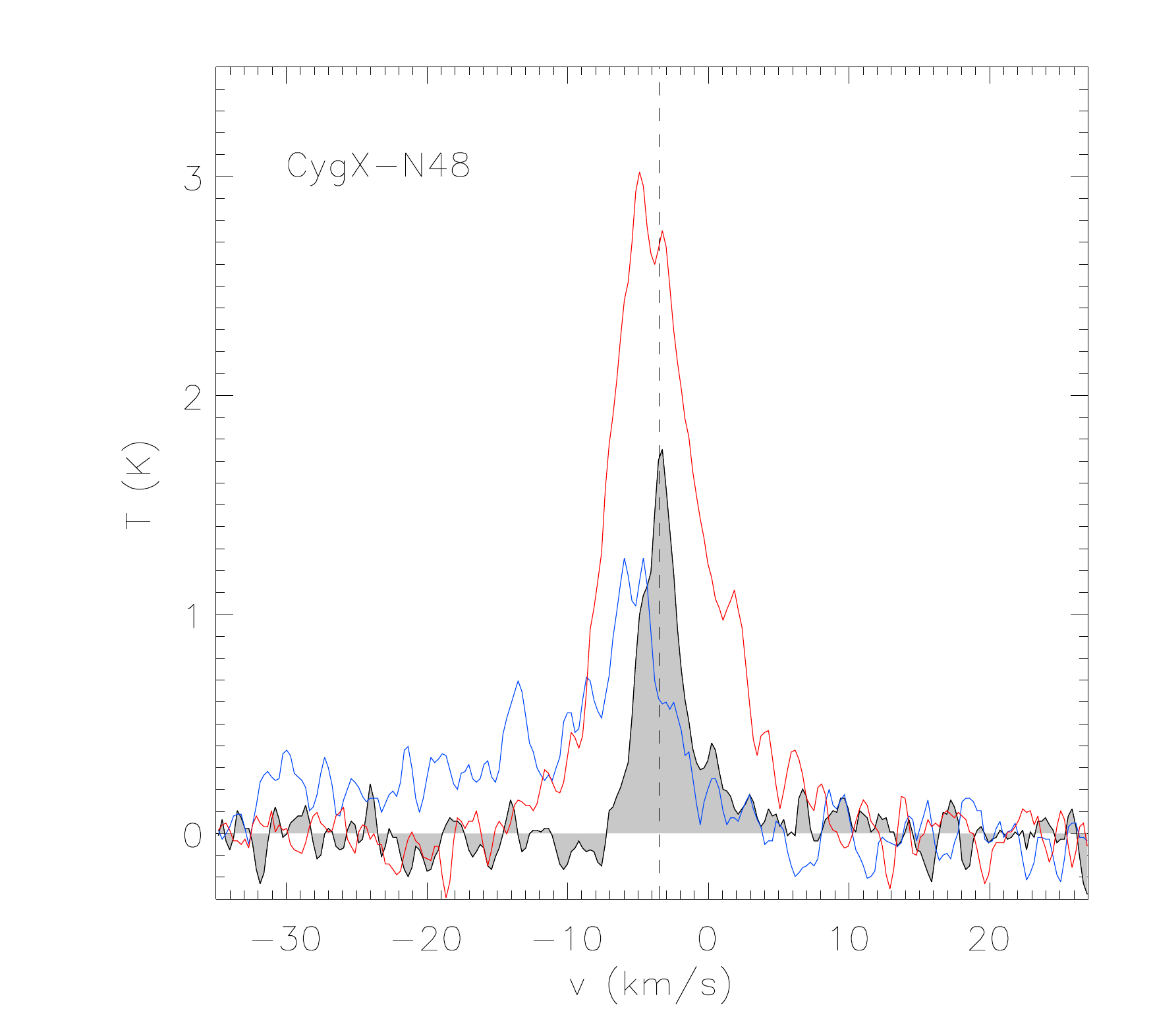}\\
	\includegraphics[width=0.27\textwidth]{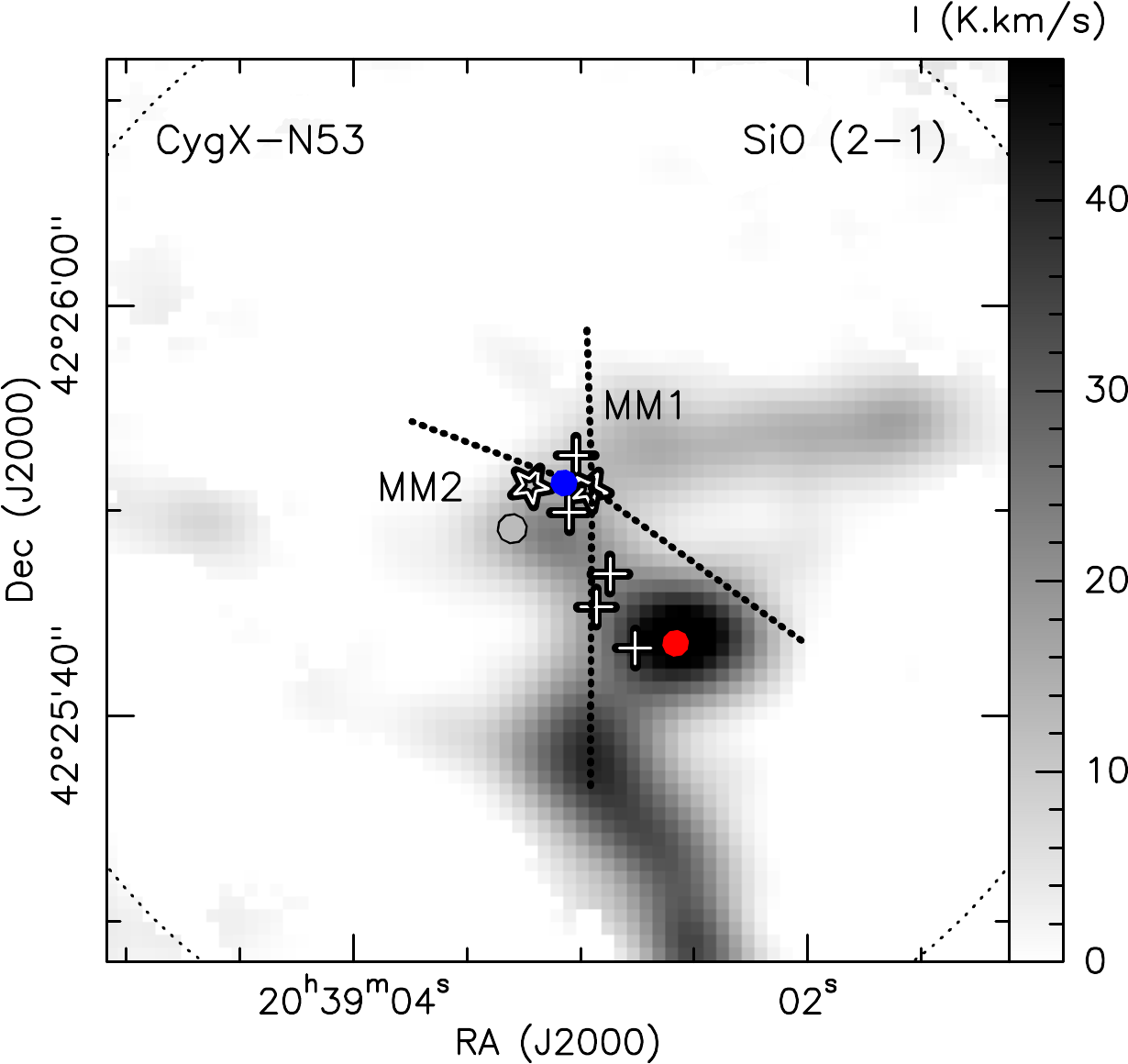}
	\hspace{-0.6cm}
	\includegraphics[width=0.245\textwidth]{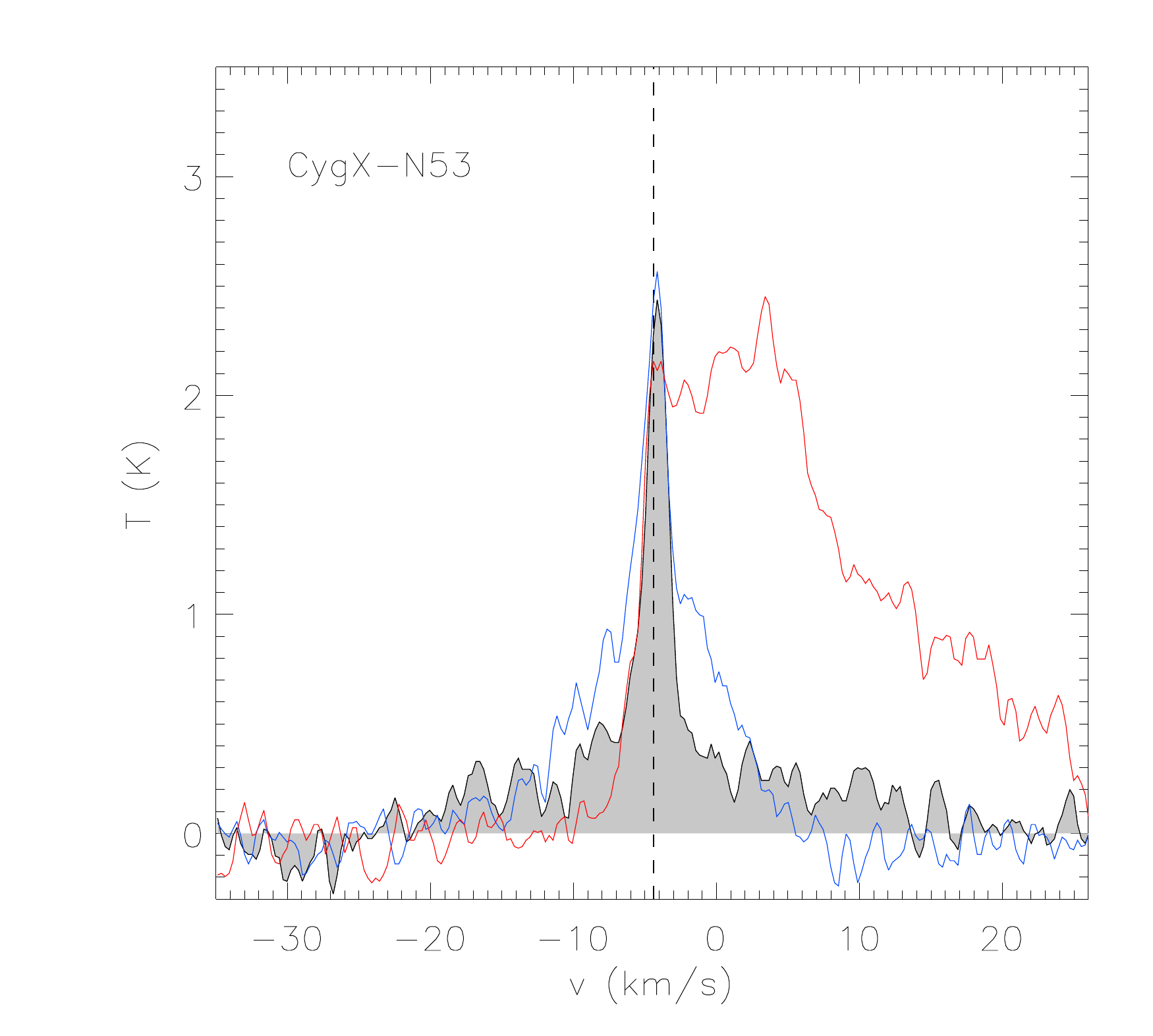}
	\hfill
	\includegraphics[width=0.27\textwidth]{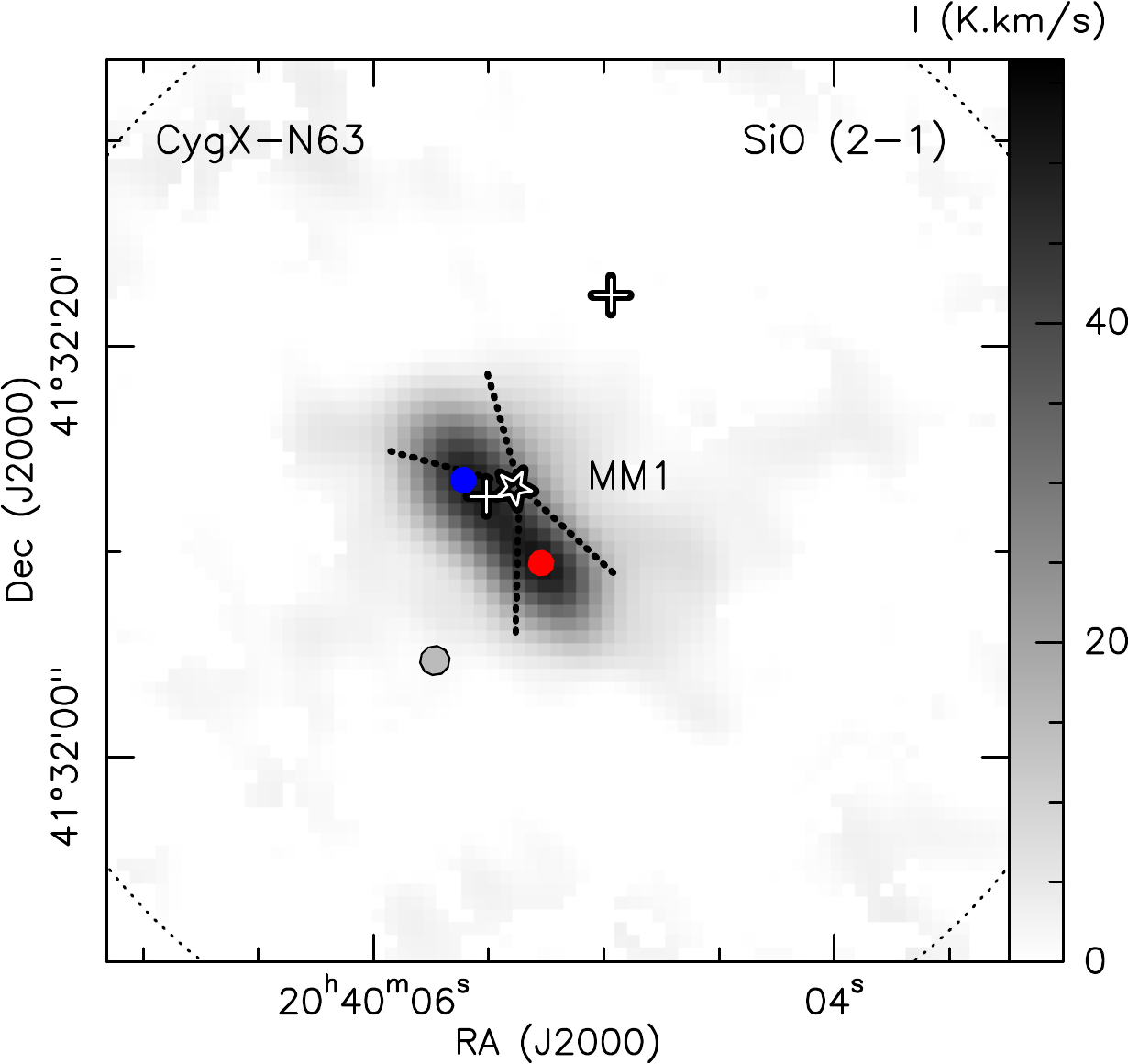}
	\hspace{-0.6cm}
	\includegraphics[width=0.245\textwidth]{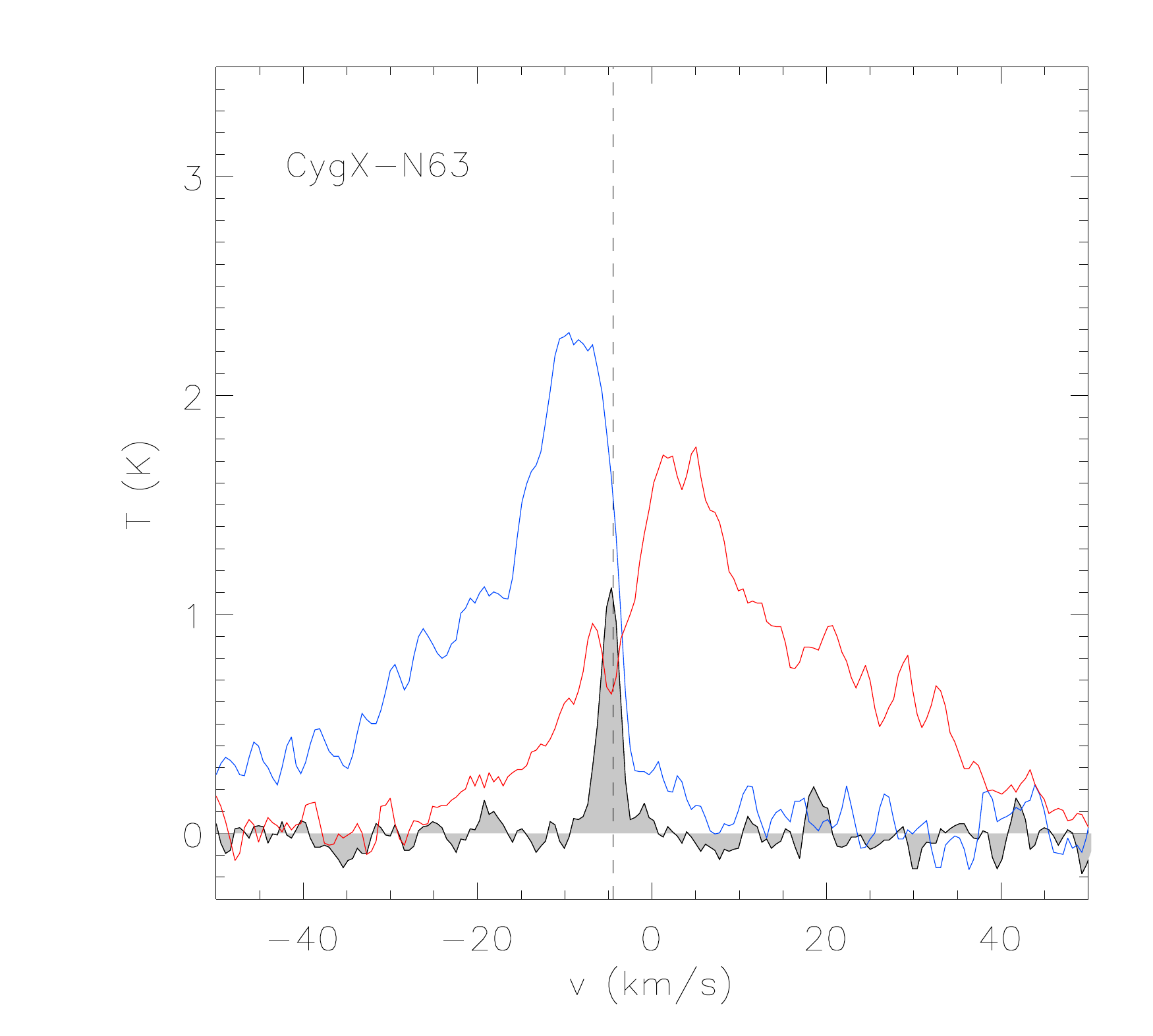}\\
	\caption[]{\small{SiO emission (in units of $T_{\rm mb}$) of the six MDCs studied. For each MDC we show, on the left, a greyscale showing the total integrated SiO emission and, on the right, three overlaid spectra extracted from the datacubes. The grey, blue, and red circles in the left-hand panels show the positions of the grey, blue, and red spectra displayed on the right. The grey spectra exemplify the narrow profile of the SiO emission, and the blue and red spectra are taken at positions where there is significant wing emission. Sources and outflow directions are as labelled in Figs.~\ref{fig:sio_co_highv} and \ref{fig:sio_co_highv2}.}}
	\label{fig:sio_indiv_spectra}}
\end{figure*}

\end{appendix}

\end{document}